\documentstyle[12pt,epsf,epsfig,fleqn]{article}
\newcommand{\fips}[1]{\epsffile{#1.eps}}
\newcommand{\be}{\begin{equation}}
\newcommand{\ee}{\end{equation}}
\newcommand{\lee}[1]{\label{#1} \end{equation}}
\newcommand{\bea}{\begin{eqnarray}}
\newcommand{\leea}[1]{\label{#1} \end{eqnarray}}
\newcommand{\eea}{\end{eqnarray}}
\newcommand{\nn}{\nonumber}
\newcommand{\eq}[1]{eq.~(\ref{#1})}
\newcommand{\Eq}[1]{Eq.~(\ref{#1})}
\newcommand{\eqs}[2]{eqs.~(\ref{#1}) and (\ref{#2})}

\newcommand{\fig}[1]{fig.~(\ref{#1})}
\newcommand{\loadeps}[1]{\epsfig{file=#1.eps,width=45mm}}
\newcommand{\floadeps}[1]{\epsfig{file=#1.eps,width=20mm}}
\newcommand{\diagform}[2]{\put(60,#1){\makebox(110,44.61)[l]{
  \begin{minipage}{110mm} #2 \end{minipage} }} }
\newcommand{\al}{\alpha}

\newcommand{\ga}{\gamma}
\newcommand{\de}{\delta}

\newcommand{\varep}{\varepsilon}

\newcommand{\et}{\eta}

\newcommand{\la}{\lambda}

\newcommand{\si}{\sigma}

\newcommand{\th}{\theta}

\newcommand{\ph}{\phi}

\newcommand{\ch}{\chi}

\newcommand{\om}{\omega}

\newcommand{\Ga}{\Gamma}
\newcommand{\De}{\Delta}
\newcommand{\La}{\Lambda}
\newcommand{\Si}{\Sigma}

\newcommand{\Om}{\Omega}
\def\bbbone{{\mathchoice {\rm 1\mskip-4mu l} {\rm 1\mskip-4mu l}
 {\rm 1\mskip-4.5mu l} {\rm 1\mskip-5mu l}}}

\begin{document}
\title{Modified similarity renormalization of Hamiltonians. \\
Positronium on the light front \thanks{Based on the talk presented
at International Workshop "New Nonperturbative Methods and Quantization
on the Light Cone", Les Houches, France, Feb.24-March 7, 1997}
\vspace*{.5cm}}
\author{E.~L.~Gubankova\thanks{E-mail address: 
elena@hal5000.tphys.uni-heidelberg.de},
 F.~Wegner \vspace*{.5cm}\\
\normalsize\it Institut f\"ur Theoretische Physik der Universit\"at 
Heidelberg \\
\normalsize\it Philosophenweg 19, D69120 Heidelberg, FRG}

\maketitle

\vspace*{1cm}
\begin{abstract}
Modified similarity renormalization (MSR) of Hamiltonians is proposed,
that performes by means of flow equations the similarity transformation 
of Hamiltonian in the particle number space. This enables to 
renormalize in the energy space the field theoretical Hamiltonian
and makes possible to work in a severe trancated Fock space for
the renormalized Hamiltonian.

Original works of Wegner in solid state physics has served us as the 
guiding principle in constructing the MSR scheme.

The renormalized to the second order effective $QED_{3+1}$
Hamiltonian on the light front is obtained. This Hamiltonian
reproduces in $|e\bar{e}>$ sector the standard singlet-triplet
splitting of positronium and recorves rotational symmetry
of the canonical theory. The lowest and next-to-lowest energy
states of positronium are almost independent on the cutoff
when in both cases the 'same' (state and sector independent) 
counterterms are included.

The electron (photon) mass counterterms are IR (collinear) finite
if all diagrams to the second order, arising from the flow equations
and normal-ordering Hamiltonian, are taken into account,
and vary with UV cutoff in accordance with $1$-loop
renormalization group equations.

Both approximations of perturbation theory and Fock space trancation
are under the control and can be improved systematically
within the proposed renormalization scheme.

\end{abstract}

\newpage
\section{Introduction}

About two years ago Glazek and Wilson proposed a new renormalization scheme 
for Hamiltonians, called similarity renormalization. The main idea of their
approach is to transform the regularized bare Hamiltonian, with a large cutoff
$\La$, into an effective Hamiltonian that has no matrix elements 
with energy jumps larger compared to the small cutoff $\la<\La$.

The initial field theoretical Hamiltonian
contains usually fluctuations (or states) of all 
multiple energy scales, that couple each other.
The effective Hamiltonian of Glazek and Wilson has by 
construction a band-diagonal form in the energy space.
This means that the different energy scales present in the initial 
Hamiltonian are decoupled in the effective Hamiltonian. This is the
main idea of renormalization group approach to take under the control
the effects of high-energy states, that gives rise to the effective 
low-energy theory.

In a relativistic field theory the Hamiltonian usually contains 
interactions which change the particle number of a state.
In the Hamiltonian matrix in Fock space representation many-body
states couple to few-body states. This means that relativistic Hamiltonian
is infinite in particle number space. The same is true 
for the renormalized in the energy space effective Hamiltonian
of Glazek and Wilson as far as it contains the particle number
changing interactions.

What kind of problems arise in solving the relativistic bound state problem
by diagonalizing finite dimensional Hamiltonian matrices numerically.
Suppose, that the energy renormalization is performed in the frame 
of Glazek,Wilson and we are left with the finite in energy space
but infinite in particle number space effective Hamiltonian.

First the calculation of effective Hamiltonian so far done are always
perturbative \cite{JoPeGl},\cite{BrPeWi}. This means that a kind 
of Tamm-Dancoff (TD) trancation,
which restricts the total number of particles in any intermediate
states, is performed. There are still effective induced interactions, 
together with the canonical one, which change the particle number by two
in the band-diagonal effective Hamiltonian. To restrict the Fock space
in this case
is to throw away the interactions corresponding to the diagrams with
more particles in the intermediate states. Therefore the narrow
range of the cutoff (band) $\la$, $m\al^2<\la<m\al$ for QED \cite{JoPeGl},
is always introduced in Glazek,Wilson aproach, where 
the correct and cutoff independent values of physical masses
are obtained. As far as the cutoff $\la$ shifted below this range,
the perturbative theory blows up and the contribution of higher orders
in coupling constant diagrams is needed to cure the situation \cite{JoPeGl}.
At least for the theories well defined in IR domain as QED
we need to elaborate the systematic procedure that avoids these 
difficulties. As was mentiond above the source of these difficulties
are the interactions which change the number of particles in a state.

Second, in Glazek,Wilson approach the elimination of far-off diagonal
matrix elements is performed in each particle number sector,
including the sectors that conserve the number of particles.
In particle number conserving sector the interactions are induced,
corresponding to the marginal relevant operators of the theory.
The elimination of such terms can cause the convergency problems \cite{Mi}.
Within the framework of Wegner's flow equations, this has already
been observed for the one-dimensional problem and solved by the proper
choice of the generator of unitary transformation \cite{We}.
This has served us as a guiding principle in constructing the MSR scheme.

The source of all these difficulties is the presence of 
interactions in the effective Hamiltonian, which change the number of 
particles in a state and mix different Fock components.
This causes also the general problems of the so-called
'sector-dependent' (occuring in a naive TD approximation \cite{HaOk})
and 'state-dependent' (occuring in the method of iterrative
resolvents \cite{TrPa}) counterterms.
Namely, it means,that the counterterms are needed to be added
which depend on the Fock sector \cite{HaOk} (and physical state \cite{TrPa})
to get the cutoff independent physical masses of states.

Therefore our aim is to modify the similarity scheme 
of Glazek, Wilson in a way to decouple different particle number
states, that means to eliminate the interactions which change
the number of particles in a state. Generally speaking,
we aim to simulate the effects not only of high-energy states
on the low-energy one, what conventional renormalization procedure
does,but also the effects of many-body states on the few-body one
by a set of effective interactions which do not change particle number
and do not couple to high-energy states either.
In this way we are able to control the approximations of the 
perturbation theory (or any other trancation made), and improve it
systematically.

\begin{figure}
$$
\fips{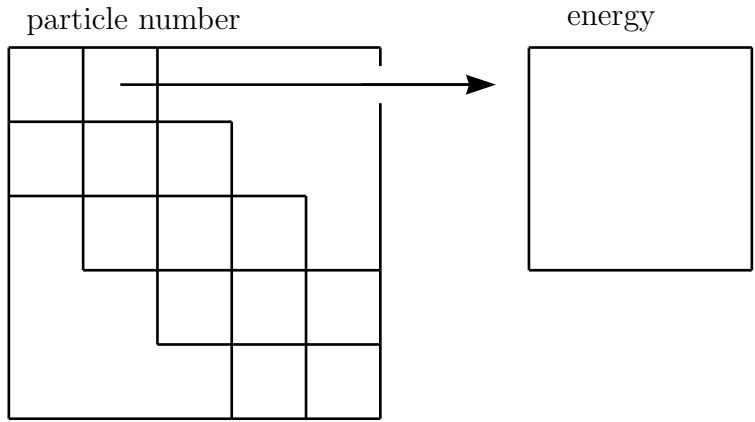}
\setlength{\unitlength}{0.240900pt}
\begin{picture}(0,0)
\put(-1000,650){\makebox(0,0){particle number}}
\put(-250,650){\makebox(0,0){energy}}
\end{picture}
$$
\caption{Pentadiagonal form of field theoretical Hamiltonian
in 'particle number' space (the case of one-body Hamiltonian).
Not squared region corresponds to zero matrix elements.}
\label{pentadiagonal}
\end{figure}

\begin{figure}
$$
\fips{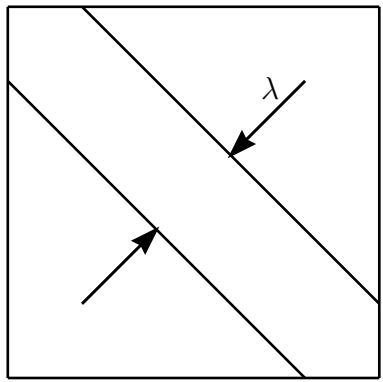}
\setlength{\unitlength}{0.240900pt}
\begin{picture}(0,0)
\put(-200,480){\makebox(0,0){$\lambda$}}
\end{picture}
$$
\caption{Similarity renormalization (SR) in 'energy space'
results to the band-diagonal form for effective Hamiltonian, where
all matrix elements are sqeezed in the energy band $|E_i-E_j|<\la$.}
\label{band-diagonal}
\end{figure}

\begin{figure}
$$
\fips{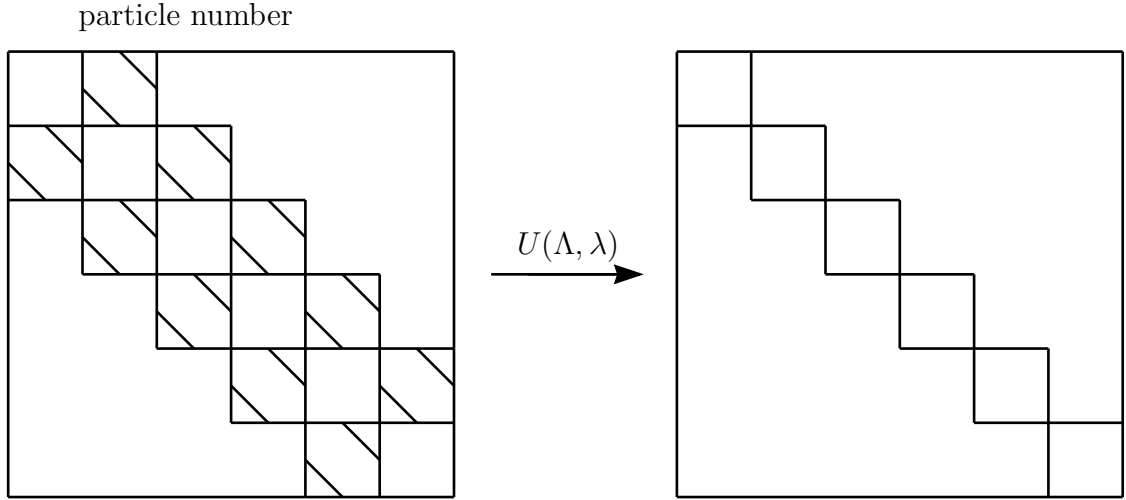}
\setlength{\unitlength}{0.240900pt}
\begin{picture}(0,0)
\put(-1500,780){\makebox(0,0){particle number}}
\put(-900,420){\makebox(0,0){$U(\La,\la)$}}
\end{picture}
$$
\caption{Modified similarity renormalization (MSR) 
of field theoretical Hamiltonians combines both similarity renormalization
in 'particle number' and 'energy' representations. For the finite value
of $\la$ (after the unitary transformation
$U(\La\rightarrow\infty,\la)$ is performed) the matrix elements 
of 'particle number changing' sectors 
are sqeezed in the energy band $|E_i-E_j|<\la$ on the left hand side 
picture and are eliminated completly as $\la\rightarrow 0$
(that corresponds to $U(\La\rightarrow\infty,\la\rightarrow 0)$)
on the right hand side picture. One ends up with the block-diagonal
in 'particle number space' renormalized effective Hamiltonian.}
\label{MSR}
\end{figure}

How does the 'modified similarity renormalization' (MSR) work?
Say we have the two-body  bare Hamiltonian, which changes the state
of at most two particles (the case of light-front QED, QCD).
The Hamiltonian is therefore the pentadiagonal block matrix 
in the representation of different number of particle-hole states.
Also there is a large number,restricted by the UV-cutoff $\La$,
of states in the energy representation within each block 
fig.\ref{pentadiagonal},
so that there are (infinitely) many 'energy states' and 'Fock components'
in the field theoretical Hamiltonian.The idea of the MSR  is to perform
similarity transformation, analogous to similarity transformation
in the 'energy space' of Glazek,Wilson, but in the 'particle number space'.
This results in effective Hamiltonian with a block-diagonal structure  
in a 'particle number space'.  How does it work technically?
We apply the flow equations in a way to eliminate
the far-off diagonal elements in the energy space (the matrix
elements with large energy jumps $|E_i-E_j|>\la$) only for those 
blocks that change the number of particles. This is to be done
continuously in a differential way for all cutoffs from $\la$
equal to bare cutoff up to $\la$ tending to zero. For the value
of $\la=0$ all interactions, canonical and induced, which change 
the number of particles are completly eliminated (except for diagonal
in the 'energy space' matrix elements, that do not contribute
in physical processes). 
One can say that instead of {\bf band-diagonalization} in the 'energy space'
in each 'particle number block' fig.\ref{band-diagonal}
(similarity renormalization (SR) of
Glazek, Wilson) we perform {\bf block-diagonalization} in 
'particle number space', where the elimination of the blocks, which change
the number of particles, is going in the 'energy space' 
fig.\ref{MSR}
(modified similarity renormalization (MSR)).

Such construction of MSR brings several advantages with respect to
the known methods to get the effective theories.
First, we are able to treat different energy scales in sequence,
what is usual for the renormalization approach. This solves the problem
with divergencies, appearing when single unitary transformation (by one step)
is performed and all energy scales are treated at once \cite{HaOk},
\cite{Fr}.
Second,in MSR we hope to control both effects of high-energy
and many-body states.

One can distiguish then two 'renormalizations', in 'energy' and
'particle number' space, that are closely related to each other
in a complicated way.

What are the main differences between two approaches (SR and MSR)
and  consequences for the solution of bound state problem?
First,in MSR there are still matrix elements with large energy jumps
in particle number conserving sectors.
Second, it is possible in MSR to eliminate {\bf completly}
the interactions which change the number of particles, i.e.
the limit of $\la\rightarrow 0$ for the effective Hamiltonian 
is well defined (as $\la\rightarrow 0$ no patalogies occure in MSR
in contrast to the case of band-diagonal effective Hamiltonian 
in SR).

We define the effective Hamiltonian
\be
H_{eff}(\La)=\lim_{\la\rightarrow 0}H_B(\la,\La)
\ee
where $H_B(\la,\La)$ is obtained by the unitary transform of
the bare Hamiltonian
\be
H_B(\la,\La)=U(\la,\La)H_B(\La)U^+(\la,\La)
\ee
The unitary transformation $U(\la,\La)$ is specified 
in the flow equation method below.
The renormalized effective Hamiltonian is defined
\be
H_{ren}=\lim_{\La\rightarrow\infty}H_{eff}(\La)
\ee
Accepting the above definitions we obtain the renormalized effective
Hamiltonian
in the 'particle number conserving' sectors, which do not
depend explicitly on the UV-cutoff.
 Implicit dependence on the cutoff
due to the renormalization group running of coupling constants
and masses is present. We leave aside in this work the question
on the new types of counterterms,probably appeared due to the new generated
interactions in effective Hamiltonian and which can not be absorbed
by the running of relevant and marginal operators of the canonical
theory \cite{Pe}.      
This means that the physical results become insensitive
to the cutoffs.

The only problem that can arise now by numerical diagonalization
of the effective Hamiltonian matrix, that depend on the large UV-cutoff
$\La$ as size of the matrix and does not depend on the artificial
cutoff $\la<\La$, is the neccesity to introduce the 'sector-dependent'
\cite{HaOk} or 'state-dependent' \cite{TrPa} counterterms to get the cutoff
independent physical masses of states. This problem is solved
in the method of MSR, where the resulting effective Hamiltonian
is diagonal in 'particle number space' and different particle number
(Fock) states are completly decoupled.
Explicitly, the lowest and next-to-lowest energy states
are almost independent on the cutoff when in both cases
the same counterterms are included. 
This is true despite that we perform the calculations perturbatively and 
work in a severly trancated space \cite{GuTrWe}.

In this work we give the main points of MSR and consider
QED on the light front to illustrate the method.
The key ingredient of MSR are flow equations that perform
infinitesimal unitary transformation in a continuous way \cite{We}.
On the other hand Wegner's flow equations can be used
in SR scheme to diagonalize a given Hamiltonian in the energy space
approximately. If in that case one performs the integration
of the flow equations only to a finite value of the flow parameter,
one obtains a Hamiltonian with a band-diagonal structure \cite{GuWe}.
In MSR scheme we use the flow equations to transform 
the Hamiltonian into a block-diagonal form in the particle number space.
Both approaches (SR and MSR) are usefull depending on the problem
one wants to treat. The purpose of this paper is to illustrate
the advantages of MSR manifest in solving the bound state problem.

\section{Flow equations and MSR}.

The flow equation is written
\be
\frac{dH(l)}{dl}=[\eta(l),H(l)]
\ee
This is the differential form of the continuous unitary transformation
acting on the Hamiltonian in general; $\eta(l)$ is the generator
of the transformation and $l$ is the flow parameter.
The aim of the Wegner's flow equations is to bring the initial
Hamiltonian matrix $H(l=0)$, written in the energy space,
to a diagonal (or block-diagonal where exact diagonalization
is not possible) form. The finite values of the flow parameter
$l$ correspond to intermediate stages of diagonalization
with the band-diagonal structure of Hamiltonian (SR) \cite{GuWe}.
The result of the procedure at $l\rightarrow\infty$ is the
'approximately' diagonal Hamiltonian, where off-diagonal matrix
elements do not vanish, but they become exponentially small.
Also the transformation is designed in a way to avoid
small energy denominators usually present in the perturbation theory.

What is the choice of the generator $\eta(l)$ that performs such 
transformation? Break the Hamiltonian into 'diagonal' and 'rest'
parts $H=H_d+H_r$. Then the prescription of Wegner is \cite{We}
\be
\eta(l)=[H_d(l),H_r(l)]
\ee
At this step the unitary transformation is defined completly. The only 
freedom left is the princip of separation into 'diagonal' and 
'rest' parts. It depends on the problem one wants to treat,
how to separate the relevant information (degrees of freedom),
carried by $H_d$, and irrelevant one, put in $H_r$ and eliminated
by the unitary transformation.

Our goal is to transform the Hamiltonian into the blocks with the 
same number of~~~~ (quasi)particles (MSR). 
This means, that we define the 'diagonal' part $H_d$ as conserving
the number of particles part before and after interaction,
and the 'rest' $H_r$ as particle number changing part.
In the case of QED(QCD),
where the electron-photon (quark-gluon) coupling is present,
the number of photons (gluons) is conserved in each block of the 
effective Hamiltonian.

As a result of unitary transformation the new interactions are induced
(see further). 
They correspond to the marginal relevant operators of the theory,
i.e. are irrelevant at $l=0$ and manifest themself only at finite value 
of $l$. This means, that these terms must be added to the canonical
Hamiltonian at $l\neq 0$, and therefore give rise to the new terms 
in the generator of transformation $\eta(l)$. This in its turn
generates new interactions again.

To be able perform the calculations analytically
we proceed further in perturbative frame and trancate
this series assuming the coupling constant is small.

As illustration of the method we consider in the work
QED on the light front, therefore we proceed further in application
to it. For any finite value of $l$ one has
\be
H(l)=H_{0d}+H_r^{(1)}+H_d^{(2)}+H_r^{(2)}+...
\ee
where the subscript above denotes the order in bare coupling constant,
$H^{(n)}\sim e^n$; indices 'd'-diagonal, 'r'-rest parts correspondingly.
The part $H_{0d}$ is the free Hamiltonian, corresponding to the single
particle energies with the structure in secondary quantization
$a^+a,b^+b (d^+d)$; 
$H_r^{(1)}$ denotes electron-photon coupling (of the type $a^+b^+b$);
$H_d^{(2)}$ is the second order diagonal part of the Hamiltonian,
having the structure $b^+d^+bd, b^+b^+bb, d^+d^+dd$ (in the light front
this corresponds to the canonical instantaneous (seegull) and new generated
to the second order interactions in 'diagonal' sector).

Note, that the diagonal part in MSR scheme  is not only the free
Hamiltonian but the particle number conserving part of the 
effective Hamiltonian, which contains also corresponding interactions.
The choice of only $H_0$ for the diagonal part give rise
to the band-diagonal structure of the effective Hamiltonian
in each Fock sector in SR scheme \cite{GuWe}.

The generator of transformation is
\be
\eta(l)=[H_d,H_r]=[H_{0d},H_r^{(1)}]+[H_{0d},H_r^{(2)}]+...=
\eta^{(1)}+\eta^{(2)}+...
\ee
To the second order the flow equation is written
\be
\frac{dH(l)}{dl}=[\eta,H]=
[[H_{0d},H_r^{(1)}]H_{0d}]+[[H_{0d},H_r^{(1)}]H_r^{(1)}]+
[[H_{0d},H_r^{(2)}]H_{0d}]+...
\ee
The new terms of higher orders $e^n$ than present in the canonical
Hamiltonian are generated by flow equations.

We choose for the basis the single particle states
\be
H_{0d}|i>=E_i|i>
\ee
Then in matrix form one has
\bea
&& \eta_{ij}=(E_i-E_j)H_{rij}^{(1)}+(E_i-E_j)H_{rij}^{(2)}+...\nn\\
&& \frac{dH_{ij}}{dl}=-(E_i-E_j)^2H_{rij}^{(1)}
+[\eta^{(1)},H_r^{(1)}]_{ij}-(E_i-E_j)^2H_{rij}^{(2)}+...
\eea
Neglecting the dependence of the single particle energies on the flow
parameter $E_i(l)$, one has to the first order
\bea
&& \frac{dH_{rij}^{(1)}}{dl}=-(E_i-E_j)^2H_{rij}^{(1)}\nn\\
&& H_{rij}^{(1)}(l)=H_{rij}^{(1)}(l=0){\rm e}^{-(E_i-E_j)^2l}
=H_{rij}^{(1)}(\la=\La\rightarrow\infty){\rm e}^{-\frac{(E_i-E_j)^2}{\la^2}}
\eea
Here we have used the physical sence of the flow parameter $l$.
Namely, in SR scheme, it defines the size of the band $\la$,
corresponding to the UV-cutoff, where the effective
Hamiltonian is defined ($|E_i-E_j|<\la$) \cite{GuWe}.
The connection between these two values is
\be
l=\frac{1}{\la^2}
\ee
In the MSR scheme only the interactions, which change the number
of particles, are defined in band size $\la$, while the particle number
conserving part of the effective Hamiltonian exist everywhere
(matrix elements with all energy differences are present).

For the following analyse it is useful to introduce the similarity
function
\be
f_{ij}(l)=e^{-(E_i-E_j)^2l}={\rm e}^{-\frac{(E_i-E_j)^2}{\la^2}}
\ee
which characterizes the behavior (fall off) of the leading order
'rest' interaction with the cutoff $\la$.

We define the initial Hamiltonian at the bare UV-cutoff 
$\La\rightarrow\infty$ (corresponding to the flow parameter $l=0$)
and scale it by flow equations down to the cutoff $\la<\La$.
The resulting effective Hamiltonian is defined in the limit
$\la\rightarrow 0$ ($l\rightarrow\infty$), where off-diagonal
matrix elements of the particle number changing interactions are
completly eliminated. It turns out that the diagonal elements
of these interactions contribute the trivial terms to the 
'diagonal' sectors in particle space (see further).
Therefore one can assume that the effective Hamiltonian at
$\la\rightarrow 0$ has only particle number conserving interactions,
i.e. has the block-diagonal form in particle space.   

To the second order one has to distinguish the 'diagonal' and 'rest'
terms. For the 'rest' part one has
\be
\frac{dH_{rij}^{(2)}}{dl}=[\eta^{(1)},H^{(1)}]_{rij}
-(E_i-E_j)^2H_{rij}^{(2)}
\ee
where index 'r' by $[\eta^{(1)},H^{(1)}]_{r}$ defines
the particle number changing part of the commutator. Introduce
\be
H_{rij}^{(2)}(l)={\rm e}^{-(E_i-E_j)^2l}\tilde{H}_{rij}^{(2)}(l)
\ee
i.e. the 'rest' part is defined in the energy band $\la=1/\sqrt{l}$.
Then the solution reads
\be
\tilde{H}_{rij}^{(2)}(l)=\tilde{H}_{rij}^{(2)}(l=0)
+\int_0^l dl'{\rm e}^{(E_i-E_j)^2l'}[\eta^{(1)},H^{(1)}]_{rij}(l')
\ee
For the 'diagonal' part one has
\be
\frac{dH_{dij}^{(2)}}{dl}=[\eta^{(1)},H^{(1)}]_{dij}
\ee
and the solution is
\be
H_{dij}^{(2)}(l)=H_{dij}^{(2)}(l=0)
+\int_0^l dl'[\eta^{(1)},H^{(1)}]_{dij}(l')
\ee
Note, that though in general the commutator $[[H_{0d},H_d^{(2)}]H_{0d}]$
is not zero, it is not present in the flow equation due to the definition 
of the diagonal part. The corresponding commutator 
$[[H_{0d},H_r^{(2)}]H_{0d}]$ in the 'nondiagonal' sector insures the band-diagonal
form for the 'rest' interaction and also gives rise to the different
structure of generated interaction (the integral term) in 'rest' 
and 'diagonal' sectors. 

The commutator $[\eta^{(1)},H^{(1)}]$ gives rise to the new terms 
to the second order in bare coupling $e$. In the case of QED it induces
the new types of interactions, corresponding to the marginal relevant
operators of the theory, and generates the renormalization group corrections
to the electron (photon) masses. The coupling constant starts to run to
the third order in $e$.

Further we consider in 'diagonal' sector all contributions
to the second order to protect gauge invariance, though
in practical calculations we focus separately on the problem
of new generated interactions and calculation of corresponding
counterterms.

Note, that the MSR scheme (and also SR) enables to build the effective
low energy Hamiltonian together with all, 'canonical' and possible to
appear 'new', counterterms, found (from the coupling coherence condition 
\cite{Pe},\cite{PeWi})
order by order in coupling constant $e$. This defines the renormalized 
Hamiltonian used for the solution of bound state problem numerically.

\section{Renormalized effective Hamiltonian}

Using the flow equations, we derive in this section the renormalized
to the second order in bare coupling effective Hamiltonian 
for $QED_{3+1}$ on the light front.

\subsection{Canonical light-front $QED_{3+1}$ Hamiltonian}

We start with the canonical light-front QED Hamiltonian $H_{can}$,
devided into free and interacting parts 
\be
P^-=H_{can} = \int dx^-d^2x^{\bot}({\cal H}_0+{\cal H}_I)
\; . \lee{ch1}
In light-front gauge $A^+=A^0+A^3=0$, the constrained degrees of freedom
$A^-$ and $\psi _-$ ($\psi=\psi _+ +\psi_- ,\psi_{\pm}=\Lambda_{\pm}\psi$)
can be removed explicitly; this gives the light-front gauge Hamiltonian
defined through the independent physical fields 
$A_{\bot}$ and $\psi_+$ only \cite{ZhHa}
\be
{\cal H}_0=\frac{1}{2}(\partial ^iA^j)(\partial ^iA^j)+
\xi^+ \left( \frac{-\partial _{\bot}^2+m^2}{i\partial ^+} \right) \xi
\; , \lee{ch2}
\be
{\cal H}_I
 = {\cal H}_{ee \gamma} + {\cal H}_{ee \gamma \gamma} + {\cal H}_{eeee}
\lee{ch3}
and 
\be
{\cal H}_{ee \gamma}=e\xi^+ \left[ -2(\frac{\partial ^{\bot}}{\partial ^+}
\cdot A^{\bot})+\sigma \cdot A^{\bot}\frac{\sigma \cdot\partial ^{\bot}+
m}{\partial ^+}+\frac{\sigma \cdot \partial ^{\bot}+m}{\partial ^+}
\sigma \cdot A^{\bot} \right] \xi
\; , \lee{ch4}
\be
{\cal H}_{ee \gamma \gamma}= -ie^2 \left[ \xi ^+ \sigma\cdot A^{\bot}
\frac{1}{\partial ^+}(\sigma \cdot A^{\bot}\xi) \right]
\; , \lee{ch5}
\be
{\cal H}_{eeee}=2e^2 \left[ \left( \frac{1}{\partial ^+}(\xi ^+ \xi) \right)
 \left( \frac{1}{\partial ^+}(\xi ^+ \xi) \right) \right]
\; , \lee{ch6}
where $\{\sigma ^i\}$ are the standard $2 \times 2$ Pauli matrices, and
$\partial ^+=2\partial _- = 2\frac{\partial}{\partial x^-}$.
We have used the two-component representation for fermion fields introduced
by Zhang and Harindranath \cite{ZhHa} $\psi_+={\xi \choose 0}$.
To simplify the calculations we rewrite all interactions through creation
and annihilation operators. This turns out to be useful in the flow equations
formalism, \cite{Mi}.

Following standard quantum field theory procedure
we use the momentum-space representation for the field operators, \cite{PeWi}
and \cite{ZhHa},
\bea
\xi(x) &=& \sum_s \chi_s \int\frac{dp^+d^2p^{\bot}}{2(2\pi)^3}\theta(p^+)
(b_{p,s}e^{-ipx}+d_{p,\bar{s}}e^{ipx}) \nn \\
A^i(x) &=& \sum_{\lambda}\int \frac{dq^+ d^2q^{\bot}}{2(2\pi)^3}
\frac{\theta (q^+)}{\sqrt{q^+}}(\varepsilon _{\lambda}^i a_{q,\lambda}
e^{-iqx}+h.c.)
\; , \leea{ch8}
where spinors are $\chi _{1/2}^{tr}=(1,0)$, $\chi_{{-1/2}^{tr}}=(0,1)$,
with $\bar{s}=-s$
and polarization vectors $\varepsilon_1^i=\frac{-1}{\sqrt{2}}(1,i)$,
$\varepsilon_{-1}^i=\frac{1}{\sqrt{2}}(1,-i)$;
the integration running over the $p^+\ge 0$ only these
states, that are allowed the light-front theory.  

The corresponding (anti)commutation relations are
\bea
& \{ b_{p,s},b_{p',s'}^+\}=\{d_{p,s},d_{p',s'}^+\}=\bar{\de}_{p,p'}\delta_{ss'} & \nn \\
& [a_{q,\lambda},a_{q',\lambda '}^+]=\bar{\de}_{q,q'} \delta_{\lambda,\lambda '} &
\; , \leea{ch10}
where 
\be
\bar{\de}_{p,p'}\equiv 2(2\pi)^3\delta(p^+-p'^+)\delta^{(2)}
(p^{\bot}-p'^{\bot}) 
\; . \lee{ch11}
The light-front vacuum has trivial structure for both boson and fermion
sectors, namely $a_q|0>=0$; $b_p|0>=0$, simpifying the 
analitical calculations.
The normalization of states is according to
\be
<p_1,s_1|p_2,s_2>=\bar{\de}_{p_1,p_2}\delta_{s_1,s_2}
\; , \lee{ch12}
where $b_{p,s}^+|0>=|p,s>$.

Making use of the field representation \eq{ch8},
we have the following Fourier transformed for

\noindent
the {\bf free} Hamiltonian 
\be
H_0=\sum _s\int\frac{dp^+ d^2p^{\bot}}{2(2\pi)^3}\theta (p^+)\\
\frac{p^{\bot 2}+m^2}{p^+} (b_{p,s}^+b_{p,s}+d_{p,s}^+d_{p,s})+\\
\sum_{\lambda}\int\frac{dq^+d^2q^{\bot}}{2(2\pi)^3}\theta(q^+)\\
\frac{q^{\bot 2}}{q^+}a_{q,\lambda}^+a_{q,\lambda}
\; , \lee{ch13}

\noindent
the leading order $O(e)$-{\bf the electron-photon coupling} 
\bea
H_{ee\gamma}&=&\sum_{\lambda s_1s_2}\int_{p_1p_2q} \!\!\!
  [g_{p_1p_2q}^*(l) \varepsilon_{\lambda}^i\tilde{a}_{q}
 + g_{p_1p_2q}(l) \varepsilon_{\lambda}^{i *}\tilde{a}_{-q}^+]
(\tilde{b}_{p_2}^+\tilde{b}_{p_1} +\tilde{b}_{p_2}^+\tilde{d}_{-p_1}^+ +
\tilde{d}_{-p_2}\tilde{b}_{p_1} +\tilde{d}_{-p_2}\tilde{d}_{-p_1}^+)\nonumber\\
& &\times\chi_{s_2}^+\Gamma_l^i(p_1,p_2,-q)\chi_{s_1} \bar{\de}_{q,p_2-p_1}
\; , \leea{ch14}
where
\be
\Gamma_l^i(p_1,p_2,q)=2\frac{q^i}{q^+}-
\frac{\sigma\cdot p_2^{\bot}-im}{p_2^+}\sigma^i-
\sigma^i\frac{\sigma\cdot p_1^{\bot}+im}{p_1^+}
\; , \lee{ch15}
The $l$-dependence of the vertex $\Ga_l^i$ arises from the dependence of
light-front energies (masses) on the flow parameter.
Further we have for the {\bf instantaneous} interactions of the order $O(e^2)$
\bea
H_{eeee}^{inst}&=&\sum_{s_1s_2s_3s_4}\int_{p_1p_2p_3p_4}
\!\!\!\!\!g_{p_1p_2p_3p_4}^{eeee}(l)
(\tilde{b}_{p_3}^+ +\tilde{d}_{-p_3})(\tilde{b}_{p_4}^+ +\tilde{d}_{-p_4})
(\tilde{b}_{p_1} +\tilde{d}_{-p_1}^+)(\tilde{b}_{p_2} +\tilde{d}_{-p_2}^+)
\nonumber \\
& &\times\chi_{s_3}^+\chi_{s_4}^+\frac{4}{(p_1^+-p_3^+)^2}\chi_{s_1}\chi_{s_2}
\bar{\de}_{p_3+p_4,p_1+p_2}
\leea{ch16}
and
\bea
H_{ee\gamma\gamma}^{inst}&=&\sum_{s_1s_2\lambda_1\lambda_2}
\int_{p_1p_2q_1q_2}\!\!\!\!\!g_{p_1p_2q_1q_2}^{ee\gamma \gamma}(l)
(\varepsilon_{\lambda_1}^{i *}\tilde{a}_{q_1}^+ +
\varepsilon_{\lambda_1}^i\tilde{a}_{-q_1})
(\varepsilon_{\lambda_2}^j\tilde{a}_{q_2}+
\varepsilon_{\lambda_2}^{j *}\tilde{a}_{-q_2}^+)
(\tilde{b}_{p_2}^+ +\tilde{d}_{-p_2})(\tilde{b}_{p_1}+\tilde{d}_{-p_1}^+)
\nonumber\\
& &\times\chi_{s_2}^+\frac{\sigma^j\sigma^i}{(p_1^+-q_1^+)}\chi_{s_1}
\bar{\de}_{p_1+q_2,q_1+p_2}
\; ; \leea{ch17}
here 
\bea
& \tilde{a}_q\equiv a_{q,\lambda}\frac{\theta(q^+)}{\sqrt{q^+}}, \qquad
 \left[ \tilde{a}_{-q}\equiv a_{-q,\lambda}\frac{\theta(-q^+)}{\sqrt{-q^+}} \right]
 \; , & \nn \\
& \tilde{b}_p\equiv b_{p,s}\theta(p^+), \qquad
 \tilde{d}_p\equiv d_{p,\bar{s}}\theta(p^+)
\; , \leea{ch19}
and the $\bar{\de}$-simbol stands for the function defined in \eq{ch11}, the
short notation for the integral is
\be
\int_p\equiv\int\frac{dp^+d^2p^{\bot}}{2(2\pi)^3}
\; . \lee{ch20}

In the formulas above we write explicitly the momentum dependence of 
the coupling constants as long as $l\neq 0$. The initial conditions 
for the couplings are defined at the value of the bare cutoff 
$\La\rightarrow\infty$($l_{\La}=0$), namely   
\be
\lim_{\La\rightarrow\infty}g^{ee\gamma}(l_{\La})=e
\lee{ch21}
and for both instantaneous interaction couplings
\be 
\lim_{\La\rightarrow\infty}g^{inst}(l_{\La})=e^2
\; ; \lee{ch22}
this corresponds to the couplings of the canonical theory.

\subsection{Flow equations for $QED_{3+1}$ Hamiltonian
on the light front} 

\subsubsection{\label{3.2.1}Generated interaction in $|e\bar{e}>$ sector}

Following the procedure outlined in the second section, the leading order
generator of the unitary transformation is
\bea
\eta^{(1)}(l)&=&\sum_{\lambda s_1s_2}\int_{p_1p_2q}\!\!\!(\eta_{p_ip_f}^*(l)
\varepsilon_{\lambda}^i\tilde{a}_q+
\eta_{p_ip_f}(l)\varepsilon_{\lambda}^{i *}\tilde{a}_{-q}^+)
(\tilde{b}_{p_2}^+\tilde{b}_{p_1}+\tilde{b}_{p_2}^+\tilde{d}_{-p_1}^+ +
\tilde{d}_{-p_2}\tilde{b}_{p_1}+\tilde{d}_{-p_2}\tilde{d}_{-p_1}^+)\nonumber\\
& &\times\chi_{s_2}^+\Gamma_l^i(p_1,p_2,-q)\chi_{s_1} \bar{\de}_{q,p_2-p_1}
\; , \leea{gi1}\nn\\
\be
\eta_{p_ip_f}(l)=-\Delta_{p_ip_f}g_{p_ip_f}=
\frac{1}{\Delta_{p_ip_f}}\cdot\frac{dg_{p_ip_f}}{dl}
\; . \lee{gi2}
where $p_i$ and $p_f$ stand for the set of initial and final momenta,
respectively, and $\De_{p_ip_f}=\sum p_i^--\sum p_f^-$,
and the light-front fermion energy is 
\mbox{$p^- = \frac{p^{\bot 2} + m^2}{p^+}$},
the photon one \mbox{$q^- = \frac{q^{\bot 2}}{q^+}$}.
In previous notations $g_{p_ip_f}=g_{p_1p_2q}$.
Further we calculate the bound states of positronium.
In what follows we consider in $|e\bar{e}>$ sector

\noindent
the {\bf generated interaction} to the first nonvanishing order
\be
H_{e\bar{e}e\bar{e}}^{gen}=\sum_{s_1\bar{s}_2s_3\bar{s}_4}
\int_{p_1p_2p_3p_4}
V_{p_ip_f}^{gen}(l)b_{p_3}^+d_{p_4}^+d_{p_2}b_{p_1}
\chi_{s_3}^+\chi_{\bar{s}_4}^+\chi_{\bar{s}_2}\chi_{s_1}
\bar{\de}_{p_1+p_2,p_3+p_4}
\; , \lee{gi3}
with the initial condition 
$\lim_{\La\rightarrow\infty}V_{p_ip_f}^{gen}(l_{\La})=0$,

\noindent
and the {\bf instantaneous interaction}
\be
H_{e\bar{e}e\bar{e}}^{inst}=\sum_{s_1\bar{s}_2s_3\bar{s}_4}
\int_{p_1p_2p_3p_4}
V_{p_ip_f}^{inst}(l)b_{p_3}^+d_{p_4}^+d_{p_2}b_{p_1}
\chi_{s_3}^+\chi_{\bar{s}_4}^+\chi_{\bar{s}_2}\chi_{s_1}
\bar{\de}_{p_1+p_2,p_3+p_4}
\; , \lee{gi4}
where 
\bea
&& V_{p_ip_f}^{inst}(l) = g_{p_ip_f}^{inst}(l) \, 
\frac{4}{(p_1^+-p_3^+)^2}\nn\\
&& \lim_{\La\rightarrow\infty}g_{p_ip_f}^{inst}(l_{\La})=e^2
\; , \leea{gi5}
The order of the field operators in both interactions
satisfies the prescription of standard Feynmann rules in the
$|e\bar{e}>$ sector.

To the leading order we neglect the $l$ dependence of light-front energies
in the interactions, that
enables to write the flow equations for the corresponding couplings.
 
The flow equations to the first (for the electron-photon coupling) and
second (for the instantaneous and generated interactions) orders are 
\bea
\frac{dg_{p_ip_f}(l)}{dl}&=&-\Delta_{p_ip_f}^2g_{p_ip_f}(l)\nonumber\\
\frac{dg_{p_ip_f}^{inst}(l)}{dl}&=& 0\\
\frac{dV_{p_ip_f}^{gen}(l)}{dl}&=&<[\eta^{(1)}(l),H_{ee\gamma}]>_{|e\bar{e}>}
\nonumber
\; , \leea{gi6}
where
\be
\Delta_{p_ip_f} = \sum p_i^- - \sum p_f^-
\lee{gi7}

The matrix element \mbox{$<[\eta^{(1)}(l) , H_{ee\gamma}]>_{|e\bar{e}>}$}
is understood as the corresponding commutator between
the free electron-positron states, namely
\mbox{$<p_3 s_3, p_4 \bar{s}_4|...|p_1 s_1, p_2 \bar{s}_2>$}.   

Note, that the instantaneous and generated interactions are changing
with the flow parameter $l$ according to the flow equations
in the 'particle number conserving' sector.

Renormalization group running of both (instantaneous and generated)
interactions starts to the order $O(e^4)$, and the electron-photon
coupling starts to run to the order $O(e^3)$.

Neglecting the dependence of the light-front energies on 
the flow parameter $l$, the solution reads 
\bea
g_{p_ip_f}(l)&=&f_{p_ip_f}\cdot e+O(e^3)\nonumber\\
g_{p_ip_f}^{inst}(l)&=&g_{p_ip_f}^{inst}(l_{\La}=0)=e^2+O(e^4)\\
V_{p_ip_f}^{gen}(l)&=&\int_0^l dl'
<[\eta^{(1)}(l'),H_{ee\gamma}(l')]>_{|e\bar{e}>}+O(e^4)
\nonumber\\
f_{p_ip_f}&=&{\rm e}^{-\De^2_{p_ip_f}l}=
{\rm e}^{-\frac{\De^2_{p_ip_f}}{\la^2}}
\nonumber
\; , \leea{gi8}
where the subscript $|e\bar{e}>$ means, that the commutator is considered 
in the electron-positron sector.
The electron-photon interaction exists in the band of size $\la$
($|\De_{p_ip_f}|<\la$), whereas the matrix elements of instantaneous and
generated interactions in $|e\bar{e}>$ sector are defined for all
energy differences.

We give further the explicit expressions for the generated interaction,
and details of calculations can be found in Appendix \ref{B}.
In what follows we use the notations of this Appendix.
 
The matrix elements  of the commutator $[\eta^{(1)},H_{ee\gamma}]$
in the exchange and annihilation channels are 
\be
<[\eta^{(1)},H_{ee\gamma}]>/\delta_{p_1+p_2,p_3+p_4} = 
\left\{ \begin{array}{l}
 M_{2ii}^{(ex)}  \frac{1}{(p_1^+-p_3^+)}(\eta_{p_1,p_3}g_{p_4,p_2}+
\eta_{p_4,p_2}g_{p_1,p_3}) \; , \\
\\
-M_{2ii}^{(an)}  \frac{1}{(p_1^++p_2^+)}(\eta_{p_1,-p_2}g_{p_4,-p_3}+
\eta_{p_4,-p_3}g_{p_1,-p_2}) \; ,
\end{array} \right. 
\lee{gi9}
\noindent
where
\bea
&& \eta_{p_1,p_2}(l) = e\cdot\frac{1}{\Delta_{p_1p_2}}
 \frac{df_{p_1,p_2}(l)}{dl} \nn \\
&&g_{p_1,p_2}(l) = e\cdot f_{p_1,p_2}(l)
\leea{gi10}
and $\Delta_{p_1,p_2} = p_1^- - p_2^- - (p_1-p_2)^-$.
The matrix elements $M_{2ii}$ between the corresponding spinors
in both channels are
\bea
M_{2ij}^{(ex)}&=&[\chi_{s_3}^+\Gamma_l^i(p_1,p_3,p_1-p_3)\chi_{s_1}]\,
[\chi_{\bar{s}_2}^+\Gamma_l^j(-p_4,-p_2,-(p_1-p_3))\chi_{\bar{s}_4}]
\nonumber\\
\\
M_{2ij}^{(an)}&=&[\chi_{s_3}^+\Gamma_l^i(-p_4,p_3,-(p_1+p_2))\chi_{\bar{s}_4}] \,
[\chi_{\bar{s}_2}^+\Gamma_l^j(p_1,-p_2,p_1+p_2)\chi_{s_1}]\nonumber
\leea{gi11}
that determine the spin structure of the generated interaction.

We plug the formulas for commutator $[\eta^{(1)},H_{ee\ga}]$ together
with the generator $\eta(l)$ and coupling constant $g(l)$, expressed
through the similarity function $f(l)$, into the formula for 
generated interaction. This gives rise in both channels

\bea
V_{gen}^{(ex)}(\la)&=&-e^2M_{2ii}^{(ex)}\frac{1}{(p_1^+-p_3^+)}
\left(\frac{\int_{\la}^{\infty}\frac{df_{p_1,p_3,\la'}}{d\la'}f_{p_4,p_2,\la'}d\la'}
{\De_{p_1,p_3}}+
\frac{\int_{\la}^{\infty}\frac{df_{p_4,p_2,\la'}}{d\la'}f_{p_1,p_3,\la'}d\la'}
{\De_{p_4,p_2}}\right)
\nonumber\\ 
\\
V_{gen}^{(an)}(\la)&=&e^2M_{2ii}^{(an)}\frac{1}{(p_1^++p_2^+)}
\left(\frac{\int_{\la}^{\infty}\frac{df_{p_1,-p_2,\la'}}{d\la'}f_{p_4,-p_3,\la'}d\la'}
{\De_{p_1,-p_2}}+
\frac{\int_{\la}^{\infty}\frac{df_{p_4,-p_3,\la'}}{d\la'}f_{p_1,-p_2,\la'}d\la'}
{\De_{p_4,-p_3}}\right)
\nonumber 
\; . \leea{gi13}
where in the integral we have neglected the dependence of light-front energies
on the cutoff $\la$ (that is the correction of order $O(e^2)$),
and the connection between flow parameter and cutoff ($l=1/\la^2$) is used.

Other unitary transformations can be performed, that aim to bring the field
theoretical Hamiltonian to the block-diagonal form in 'particle number space'.
The transformations we discuss further, performed in the frame of MSR 
(section \ref{3.3}),
act also in the energy space and differ from the flow equations mainly of
how fast the 'particle number changing' interactions are eliminated,
i.e. by the convergency of the procedure.

Here we note, that the form of second order generated interaction,
induced in MSR and written through the similarity functions $f_{p_ip_f}$,
is universal for all transformations, which will be discussed.
Specifying the similarity function we obtain the explicit form 
of generated interaction, induced by different unitary transformations 
(section \ref{3.3}).

Following the flow equation presciption we use here
\be
f_{p_1,p_2,\la}={\rm e}^{-\frac{\De_{p_1p_2}^2}{\la^2}}
\; . \lee{gi14}  
that gives for the {\bf generated interaction} in both channels
fig.\ref{feynrules}
\bea
V_{gen}^{(ex)}(\la) &\hspace{-5em}=\hspace{-5em}& 
-e^2M_{2ii}^{(ex)}\frac{1}{(p_1^+-p_3^+)} \,
\frac{\De_{p_1,p_3}+\De_{p_4,p_2}}{\De_{p_1,p_3}^2+\De_{p_4,p_2}^2}
\cdot(1-f_{p_1,p_3,\la}f_{p_4,p_2,\la})\nn \\
\\
V_{gen}^{(an)}(\la) &\hspace{-5em}=\hspace{-5em}& 
e^2M_{2ii}^{(an)}\frac{1}{(p_1^++p_2^+)} \,
\frac{\De_{p_1,-p_2}+\De_{p_4,-p_3}}{\De_{p_1,-p_2}^2+\De_{p_4,-p_3}^2}
\cdot(1-f_{p_1,-p_2,\la}f_{p_4,-p_3,\la})\nn 
\; , \leea{gi14}
By definition, given in the introduction, the renormalized effective
Hamiltonian is obtained in the limit of cutoff tending to zero 
($\la\rightarrow 0$). In this limit the electron-photon coupling,
present in generated interaction through the similarity functions $f_{p_ip_f}$,
is completly eliminated ~~~~$f_{p_ip_f}(\la\rightarrow 0)$ for 
$\De_{p_ip_f}\neq 0$, and generated interaction is given by corresponding
to $1$ in bracket expression that does not depend explicitly on the cutoff
$\la$. The implicit dependence as renormalization group running
of coupling constant $e$ and light-front energies (masses)
to the next orders is present.

The modified similarity transformation is constructed to avoid
divergencies in the form of small energy denominators,
present in the second order perturbative approach.
Namely, one has for the generated interaction in the exchange channel
in the case of energy conserving process ($\De_{p_ip_f}=0$)
\be
\frac{\De_{p_1,p_3}+\De_{p_4,p_2}}{\De_{p_1,p_3}^2+\De_{p_4,p_2}^2}
=\frac{\De_{p_1,p_3}+\De_{p_4,p_2}}{\De_{p_ip_f}+
2\De_{p_1,p_3}\De_{p_4,p_2}}
\rightarrow \frac{1}{2}
\left(\frac{1}{\De_{p_1,p_3}}+\frac{1}{\De_{p_4,p_2}}\right)
\lee{gi14a}
where  
\be
\Delta_{p_ip_f}\equiv p_1^-+p_2^--p_3^--p_4^-=
\Delta_{p_1,p_3}-\Delta_{p_4,p_2}
=\Delta_{p_1,-p_2}-\Delta_{p_4,-p_3}
\lee{gi14b}
due to the total momentum conservation in $'+'$ and 'transversal' directions.
The divergent contribution in the generated interaction 
when $\De_{p_1,p_3}\sim\De_{p_4,p_2}\sim 0$
is effectively cancelled  
by the factor in bracket containing similarity functions 
$(1-f_{p_1,p_3,\la}f_{p_4,p_2,\la})$.

Also it is obvious from \eq{gi14a}, that any energy differences
(i.e. $\forall \De_{p_ip_f}$) are permitted for matrix elements of generated
interaction in 'diagonal' sector.

For the further analyses we write the effective Hamiltonian at the finite
value of $\la$, performing at the end the limit $\la\rightarrow 0$.

We rewrite the generated to the second order interaction
in the form fig.\ref{reneebarint}
\bea
\tilde{V}_{gen}^{(ex)}(\la) &=& -e^2 N_{1,\la} \,
\frac{\tilde{\De}_1+\tilde{\De}_2}{\tilde{\De}_1^2+\tilde{\De}_2^2}
\cdot
\left( 1 - {\rm e}^{-\frac{\De_1^2+\De_2^2}{\la^4}}\right)\nn\\
\\
\tilde{V}_{gen}^{(an)}(\la) &=& e^2 N_{2,\la} \,
\frac{M_0^2+M_0^{'2}}{M_0^4+M_0^{'4}}
\cdot
\left( 1 - {\rm e}^{-\frac{M_0^4+M_0^{'4}}{\la^4}}\right)\nn
\; , \leea{fgi15}
where we have introduced
\bea
& P^{+ 2}M_{2ii,\la}^{(ex)} = - N_1 \quad;\qquad 
P^{+ 2}M_{2ii,\la}^{(an)} = N_2 & \nn \\
\nn \\
& \De_{p_1 p_3} = \frac{\De_1}{P^+} = \frac{\widetilde{\De}_1}
{(x' - x) P^+} \quad;\qquad
 \De_{p_4 p_2} = \frac{\De_2}{P^+} = \frac{\widetilde{\De}_2}
{(x' - x) P^+}; & \nn \\
\nn \\
& \De_{p_1,-p_2} = \frac{M^2_0}{P^+} \quad;\qquad
 \De_{p_4,-p_3} = \frac{{M'}^2_0}{P^+} & \nn \\
\leea{gi16}
(see Appendix \ref{B} for the explicit definition of these quantities
in the light-front frame).

The expression \eq{fgi15} is written for the rescaled value
of the potential, i.e. $V_{\la}=\tilde{V}_{\la}/P^{+2}$,
and the cutoff is defined in units of the total momentum $P^+$,
i.e. $\la\rightarrow\frac{\la^2}{P^+}$, with $l=1/\la^2$.
The spin structure of the interaction is carried by
the matrix elements $M_{2ii}$, defined in Appendix \ref{B}.

We summarize the {\bf instantaneous interaction} in both channels
to the order ${\bf O(e^2)}$ \fig{feynrules}
\bea
&& V_{inst}^{(ex)}(\la) = -\frac{4e^2}{(p_1^+-p_3^+)^2} \;
\delta_{s_1s_3} \delta_{s_2s_4} \; \nn \\
\\
&& V_{inst}^{(an)} = \frac{4e^2}{(p_1^++p_2^+)^2} \;
\delta_{s_1\bar{s}_2} \delta_{s_3\bar{s}_4} \;
\nonumber
\; , \leea{gi17}
where we have used
\mbox{$\chi_{s_3}^+ \chi_{\bar{s}_2}^+ \bbbone \chi_{s_1} \chi_{\bar{s}_4} =
\delta_{s_1 s_3} \delta_{s_2 s_4} + \delta_{s_1 \bar{s}_2} \delta_{s_3 \bar{s}_4}$}.
For the rescaled potential in the light-front frame Appendix \ref{B}
\eq{b14} we have 
\bea
&& \tilde{V}_{inst,\la}^{(ex)} = -\frac{4e^2}{(x-x')^2} \;
\delta_{s_1s_3} \delta_{s_2s_4} \;\nn \\
\\
&& \tilde{V}_{inst,\la}^{(an)} = 4e^2 \;
\delta_{s_1\bar{s}_2} \delta_{s_3\bar{s}_4} \;\nn
\; . \leea{gi18}
Further we use the generated and instantaneous interactions 
in $|e\bar{e}>$ sector, obtained in this section,
to calculate positronium bound state spectrum.

\subsubsection{\label{3.2.2}Renormalization issues}

As was discussed above the commutator $[\eta^{(1)},H_{ee\ga}]$ also
contributes to the self-energy term, giving rise to the renormalization
of fermion and photon masses to the second order.
The flow equation for the electron (photon) light-cone energy $p^-$ is
\be
\frac{dp^-}{dl}=<[\eta^{(1)},H_{ee\ga}]>_{self~energy}
\; , \lee{ri1}
where the matrix element is calculated between the single
electron (photon) states $<p',s'|...|p,s>$. We drop the finite part
and define $\de p_{\la}^- = p^-(l_{\la})-<|H_0|>$. Integration over the
finite range gives
\be
\de p_{\la}^--\de p_{\La}^-=\int_{l_{\La}}^{l_{\la}}
<[\eta^{(1)},H_{ee\ga}]>_{self~energy}dl'
=-\frac{(\de\Sigma_{\la}(p)-\de\Sigma_{\La}(p))}{p^+}
\; , \lee{ri2}
that defines the cutoff dependent self energy $\de\Sigma_{\la}(p)$.
The mass correction and wave function renormalization constant
are given correspondingly, cf. \cite{ZhHa}
\bea
\de m_{\la}^2 &=& \left. p^+\de p^- \right|_{p^2=m^2}
 =-\de\Si_{\la}(m^2) \nn \\
Z_2 &=& \left. 1 + \frac{\partial \de p^-}{\partial p^-} \right|_{p^2=m^2}
\; . \leea{ri3}
The on-mass-shell condition is defined through the mass $m$ in the
free Hamiltonian $H_0$.

We show further, that to the second order $O(e^2)$ the electron and photon
masses and corresponding wave function renormalization constants
in the renormalized Hamiltonian vary in accordance with the result of $1$-loop
renormalization group equations. This can serve as evidence for the 
equivalence of the flow equations and Wilson's
renormalization. Therefore we have rewritten
the mass correction $\de m_{\la}^2$ through the self energy term,
arising in $1$-loop calculations of ordinary perturbative theory. The negative
overall sign stems from our definition of the flow parameter,
namely for $\De l>0$ we are lowering the cutoff 
\mbox{$dl=-\frac{2}{\la^3}d\la$}. 
 
We start with the bare cutoff mass \mbox{$m_{\La}^2=m^2+\de M_{\La}^{(2)}$},
where \mbox{$\de M_{\La}^{(2)}$} is the second order mass counterterm.
According to \eq{ri2},\eq{ri3} the electron (photon) mass runs
\be
m_{\la}^2 = m_{\La}^2 - [\de\Sigma_{\la}(m^2) - \de\Sigma_{\La}(m^2)]
\lee{ri4}
defining, due to renormalizability, the counterterm
\mbox{$\de M_{\La}^{(2)} =\de m_{\La}^2 = -\de\Sigma_{\La}(m^2)$}
and the dependence of the renormalized mass on the cutoff $\la$
\be
m_{\la}^2 = m^2 + \de m_{\la}^2 = m^2 - \de\Sigma_{\la}
\; . \lee{ri5}
We calculate explicitly the self-energy term.
The {\bf electron} energy correction contains several terms
\be
\de p_{\la}^-
= <p', s'|H - H_0|p, s>
= \left(\sum_{n=1}^3 \de p_{\la n}^-\right) \cdot \de^{(3)}(p-p') \de_{s s'}
\; . \lee{ri6}
The first term is induced by the flow equation in single electron
sector, namely comes from the commutator $[\eta{(1)}, H_{ee\ga}]$
\be
\de p_{1\la}^-=-\int_{l_{\la}}^{\infty}
<[\eta^{(1)},H_{ee\ga}]>_{self~energy}dl'=-\frac{\de\Sigma_{1\la}(p)}{p^+}
\; ; \lee{ri7}
it reads, cf. \eq{d9} in Appendix\ref{C},
\bea
\de p_{1\la}^- &=& e^2 \int \frac{d^2k^{\bot} dk^+}{2(2\pi)^3}
\frac{\th (k^+)}{k^+} \th (p^+-k^+) \nn \\
&& \times \Ga^i(p - k, p, -k) \Ga^i(p, p - k, k) \,
\frac{1}{p^- - k^- - (p-k)^-} \times (-R)  
\; . \leea{ri8}
This term explicitly depends on the cutoff $\la$ ($l=1/\la^2$) through the
similarity function, that plays the role of a regulator in the loop integration
\be
R_{\la}=f_{p,k,\la}^2=\exp\left\{-2\left(\frac{\De_{p,k}}{\la}\right)^2\right\}
\; . \lee{ri9}
\Eq{ri8} corresponds to the first diagram in \fig{eselfen}.

Two instantaneous diagrams, the second and third in \fig{eselfen},
contribute the cutoff independent (constant) terms.
They arise from normal-ordered instantaneous interactions
in the single electron sector and can be written as
\be
\de p_{n\la}=\de p_n(l=0)=c_n<\hat{O}\hat{O}^+>V_n^{inst}(l=0)
\ee
where $n=2,3$ corresponds to the second and third diagrams in \fig{eselfen},
$c_n$ is the symmetry factor, $<\hat{O}\hat{O}^+>$ stands for the boson
$(n=2)$ and  fermion $(n=3)$ contractions
(i.e. \mbox{$<\tilde{a}_k \tilde{a}_k^+> = \th(k^+)/k^+$} and
\mbox{$<\tilde{b}_p \tilde{b}_p^+> = \th(p^+)$}), 
and $V_n^{inst}(l=0)$ arises from normal-ordering 
of $H_{ee\ga\ga}$ for $n=2$ and of $H_{eeee}$ for $n=3$ (\eqs{ch16}{ch17}).

These two diagrams $\de p_n(l=0)$ define together with the first one 
$\de p_1(l=0)$ the initial condition for the total energy correction, \eq{ri6}.

Since the diagrams $n=2,3$ come from the normal-ordering canonical
Hamiltonian at $l=0$, they must accompany the first diagram
for any flow parameter $l$. In what follows we use for 
the instantaneous terms the same regulator $R$, \eq{ri9}
\bea
\de p_{2\la}^- &=& e^2\int \frac{d^2k^{\bot}dk^+}{2(2\pi)^3} \,
\frac{\th (k^+)}{k^+} \, \frac{\si^i\si^i}{[p^+-k^+]} \times (-R) \nn \\
\de p_{3\la}^- &=& e^2\int \frac{d^2k^{\bot}dk^+}{2(2\pi)^3} \, \th (k^+) \,
\frac{1}{2} \left( \frac{1}{[p^+ - k^+]^2} - \frac{1}{[p^+ + k^+]^2} \right) \times (-R)
\; . \leea{ri12}
We define the set of coordinates
\bea
x &=& \frac{k^+}{p^+} \nn \\
k &=& (xp^+, xp^{\bot} + \kappa^{\bot})
\; , \leea{ri13}
where \mbox{$p = (p^+, p^{\bot})$} is the external electron momentum.
Then the electron self energy diagrams, \fig{eselfen}, \eq{d13}
in Appendix\ref{C}, contribute
\bea
\hspace{-10mm} p^+\de p_{1\la}^- &=& -\frac{e^2}{8\pi^2}
 \int_0^1 dx \int d\kappa_{\bot}^2 \nn \\
&& \hspace{10mm} \times \left[
 \frac{p^2 - m^2}{\kappa_{\bot}^2 + f(x)} \left( \frac{2}{[x]} - 2 + x \right)
 -\frac{2m^2}{\kappa_{\bot}^2 + f(x)} + \left( \frac{2}{[x]^2} + \frac{1}{[1-x]} \right)
 \right]
 \times (-R) \nn \\
f(x) &=& xm^2 - x(1-x) p^2
\leea{ri14}
and
\bea
p^+ \de p_{2\la}^- &=& \frac{e^2}{8\pi^2} \int_0^{\infty}dx \int d\kappa_{\bot}^2
\left( \frac{1}{[x][1-x]} \right) \times (-R) \nn \\
&& \hspace{-7mm} \rightarrow \; \frac{e^2}{8\pi^2} \int_0^1dx \int d\kappa_{\bot}^2
\left( \frac{1}{[x]} \right) \times (-R) \nn \\
p^+ \de p_{3\la}^- &=& \frac{e^2}{8\pi^2} \int_0^{\infty}dx \int d\kappa_{\bot}^2
\left( \frac{1}{[1-x]^2} - \frac{1}{(1+x)^2} \right) \times (-R) \nn \\
&& \hspace{-7mm} \rightarrow \; \frac{e^2}{8\pi^2} \int_0^1dx \int d\kappa_{\bot}^2
\left( \frac{2}{[x]^2} \right) \times (-R)
\; ; \leea{ri15}
for details we refer to Appendix\ref{C}. 
Note, that the transformation in the integrals over $x$
is performed before the regulator is taken into account \cite{ZhHa}.
(In the second integral the electron momentum is replaced by the gluon
one due to momentum conservation). The brackets '\mbox{\boldmath{$[\;]$}}'
denote the principle value prescription, defined further in \eq{ri21}.

The loop integral over $k$ \eqs{ri14}{ri15}
contains two types of divergencies: UV in the transversal
coordinate $\kappa^{\bot}$ and IR in the longitudinal component $k^+$.
The physical value of mass must be IR-finite.
We show, that the three relevant diagrams together 
in fact give an IR-finite value for the renormalized mass; this enables to
determine counterterms independent of longitudinal momentum. 
In the wave function renormalization constant, however, the IR-singularity
is still present.
    
Define 
\be
\de_1=\frac{p^+}{P^+}
\; , \lee{ri16}
where $P=(P^+,P^{\bot})$ is the positronium momentum, $p$ the electron momentum. 
The transversal UV divergency is regularized through the unitary
transformation done, i.e. by the regulator $R$, \eq{ri9}
\be
R_{\la} =
\exp\left\{ -\left( \frac{\tilde{\De}_{p,k}}{\la^2\de_1} \right)^2 \right\}
\; \approx \;
\th(\la^2 \de_1 - |\tilde{\De}_{p,k}|)
\; , \lee{ri17}
where the cutoff is rescaled and defined in units of the
positronium momentum $P^+$, namely \mbox{$\la\rightarrow \sqrt{2}\la^2/P^+$},
and \mbox{$\De_{p,k}=p^--k^--(p-k)^-=\tilde{\De}_{p,k}/p^+$}. The rude approximation
for the exponential through a $\th$-function changes the numerical coefficient
within a few percent; nevertheless it is useful to estimate the integrals
in \eqs{ri14}{ri15} in this way analitically.
From \eq{ri17} we have for the sum of intermediate (electron and photon) state momenta 
(the external electron is on-mass-shell \mbox{$p^2=m^2$}) 
\be
\frac{\kappa^{\bot 2}}{[x]}+\frac{\kappa^{\bot 2}+m^2}{[1-x]} \; \leq \;
\la^2\de_1+m^2
\lee{ri18}
giving for the regulator
\bea
R_{\la} &=& \th(\kappa^{\bot 2}_{\la max}-\kappa^{\bot 2}) \,
\th(\kappa^{\bot 2}_{\la max})\nn\\
\kappa^{\bot 2}_{\la max}&=&x(1-x)\la^2\de_1-x^2m^2
\leea{ri19}
and \mbox{$\th(\kappa^{\bot 2}_{\la max})$} leads
to the additional condition for the longitudinal momentum
\bea
&& 0\leq x\leq x_{max} \nn \\
&& x_{max}=\frac{1}{1+m^2/(\la^2 \de_1)}
\leea{ri20}
implying that the singularity of the photon longitudinal momentum
for $x\rightarrow 1$ is regularized by the function $R_{\la}$.
This is the case due to the nonzero fermion mass present in \eq{ri18}
for the intermediate state with $(1-x)$ longitudinal momentum.
The IR-singularity when $x\rightarrow 0$ is still present; it
is treated by the principle value prescription \cite{ZhHa}
\be
\frac{1}{[k^+]}=\frac{1}{2}\left( \frac{1}{k^++i\varep P^+}
+\frac{1}{k^+-i\varep P^+}\right)
\; , \lee{ri21}
where $\varep=0_+$, and $P^+$ is the longitudinal part of the positronium
momentum (used here as typical momentum in the problem).
This defines the bracket '\mbox{\boldmath{$[\;]$}}'
in \eqs{ri14}{ri15}
\be
\frac{1}{[x]}=\frac{1}{2} \left( \frac{1}{x+i\frac{\varep}{\de_1}}+\\
\frac{1}{x-i\frac{\varep}{\de_1}} \right)
\; . \lee{ri22}
Making use of both regularizations for 
transversal and longitudinal components, we have for the first diagram,
\eq{ri14},
\bea
\de m_{1\la}^2 &\hspace{-1em}=\hspace{-1em}& p^+ \de p^- |_{p^2=m^2} \nn \\
\de m_{1\la}^2 &\hspace{-1em}=\hspace{-1em}&-\frac{e^2}{8\pi^2}
\left\{ 3m^2 \ln \left( \frac{\la^2\de_1+m^2}{m^2} \right)
+ \frac{\la^2 \de_1}{\la^2\de_1+m^2} \left( \frac{3}{2}\la^2\de_1+m^2 \right)
-2 \la^2 \de_1 \ln \left( \frac{\la^2\de_1}{\la^2\de_1+m^2} \frac{\de_1}{\varep} \right)
\right\} \nn \\
\leea{ri23}
Note, that the third term has the mixing UV and IR divergencies.
Combining the three relevant diagrams, \fig{eselfen}, and integrating
with the common regulator, one obtains for the {\bf electron mass correction} 
\bea
\de m_{\la}^2 &=& p^+(\de p_1 + \de p_2 + \de p_3)|_{p^2 = m^2} =
-\de \Sigma_{\la}(m^2) \nn \\
\de m_{\la}^2 &=& -\frac{e^2}{8\pi^2}
\left\{ 3m^2 \ln \left( \frac{\la^2\de_1+m^2}{m^2} \right)
-\frac{\la^2 \de_1 m^2}{\la^2 \de_1 + m^2} \right\}
\; . \leea{ri24}
The mass correction is IR-finite (that gives rise to IR-finite
counterterms) and contains only a logarithmic UV-divergency. Namely,
when \mbox{$\la\de_1 \rightarrow\La \gg m$}
\be
\de m_{\La}^2=-\frac{3e^2}{8\pi^2}m^2 \ln \frac{\La^2}{m^2}
\; . \lee{ri25}
It is remarkable that we reproduce with the cutoff condition of \eq{ri18}
the standard result of covariant perturbative theory calculations  
including its global factor $3/8$. As was mentioned above, the difference
in sign, as compared with the $1$-loop renormalization group result, 
comes from scaling down from high to low energies 
in the method of flow equations.     

The similar regularization for the intermediate state momenta
in the self-energy integrals, called 'global cutoff scheme', 
was introduced by W.~M.~Zhang and A.~Harindranath \cite{ZhHa}.
In our approach the UV-regularization, that defines the concrete form of 
the regulator $R$, arises naturally from the method of flow equations, 
namely from the unitary transformation performed, where the generator
of the transformation is chosen as the commutator $\eta=[H_d,H_r]$.
Note also, that the regulator $R$, \eq{ri17}, in general
is independent of the electron momentum $p^+$ (rescaled cutoff
\mbox{$\la\de_1\longrightarrow\la$}), and therefore is boost invariant.

For the wave function renormalization constant, \eq{ri3}, one has
\be
\left. \frac{\partial\de p^-}{\partial p^-} \right|_{p^2=m^2} =
-\frac{e^2}{8\pi^2}\int_0^1\int d\kappa_{\bot}^2 \left[
\frac{2\frac{1}{x}-2+x}{\kappa_{\bot}^2+f(x)}
-\frac{x(1-x)2m^2}{(\kappa_{\bot}^2+f(x))^2}\right]_{p^2=m^2}
\times(-R)
\; , \lee{ri26}
that together with the regulator $R$, \eq{ri17}, results
\bea
Z_2 &\hspace{-0.5em}=\hspace{-0.5em}& 1 - \frac{e^2}{8\pi^2} \left\{
\ln\frac{\la^2\de_1}{m^2} \cdot \left( \frac{3}{2}-2 \ln\frac{\de_1}{\varep} \right)
+ \ln\frac{\de_1}{\varep} \cdot \left( 2 - \ln\frac{\de_1}{\varep} \right)
+ F \left( \ln\frac{\la^2\de_1}{\la^2\de_1+m^2}; \frac{\la^2\de_1}{\la^2\de_1+m^2} \right)
\right\} \nn \\
F &\hspace{-0.5em}=\hspace{-0.5em}& \ln\frac{\la^2\de_1}{\la^2 \de_1+m^2} \left(
\frac{1}{2} - \ln\frac{\la^2 \de_1}{\la^2 \de_1+m^2} \right)
+\frac{1}{2} \frac{\la^2 \de_1}{\la^2 \de_1 + m^2}
- 2 + 2 \int_0^{x_{max}}dx \frac{\ln x}{x-1}
\; . \leea{ri27}
As \mbox{$\la\de_1 \rightarrow \La \gg m$} the function $F$ tends to a constant
\be
F|_{\La \gg m} = C = -\frac{3}{2} +\frac{\pi^2}{3}
\; . \lee{ri28}
Therefore, by dropping the finite part, we obtain
\bea
Z_2 &=& 1 - \frac{e^2}{8\pi^2}
\left\{\ln\frac{\La^2}{m^2} \cdot \left( \frac{3}{2}
-2 \ln\frac{1}{\varep} \right)
+ \ln\frac{1}{\varep} \left( 2-ln\frac{1}{\varep} \right) \right\}
\; , \leea{ri29} 
where we have rescaled 
\mbox{$\frac{\varepsilon}{\de_1}\rightarrow\varepsilon$}. 
The electron wave function renormalization constant contains
logarithmic UV and IR divergencies mixed, together with pure
logarothmic IR divergencies. 
We mention, that the value of $Z_2$ is not sensitive to the form of regulator
applied; the same result for $Z_2$ was obtained with another choice of
regulator \cite{ZhHa}.

We proceed with renormalization to the second order in the {\bf photon} sector.
The diagrams that contribute to the photon self energy are shown in 
\fig{photselfen}.
The commutator \mbox{$[\eta^{(1)},H_{ee\ga}]$}, 
corresponding to the first diagram,
gives rise to (\eq{d21} in Appendix\ref{C})
\bea
\de q_{1\la}^ - \, \de^{ij} &=&
\frac{1}{[q^+]} e^2 \int \frac{d^2k^{\bot}dk^+}{2(2\pi)^3} \,
\th(k^+) \th(q^+-k^+) \\
&&\hspace{3em} \times \, Tr\left[\Ga^i(k,k-q,q) \Ga^j(k-q,k,-q)\right] \,
\frac{1}{q^- - k^- - (q-k)^-} \times(-R) \nn
\; , \leea{ri30}
where momenta are given in \fig{photselfen}, and the regulator is
\be
R_{\la}=f_{q,k,\la}^2=\exp\left\{-2 \left(\frac{\De_{q,k}}{\la} \right)^2
\right\}
\; . \lee{ri31}
In full analogy with the electron self energy this also defines the regulator
for the second diagram with the instantaneous interaction, see 
\fig{photselfen},
\be
\de q_{2\la}^- \, \de^{ij} = \frac{1}{[q^+]} e^2 \int \frac{d^2k^{\bot}dk^+}{2(2\pi)^3}
\th(k^+) \, Tr(\si^i \si^j)
\left( \frac{1}{[q^+-k^+]} - \frac{1}{[q^++k^+]} \right) \times(-R)
\; . \lee{ri32}
We define the set of coordinates
\bea
\frac{(q-k)^+}{q^+} &=& x  \nn \\
k &=& ((1-x)q^+, (1-x)q^{\bot}+\kappa^{\bot}) \nn \\
(q - k) &=& (xq^+, xq^{\bot} - \kappa^{\bot})
\; , \leea{ri33}
where $q = (q^+, q^{\bot})$ is the external photon momentum. Then two diagrams
contribute (for details see Appendix\ref{C}, \eq{d25}):
\bea
q^+ \, \de q_1^- &=&
-\frac{e^2}{8\pi^2} \int_0^1dx \int d\kappa_{\bot}^2 \nn \\
&& \times \left\{
 \frac{q^2}{\kappa_{\bot}^2+f(x)} \left( 2x^2 - 2x + 1 \right) +
 \frac{2m^2}{\kappa_{\bot}^2+f(x)}
 + \left( -2 + \frac{1}{[x][1-x]} \right)
\right\} \times (-R) \nn \\
f(x) &=& m^2 - x(1-x)q^2 \nn \\
q^+ \, \de q_2^- &=& \frac{e^2}{8\pi^2} \int_0^\infty dx \int d\kappa_{\bot}^2
\left( \frac{1}{[1-x]}-\frac{1}{1+x} \right) \times (-R) \nn \\
&&\hspace{-1.5em} \rightarrow \;
-\frac{e^2}{8\pi^2} \int_0^1dx \int d\kappa_{\bot}^2 \frac{2}{[x]} \times(-R)
\; . \leea{ri34}
Note, that the transformation in the second integral is done before
the regularization (by regulator the $R$) is performed \cite{ZhHa}.

Making use of the same approximation for the regulator as in the electron
sector, we obtain for the sum of intermediate (two electron) state momenta
\bea
&& \frac{\kappa_{\bot}^2+m^2}{x} + \frac{\kappa_{\bot}^2+m^2}{1-x}
 \leq \la^2\de_2 \nn \\
&& \de_2 = \frac{q^+}{P^+}
\; , \leea{ri35}
where the photon is put on mass-shell $q^2=0$ and the rescaled
cutoff \mbox{$\la \rightarrow \sqrt{2} \la^2/P^+$} has been used.
This condition means for the transversal integration
\bea
R_{\la} &=& \th(\kappa_{\la max}^{\bot 2}-\kappa^{\bot 2}) \,
\th(\kappa_{\la max}^{\bot 2}) \nn \\
\kappa_{\la max}^{\bot 2} &=& x(1-x)\la^2 \de_2-m^2
\; , \leea{ri36} 
and for the longitudinal integration
\bea
&& x_1 \leq x \leq x_2 \nn \\
&& x_1 = \frac{1-r}{2} \approx \frac{m^2}{\la^2\de_2} \nn \\
&& x_2 = \frac{1+r}{2} \approx 1 - \frac{m^2}{\la^2\de_2} \nn \\
&& r = \sqrt{1 - \frac{4m^2}{\la^2\de_2} }
\; , \leea{ri37}
where the approximate value is used when $m \ll \la$.
This shows that the condition of \eq{ri35} for two electrons with masses $m$ 
removes the light-front infrared singularities from \mbox{$x \rightarrow 0$}
and \mbox{$x\rightarrow 1$}. Thus, both UV and IR divergencies are regularized
by the regulator R, \eq{ri36}.

The mass correction arising from the first diagram, \eq{ri34}, is
\be
\de m_{1\la}^2 = \frac{e^2}{8\pi^2} \, \frac{2}{3} \, \la^2 \de_2
\left( 1-\frac{4m^2}{\la^2\de_2} \right)^{3/2}
\; . \lee{ri38}
Combining together both diagrams with the same regulator, \eq{ri34}, we obtain
\be
\de m_{\la}^2 = \frac{e^2}{8\pi^2} \, \left( \frac{5}{3} \la^2 \de_2 \, r
- \frac{8}{3} m^2 \, r - 2m^2 \, \ln\frac{1+r}{1-r} \right)
\; , \lee{ri39}
where $r$ is defined in \eq{ri37}.
The result shows that the mass correction involves the quadratic
and logarithmic UV divergencies, i.e. as $\la\de_2\rightarrow\La\gg m$
\be
\de m_{\La}^2 = \frac{e^2}{8\pi^2} \left( \frac{5}{3} \La^2 
-2m^2 \, \ln\frac{\La^2}{m^2} \right)
\; . \lee{ri40}
The wave function renormalization constant is defined through 
\be
\left. \frac{\partial\de q^-}{\partial q^-} \right|_{q^2=0} = 
- \frac{e^2}{8\pi^2} \int_0^1dx \int d\kappa_{\bot}^2
\left. \left\{ \frac{2x^2-2x+1}{\kappa_{\bot}^2+f(x)}
+ \frac{2m^2x(1-x)}{(\kappa_{\bot}^2+f(x))^2}
\right\} \right|_{q^2=0} \times (-R)
\; , \lee{ri41}
that, with the regulator $R$, \eq{ri36}, results
\be
Z_2 = 1 - \frac{e^2}{8\pi^2} \left( -\frac{2}{3}\ln\frac{1+r}{1-r}
+ \frac{10}{9} \, r + \frac{8}{9} \frac{m^2}{\la^2\de_2}\, r \right)
\; . \lee{ri42}
The photon wave function renormalization constant contains only
logarithmic UV divergency, indeed as \mbox{$\la\de_2\rightarrow\La\gg m$} 
\bea
&& Z_2=1-\frac{e^2}{8\pi^2}(-\frac{2}{3}\ln\frac{\La^2}{m^2})
\leea{ri43}
and is free of IR divergencies (as is expected from the form of the
regulator $R$, \eq{ri35}).

\subsection{\label{3.3}Modified similarity renormalization ({\bf MSR}) 
by means of different unitary transformations}

We consider in this section the general properties of the unitary 
transfromation in MSR scheme, i.e. unitary transformation performed in 
the 'energy space' aiming to bring the field theoretical Hamiltonian
to the block-diagonal form in 'particle number space'.

Basing on this analyses we construct the general form of the new 
interactions, generated in the effective Hamiltonian
to the second order in coupling by such
transformation.

\subsubsection{\label{3.3.1}Properties of the unitary transformation 
in {\bf MSR}}

In order to illustrate the main idea of MSR consider the flow equations
in 'particle number space' \cite{Gu}. Again we break the Hamiltonian into 
'diagonal' and 'rest' parts, and choose for the 'diagonal' part
the particle number conserving part of the Hamiltonian $H_c$ and as basis
-- the states with definite number of particles
\bea
&& H=H_d+H_r \nn\\
&& H_d=H_c \nn\\
&& H_d|n>=E_n|n>
\leea{pu1}
where $|n>$ is the state with particle number $'n'$.

Then the flow equations of Wegner are
\bea
&& \frac{dH(l)}{dl}=[\eta(l),H(l)]\nn\\
&& \eta=[H_d,H_r]
\leea{pu2}
One has in the matrix form in particle number space 
\bea
&& \frac{dH_{mn}}{dl}=[\eta,H_r]_{mn}-(E_m-E_n)^2H_{rmn}\nn\\
&& \eta_{mn}=(E_m-E_n)H_{rmn}
\leea{pu3}
that can be written through the similarity function as follows
\bea
&& \frac{dH_{mn}}{dl}=[\eta,H_r]_{mn}+
\frac{du_{mn}}{dl}\frac{H_{mn}}{u_{mn}}\nn\\
&& \eta_{mn}=\frac{1}{E_m-E_n}(-\frac{du_{mn}}{dl}
\frac{H_{mn}}{u_{mn}})
\leea{pu4}
where the similarity function is defined in 'particle number space'
as 
\bea
&& u_{mn}(l)=\exp(-\De_{mn}^2l)\nn\\
&& \De_{mn}=E_m-E_n
\leea{pu5}

\begin{figure}
$$
\fips{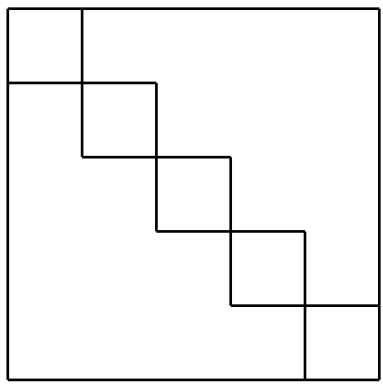}
\setlength{\unitlength}{0.240900pt}
\begin{picture}(0,0)
\end{picture}
$$
\caption{Similarity renormalization (SR) in 'particle number space'
results to the block-diagonal form for effective Hamiltonian,
where each block (sector) conserves the number of particles.}
\label{block-diagonal}
\end{figure}

According to \eq{pu4} (due to the second term) the 'rest' part,
when $m\neq n$, behaves with the flow parameter $l$
\be
H_{rmn}=u_{mn}\tilde{H}_{rmn}
\lee{pu6}
and therefore is completly eliminated as $l$ tends to infinity
($l\rightarrow\infty$, i.e. the band size $\la\rightarrow 0$).
One is left with the block-diagonal effective Hamiltonian,
where the number of particles is conserved in each block, 
\fig{block-diagonal}.

The obvious condition, following from \eq{pu5} for the similarity
function in 'diagonal' sector ($m\neq n$)
\be
u_d=1
\lee{pu7}
is used further. In MSR scheme this means that the similarity
function is equal to unity in particle number conserving sectors.
This condition is equivalent to the choice of the particle number
conserving part of Hamiltonian for the 'diagonal' part.
\be
H_d=H_c \leftrightarrow u_c=1
\lee{pu8}
The idea, that stands behind this excercise, is the 
similarity transformation of Glazek, Wilson but in 
'particle number space', which enables to decouple many-body and
few-body states in the block-diagonal effective Hamiltonian.
In the 'diagonal' sectors of the effective Hamiltonian 
the effects of many-body states are simulated on the few-body states
by a set of new interactions, induced by the similarity transformation
and corresponding to the marginal relevant operators of the theory.

The problem, that arise now, is how to simulate also in the effective
Hamiltonian the effects of high-energy states on the low-energy states.
MSR tries to take into account both effects of high-energy and many-body
states.

The idea of MSR is again to perform the similarity transformation
in the 'particle number space' to bring Hamiltonian to a block-diagonal
form, with the number of particles conserving in each block.
Technically, we use the flow equations in the 'energy space',
organized in a way to eliminate the particle number changing sectors
and to generate efective Hamiltonian in the particle number
conserving sectors. The effects of high-energy and many-body states
are simulated then by a set of effective interactions, which do not change
particle number and do not couple to high-energy states either.

The 'diagonal' sector is the particle number conserving part 
of the Hamiltonian, and (or) the similarity function in the particle
number conserving sector is equal to unity
\be
H_d=H_c \leftrightarrow u_{dij}=1
\lee{pu9}
The flow equations are written in the basis of single particle states
${|i>}$, i.e.
\be
H_0|i>=E_i|i>
\lee{pu10}
where $H_0$ is free (noninteracting) part of Hamiltonian.

We write the flow equations explicitly in the perturbative frame,
i.e. break the Hamiltonian 
\be
H=H_{0d}+\sum_n (H_d^{(n)}+H_r^{(n)})
\lee{pu11}
where $H^{(n)}\sim e^n$ and the indices 'd' ('r') correspond
to 'diagonal' ('rest') sectors (here we do not refer to the definite 
field theory). The flow equation and generator of transformation
to the 'n'-th order are  
\bea
&& \frac{dH^{(n)}}{dl}=\sum_k[\eta^{(k)},H_d^{(n-k)}+H_r^{(n-k)}]
   +\sum_k[[H_d^{(k)},H_r^{(n-k)}],H_{0d}]+[[H_{0d},H_r^{(n)}],H_{0d}]\nn\\
&& \eta^{(n)}=[H_{0d},H_r^{(n)}]+\sum_k[H_d^{(k)},H_r^{(n-k)}]
\leea{pu12}
This is written in the matrix form in the basis of states
${|i>}$ as
\bea
&& \frac{dH_{ij}^{(n)}}{dl}=\sum_k[\eta^{(k)},H_d^{(n-k)}+
H_r^{(n-k)}]_{ij}\nn\\
&& \hspace{5cm}
   -(E_i-E_j)\sum_k[H_d^{(k)},H_r^{(n-k)}]_{ij}-(E_i-E_j)^2H_{rij}^{(n)}
\leea{pu13}
In 'diagonal' and 'rest' sectors one has corresponding

\underline{'diagonal sector'}
\bea
&& \frac{dH_{dij}^{(n)}}{dl}=\sum_k[\eta^{(k)},H_d^{(n-k)}+H_r^{(n-k)}]_{dij}
+\sum_k[[H_d^{(k)},H_r^{(n-k)}]_d,H_{0d}]_{ij}
\leea{pu14}
\underline{'rest sector'}
\bea
&& \frac{dH_{rij}^{(n)}}{dl}=\sum_k[\eta^{(k)},H_d^{(n-k)}+
H_r^{(n-k)}]_{rij}\nn\\
&& \hspace{5cm}
+\sum_k[[H_d^{(k)},H_r^{(n-k)}]_r,H_{0d}]_{ij}
-(E_i-E_j)^2H_{rij}^{(n)}
\leea{pu15}
The main difference between these two sectors is the presence
of the third term in the 'rest' sector
\be
-(E_i-E_j)^2H_{rij}^{(n)}=\frac{du_{rij}}{dl}\cdot
\frac{H_{rij}^{(n)}}{u_{rij}}
\lee{pu16}
that gives rise to the band-diagonal structure in the 
'energy space' for the 'rest' part of effective Hamiltonian, i.e.
\bea
&& H_{rij}=u_{rij}\tilde{H}_{rij}\nn\\
&& u_{rij}={\rm e}^{-(E_i-E_j)^2l}
\leea{pu17}
When $l\rightarrow\infty$ the 'rest' part is completly eliminated,
except for the diagonal in 'energy space' matrix elements $i=j$,
which do not contribute to physical values (see further).
One assumes therefore, that different Fock states are decoupled
in the block-diagonal effective Hamiltonian.

In the 'diagonal' sector the matrix elements with any energy
differences are present, $u_{dij}=1$. According to \eq{pu14}
the new terms are induced in this sector, containing in the lowest
Fock sectors both the information on high-Fock components
(included due to the presence in the canonical Hamiltonian 
the interactions that mix different Fock components)
and high-energy states.

The renormalized effective Hamiltonian contains the nonlocal
new interactions, corresponding to marginal relevant operators,
and local operators -- counterterms, defined order by order 
in coupling constant. One is able therefore to perform
at a time renormalization programm and work in the lowest Fock 
sectrors to solve bound state problem.

The property of the unitary transformation in MSR scheme,
we are going to use in the next section, is the difference
between the 'diagonal' and 'rest' sectors, given by the following
expressions for similarity functions in both sectors
\bea
&& u_{dij}=1\nn\\
&& u_{rij}\neq 1
\leea{pu18}

\subsubsection{\label{3.3.2}Generated interaction in MSR}
Here we calculate the interactions in light-front $QED_{3+1}$,
generated to the second order in coupling by different unitary
transformations of MSR scheme. Flow equations and similarity
transformations of Glazek, Wilson, acting in SR scheme,
are summarized in Appendix \ref{A}.

The second order interaction, generated by flow equations in MSR
in the 'diagonal' sector, is
\bea
&& \frac{dH_{dij}^{(2)}}{dl}=[\eta^{(1)},H_r^{(1)}]_{dij}\nn\\
&& \eta_{ij}^{(1)}=-\frac{1}{E_i-E_j}\frac{dH_{rij}^{(1)}}{dl}
\leea{gi1}
where the indices 'd' and 'r' define 'diagonal' (Fock state conserving)
and 'rest' (Fock state changing) sectors corresponding. 
The commutator can be written
\be
[\eta^{(1)},H_r^{(1)}]_{dij}=
-\sum_k\left(\frac{\frac{dH_{rik}^{(1)}}{dl}H_{rjk}^{(1)}}{E_i-E_k}
+\frac{\frac{dH_{rjk}^{(1)}}{dl}H_{rik}^{(1)}}{E_j-E_k}
\right)
\lee{gi2}
Neglecting the dependence of energy on the flow parameter,
one has the following generated interaction
\bea
&& H_{dij}^{(2)}(\la)=H_{dij}^{(2)}(\la=\La\rightarrow\infty)+
\sum_k\left(\frac{\int_{\la}^{\infty}\frac{dH_{rik}^{(1)}}{d\la'}
H_{rjk}^{(1)}d\la'}{E_i-E_k}
+\frac{\int_{\la}^{\infty}\frac{dH_{rjk}^{(1)}}{d\la'}
H_{rik}^{(1)}d\la'}{E_j-E_k}\right)
\leea{gi3}
where the connection $l=1/\la^2$ is used.
Making use of the first order solution through similarity function
\be
H_{rij}^{(1)}(\la)=H_{rij}^{(1)}(\la=\La\rightarrow\infty)
\frac{f_{ij}(\la)}{f_{ij}(\la=\La\rightarrow\infty)}
\lee{gi4}
one has
\bea
&& H_{dij}^{(2)}(\la)=H_{dij}^{(2)}(\la=\La\rightarrow\infty)+
\sum_k H_{rik}^{(1)}(\la=\La\rightarrow\infty)
H_{rjk}^{(1)}(\la=\La\rightarrow\infty)\cdot\nn\\
&& \left(\frac{\int_{\la}^{\infty}\frac{df_{ik}^{(1)}(\la')}{d\la'}
f_{jk}^{(1)}(\la')d\la'}{E_i-E_k}
+\frac{\int_{\la}^{\infty}\frac{df_{jk}^{(1)}(\la')}{d\la'}
f_{ik}^{(1)}(\la')d\la'}{E_j-E_k}\right)
\leea{gi5}
Note, that first and second order flow equations \eq{gi1} in 'diagonal'
sector, i.e. when the condition $u_{dij}=1$ is implied,
coincide for both MSR \eqs{pu13}{pu14} and SR \eq{a3}{a4} of
Appendix\ref{A} schemes. Assuming $u_{dij}=1$ for 
other similarity transformations of Appendix \ref{A},
we obtain the same form of interaction, namely \eq{gi5}, 
generated to the second order
in 'diagonal' sector by any unitary transformation in MSR scheme.

We proceed along the same line in the light-front frame. Basing
on calculations of section \ref{3.2.1}, we get the following generated
interaction in MSR scheme (for exchange and annihilation channels) 
\bea
V_{gen}^{(ex)}(\la)&=&-e^2M_{2ii}^{(ex)}\frac{1}{(p_1^+-p_3^+)}
\left(\frac{\int_{\la}^{\infty}\frac{df_{p_1,p_3,\la'}}{d\la'}
f_{p_4,p_2,\la'}d\la'}{\De_{p_1,p_3}}+
\frac{\int_{\la}^{\infty}\frac{df_{p_4,p_2,\la'}}{d\la'}
f_{p_1,p_3,\la'}d\la'}{\De_{p_4,p_2}}\right)
\nonumber\\ 
\\
V_{gen}^{(an)}(\la)&=&e^2M_{2ii}^{(an)}\frac{1}{(p_1^++p_2^+)}
\left(\frac{\int_{\la}^{\infty}\frac{df_{p_1,-p_2,\la'}}{d\la'}
f_{p_4,-p_3,\la'}d\la'}{\De_{p_1,-p_2}}+
\frac{\int_{\la}^{\infty}\frac{df_{p_4,-p_3,\la'}}{d\la'}
f_{p_1,-p_2,\la'}d\la'}{\De_{p_4,-p_3}}\right)
\nonumber 
\leea{gi6}
One obtaines the explicit form of generated interaction, 
induced by one of the unitary
transformations in MSR scheme, specifying the similarity
function $f_{ij}$ in \eq{gi6} (Apendix \ref{A}). 
Namely, one has for the flow equations of Wegner \cite{We}
\bea
&& f_{p_1p_2}=u_{p_1p_2}={\rm e}^{-\frac{\De_{p_1p_2}^2}{\la^2}}
\leea{gi7}
and for the similarity transformation of Glazek,Wilson 
two alternative forms \cite{GlWi},\cite{Pe}
\bea
&& f_{p_1p_2}=u_{p_1p_2}{\rm e}^{r_{p_1p_2}}\nn\\
&& f_{p_1p_2}=u_{p_1p_2}
\leea{gi8}
where 
\bea
&& u_{p_1p_2}=\theta(\la-|\De_{p_1p_2}|)
\leea{gi9}
We use the general form of generated interaction \eq{gi6} 
in the light front frame
for the further analyses in section \ref{4.3}.

\subsection{Renormalized to the second order effective $QED_{3+1}$ Hamiltonian
on the light front}

In this section we summarize the results obtained in the previous sections.
Basing on the flow equations we have completed the renormalization of 
light front QED (LFQED) together with calculation of effective Hamiltonian
to the second order in coupling constant.

Consider first the situation when the 'rest' sector is not completly
eliminated, i.e. canonical electron-photon coupling (the matrix elements)
is still present in the energy band $|E_i-E_j|<\la$. This means that 
perturbative photons are present in the effective Hamitonian and one
is able therefore to formulate the diagramatic rules for the perturbative
expansion in coupling.

\begin{figure}
\setlength{\unitlength}{1mm}
\begin{picture}(170,219)
\put(0,185){\makebox(56,34.61){ \loadeps{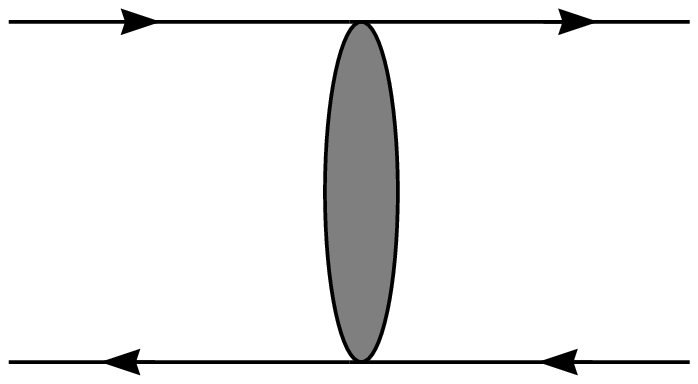} }}
  \diagform{180}{
    \bea
    &&\hspace{-17mm} -e_{\la}^2 \exp{\left\{-\frac{\De_{p_ip_f}^2}{\la^2}
      \right\}} 
      M_{2ij,\la} \, \de^{ij} 
      \frac{1}{[p_1^+ - p_3^+]} \nn \\
    &&\hspace{-12mm} 
      \times \frac{1}{2} \left( \frac{1}{\De_{p_1p_3\la}} + 
      \frac{1}{\De_{p_4p_2\la}} \right) \,
      \left( 1 - \exp{\left\{ -2 \, \frac{\De_{p_1p_3\la} \cdot 
      \De_{p_4p_2\la}}{\la^2} \right\} }
       \right) \nn \\
    && \nn \\
    &&\hspace{-7mm} M_{2ij,\la} = \biggl(\ch^+_2 \Ga^i_\la(p_1,p_2,p_1 \!-\! p_2)
      \ch_1\biggr) \nn \\
    &&\hspace{13mm} \times \biggl(\ch^+_4 \Ga^j_\la(p_3,p_4,-(p_1 \!-\! p_2)) 
      \ch_3\biggr) \nn
    \eea 
    }
\put(10,216){$p_1$}
\put(42,216){$p_3$}
\put(10,187){$p_2$}
\put(42,187){$p_4$}
\put(0,140){\makebox(56,34.61){ \loadeps{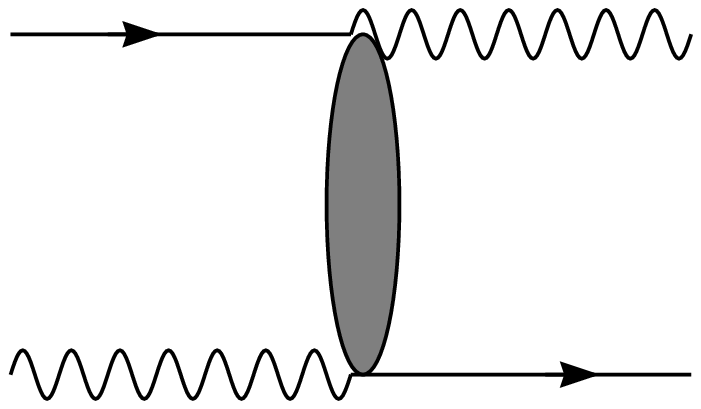} }}
  \diagform{135}{
    \bea
    &&\hspace{-17mm} e_{\la}^2\exp{\left\{-\frac{\De_{p_ip_f}^2}{\la^2}
      \right\}}
      \widetilde{M}_{2ij,\la} \,\varep^{i \ast} \varep^j \nn \\
    &&\hspace{-12mm} 
      \times \frac{1}{2} \left( \frac{1}{\De_{p_1k_1\la}} + 
      \frac{1}{\De_{p_2k_2\la}} \right) \,
      \left( 1 - \exp{\left\{ -2 \, \frac{\De_{p_1k_1\la} \cdot 
      \De_{p_2k_2\la}}{\la^2} \right\} }
       \right) \nn \\
    && \nn \\ 
    &&\hspace{-7mm} \widetilde{M}_{2ij,\la} = 
       \ch^+_2 \Ga^i_\la(p_1,p_1 \!-\! k_1,k_1) \: 
    \Ga^j_\la(p_1 \!-\! k_1,p_2,k_2) \ch_1 \nn
    \eea 
    }
\put(10,171){$p_1$}
\put(42,171){$k_1 \; (i)$}
\put(10,142){$k_2 \; (j)$}
\put(42,142){$p_2$}
\end{picture}
\vspace{-13cm}
\caption{The matrix elements of the renormalized to the second
order effective Hamitonian together with corresponding diagrams. 
The diagrams $2 - 5$
belong to the 'diagonal' sector; the $1, 6, 7$ correspond
to the 'rest' sector (the $6, 7$ diagrams are drawn schematically,
namely the corresponding momentum change must be performed
to get the real 'rest' diagrams, depicted in Table 1.)
The photon momenta are $x^+$-ordered, from left to right. The similarity
function is chosen 
\mbox{$f_{p_ip_f,\la} \!=\! \exp(-\De_{p_ip_f\la}^2 \!/\!\la^2)$},
where \mbox{$\De_{p_ip_f\la} \!=\! \Si p^-_i \!-\! \Si p^-_f$}
(the index \mbox{$`i`$} denotes initial and \mbox{$`f`$} final states)
and \mbox{$\De_{p_1p_2\la} \!=\! p_1^- \!-\! p_2^- \!-\! (p_1 \!-\! p_2)^-$},
\mbox{$p^-\!=\! (p_{\bot}^2 \!+\! m_{\la}^2)\!/\!p^+$}. } 
\label{feynrules}
\end{figure}

The matrix elements of the renormalized to the second order effective
Hamitonian, namely the diagrams in different Fock sectors, are depicted
in Table \ref{table}, and corresponding analytical expressions 
for 'diagonal' and 'rest'
sectors are listed in \fig{feynrules}.  The diagramatic rules 
are obtained by direct
calculation of matrix elements between free particle states.

Instantaneous diagrams as light front gauge artefact are not included 
in MSR procedure \cite{Pa1},\cite{Pa2}. This means that instantaneous 
terms (also
belonging to the 'rest' sectors) stay intact with the unitary transformation 
(i.e. do not obtain the similarity factor).
 
The matrix elements of the 'rest' (Fock state changing) sectors are sqeezed
in the energy band $\De_{p_ip_f}=|\sum p_i^--\sum p_f^-|<\la$; the 'diagonal'
(Fock state conserving) sectors, exist for any energy differences.
In the Table \ref{table} we denote as dot  
the zero to the second order matrix elements.

Using flow equations one has to the second order in coupling the following 
matrix elements in the {\bf 'diagonal' sectors} between the Fock states 
\bea
&& |e\bar{e}>\rightarrow |e\bar{e}>,
|e\bar{e}e\bar{e}>\rightarrow |e\bar{e}e\bar{e}>, ... \nn\\
&& -e_{\la}^2M_{2ij,\la}\de^{ij}\frac{1}{[p_1^+-p_3^+]}\left(
\frac{\int_{\la}^{\infty}\frac{df_{p_1p_3\la'}}{d\la'}f_{p_4p_2\la'}d\la'}
{\De_{p_1p_3\la}}+
\frac{\int_{\la}^{\infty}\frac{df_{p_4p_2\la'}}{d\la'}f_{p_1p_3\la'}d\la'}
{\De_{p_4p_2\la}}\right)\nn\\
&& |e\bar{e}\ga>\rightarrow |e\bar{e}\ga>,
|e\bar{e}\ga\ga>\rightarrow |e\bar{e}\ga\ga>, ...\nn\\
\nn\\
&& e_{\la}^2\tilde{M}_{2ij,\la}\varepsilon^{i*}\varepsilon^j\left(
\frac{\int_{\la}^{\infty}\frac{df_{p_1k_1\la'}}{d\la'}f_{p_2k_2\la'}d\la'}
{\De_{p_1k_1\la}}+
\frac{\int_{\la}^{\infty}\frac{df_{p_2k_2\la'}}{d\la'}f_{p_1k_1\la'}d\la'}
{\De_{p_2k_2\la}}\right)
\; , \leea{re1}
and in the {\bf 'rest' sectors} between the Fock states  
\bea
&& |e\bar{e}>\rightarrow |e\bar{e}e\bar{e}>,
|e\bar{e}e\bar{e}>\rightarrow |e\bar{e}>, ... \nn\\
&& -e_{\la}^2f_{p_ip_f\la}M_{2ij,\la}\de^{ij}\frac{1}{[p_1^+-p_3^+]}\left(
\frac{\int_{\la}^{\infty}\frac{1}{f_{p_ip_f\la'}}
\frac{df_{p_1p_3\la'}}{d\la'}f_{p_4p_2\la'}d\la'}{\De_{p_1p_3\la}}+
\frac{\int_{\la}^{\infty}\frac{1}{f_{p_ip_f\la'}}
\frac{df_{p_4p_2\la'}}{d\la'}f_{p_1p_3\la'}d\la'}{\De_{p_4p_2\la}}\right)\nn\\
\nn\\
&& |e\bar{e}>\rightarrow |\ga\ga>,
|\ga\ga>\rightarrow |e\bar{e}>, ...\nn\\
&& e_{\la}^2f_{p_ip_f\la}\tilde{M}_{2ij,\la}\varepsilon^{i*}\varepsilon^j\left(
\frac{\int_{\la}^{\infty}\frac{1}{f_{p_ip_f\la'}}
\frac{df_{p_1k_1\la'}}{d\la'}f_{p_2k_2\la'}d\la'}{\De_{p_1k_1\la}}+
\frac{\int_{\la}^{\infty}\frac{1}{f_{p_ip_f\la'}}
\frac{df_{p_2k_2\la'}}{d\la'}f_{p_1k_1\la'}d\la'}{\De_{p_2k_2\la}}\right)
\; , \leea{re2}
All momenta are given in \fig{feynrules}
We plug the similarity function
\bea
&& f_{p_ip_f\la}={\rm e}^{-\frac{\De_{p_ip_f}^2}{\la^2}} \nn\\
&& \De_{p_ip_f}=\sum p_i^--\sum p_f^-
\; , \leea{re3}
into \eq{re1},\eq{re2} to get the explicit form of interactions in both
sectors listed in \fig{feynrules}. 'Rest' diagrams are drawn schematically, 
to show
the difference between the interactions in 'diagonal' and 'rest'
sectors. Namely, for the 'rest' sectors we imply, that the 
corresponding momentum 
exchange must be done in the diagrams of \fig{feynrules} to get 
analytical expressions for diagrams depicted in Table \ref{table}.

In the diagramatic rules we write explicitly the dependence of the 
electron (photon) mass on the cutoff
\be
m_{\la}^2=m_0^2-\de\Sigma_{\la}
\; , \lee{re4}
where $\de\Sigma_{\la}$ is the self energy term to the second order
($m_0=0$ for a photon); we drop the subscript $\la$ for the polarization
vectors $\varepsilon$ and spinors $\chi$. To the next, third, order
one has $e_{\la}=e_0(1+O(e_{\la}^2))$.

The sense of the Table \ref{table} is transparent. 
Elimination to the second order
of canonical electron-photon vertex with flow equations gives rise
to one-particle and two-particle operators. The elimination of the 'rest'
sectors to the next orders generate many-particle operators. One is able
therefore to trancate the effective Hamitonian to the few-particle
sectors only in the case when the perturbative theory expansion is true.

As $\la\rightarrow 0$ the 'rest' sectors are completly
(here to the order $O(e^2)$) eliminated and Fock states in 'diagonal'
sectors are decoupled. One ends up with the renormalized effective
LFQED Hamiltonian.

The two-component LF theory, introduced by Zhang and Harindranath [5],
as compared with four-component formalism of Brodsky and Lepage,
is formulated purely in terms of physical degrees of freedom;
so that each term in the renormalized effective Hamiltonian 
corresponds to a real dynamical process (or give renormalization term).
Therefore the 'diagonal' sectors of the renormalized effective Hamiltonian
contribute for different Fock states to: $|\ga>$ - self energy photon operator,
$|e\bar{e}>$ - electron-positron bound state (or scattering),
$|\ga\ga>$ - light-light scattering, $|e\bar{e}\ga>$ - Compton scattering, 
and so on (see Table \ref{table}).

In the next section we use the diagramatic rules given in the Table \ref{table} 
(at $\la\neq 0$) to calculate the effective electron-positron interaction,
which includes generated interaction, instantaneous term and perturbative
photon exchage.

\section{Positronium bound state}

\subsection{Light front perturbative theory}

The scattering $|e\bar{e}>$ states are also needed 
in bound state calculations.  Using the propagator
techniques we include these states where required.
We exploit the perturbative theory in the coupling constant
$e$, using the diagramatic rules for the renormalized effective 
theory \fig{feynrules}.

The first order renormalized $ee\ga$-vertex \mbox{$f_{p_ip_f}H_{can}^{ee\ga}$}
contributes to the second order
to the $|e\bar{e}>$ interaction term and to the 
electron (photon) mass renormalization.
Physically, it is the perturbative photon exchange 
(photon emission and absorbtion in the case of electron mass renormalization),
with the energy widths of the photon restricted by 
the function $f_{p_ip_f,\la}$.

\subsubsection{Electron-positron interaction}

According to the light-front Feynman rules
the perturbative photon exchange gives rise to the following
second order $|e\bar{e}>$ interaction in the exchange channel 
\be
V(l)=g_1(l)g_2(l)M_{2ii} \cdot
\left\{ \frac{\th(q^+)}{q^+}\frac{1}{p_i^--p_k^-}
     + \frac{\th(-q^+)}{(-q^+)}\frac{1}{p_i^--p_k^-} \right\}
\; , \lee{epi1}
where $g_i$ stands schematically for the coupling constants in both vertices,
namely $g_{p_1p_2\la}=ef_{p_1p_2\la}$ and $M_{2ii}$ defines the spin structure of 
the interaction, coming from the corresponding structure of the $ee\ga$-vertex;
and the two terms in the curly brackets represent two different $x^+$ (time) orderings
of the photon exchange with the momentum $q$, giving rise to the two different
intermediate states with momenta $p_k$, $p_i$ corresponds to the initial state.
Explicitly one has

In the {\bf exchange channel} 
\bea
\hspace{0em} V_{PT}^{(ex)}(l) &=& -e^2(f_{-p_4,-p_2}(l)
\ch_{\bar{s}_2}^+\Ga^i(-p_4,-p_2,-q)\ch_{\bar{s}_4})
(f_{p_1,p_3}(l)
\ch_{s_3}^+ \Ga^i(p_1,p_3,q)\ch_{s_1}) \\
&&\hspace{2em} \times \left[ \frac{\th(p_1^+-p_3^+)}{(p_1^+-p_3^+)}\frac{1}{p_i^--p_3^--p_2^--q^-}
+\frac{\th(p_3^+-p_1^+)}{(p_3^+-p_1^+)}\frac{1}{p_i^--p_1^--p_4^-+q^-} \right] \,
\bar{\de}_{q,p_1-p_3} \nn
\leea{epi2}
with the initial state momentum $p_i=P=(P^+,P^{\bot})$ and momentum transfer
$q=p_1-p_3$, and
\be
P^-=\frac{P^{\bot 2}+M_N^2}{P^+}
\; , \lee{epi3}
where $M_N$ is the mass of positronium bound state.
In the light-front frame holds
\bea
-(P^--p_3^--p_2^--(p_1-p_3)^-)\th(p_1^+-p_3^+)
&=&(P^--p_1^--p_4^-+(p_1-p_3)^-)\th(p_3^+-p_1^+)\nonumber\\
&=&\frac{\tilde{\De}_3}{P^+(x-x')}
\leea{epi4}
that gives rise for the rescaled potential $V=\tilde{V}/P^{+ 2}$ 
\be
\tilde{V}_{PT,\la}^{(ex)} = -e^2N_1 \frac{1}{\tilde{\De}_3}
\exp\left( -\frac{(\De_1^2+\De_2^2)}{\la^4} \right)
\; , \lee{epi5}
where $N_1$ is defined in \eq{gi16} and
\bea
\tilde{\De}_3=(k_{\bot}-k'_{\bot})^2 + \frac{1}{2}(x-x')A+|x-x'|
\left( \frac{1}{2}(M_0^2+M_0^{'2})-M_N^2 \right)
\nonumber\\
\\
A=(k_{\bot}^2+m^2) \left( \frac{1}{1-x}-\frac{1}{x} \right)
+(k_{\bot}^{'2}+m^2) \left(\frac{1}{x'}-\frac{1}{1-x'} \right)
\nonumber
\; , \leea{epi6}
Here the cutoff $\la$ is defined in units of $P^+$, namely
$\la\rightarrow \la^2/P^+$.

Because of the absence of Z-graphen in light-front formalism (corresponding to
negative $p^+$), only one term contribute to the {\bf annihilation channel},
namely  

\bea
V_{PT}^{(an)}(l)&=&e^2(f_{p_1,-p_2}(l)
\ch_{\bar{s}_2}^+\Ga^i(p_1,-p_2,q)\ch_{s_1})
(f_{-p_4,p_3}(l)
\ch_{s_3}^+\Ga^i(-p_4,p_3,-q)\ch_{\bar{s}_4})\nonumber\\
&& \times \left[ \frac{1}{(p_1^++p_2^+)}\frac{1}{p_i^--q^-} \right]
\bar{\de}_{q,p_1+p_2}
\; , \leea{epi7}
where $p_i^-=P^-$ and the momentum transfer is $q=p_1+p_2$. This gives rise
for the rescaled potential $V=\tilde{V}/P^{+2}$ in the light-front frame,
to the expression
\be
\tilde{V}_{PT,\la}^{(an)} = e^2N_2\frac{1}{M_N^2}
\exp\left\{ -\frac{(M_0^4+M_0^{'4})}{\la^4} \right\}
\; , \lee{epi8}
where $N_2$ and the variables $\De_1, \De_2$ and $M_0^2, {M'}_0^2$
are defined in \eq{gi16}.

\subsubsection{Mass renormalization}

Following light-cone rules the perturbative energy correction
of the electron with momentum $p$, coming from the emission and 
absorption of a photon with momentum $k$, is
\bea
&& \de \tilde{p}_{1\la}^-=\int\frac{d^2k^{\bot}dk^+}{2(2\pi)^3}\frac{\th(k^+)}{k^+}
\th(p^+-k^+)g_{p-k,p,\la}\Ga^i_{\la}(p-k,p,-k)
g_{p,p-k,\la}\Ga^i_{\la}(p,p-k,k)\nn\\
&&\hspace{5em} \times\frac{1}{p^--k^--(p-k)^-}
\; , \leea{mr1}
where $g_{ee\ga}$-coupling constant restricts the energy of the
photon.
Making use of the explicit form for the coupling, one has
\bea
\de \tilde{p}_{1\la}^- &=& e^2\int
 \frac{d^2k^{\bot}dk^+}{2(2\pi)^3}\frac{\th(k^+)}{k^+}
 \th(p^+-k^+) \\
&&\hspace{3em} \times \Ga^i_{\la}(p-k,p,-k) \,
 \Ga^i_{\la}(p,p-k,k)\frac{1}{p^--k^--(p-k)^-}
 \times (R) \nn
\; , \leea{mr1a}
where $R=f_{pk\la}^2$ plays the role of regulator. 
This expression coincide up to the overall sign with
the energy correction obtained in the previous section from the flow equations method.

Two instantaneous diagramms, arising from the normal-ordering Hamiltonian,
must be added to the first term with the same regulator $R$.
Then the full perturbative energy correction
$\de\tilde{p}_{\la}^-=\de\tilde{p}_{1\la}^-+\de\tilde{p}_{2\la}^-+
\de\tilde{p}_{3\la}^-$ is
\be
\de\tilde{p}_{\la}^-=-\de p_{\la}^-
\ee
where $\de p_{\la}^-$ is defined in \eq{ri6}. This means for the perturbative mass
correction
\be
\de m_{\la}^{PT2}=\de\Sigma_{\la}
\lee{mr1b}
and the self-energy term $\de \Sigma_{\la}$ is given in \eq{ri24}.

We combine the renormalized to the second order mass, \eq{ri5},
and the perturbative correction, \eq{mr1b}, to obtain
the total physical mass to the order $O(e^2)$ 
\be
m_e^2=m_{\la}^2+\de m^2=(m^2+\de\Sigma_{\la})-\de\Sigma_{\la}
=m^2+O(e^4)
\; . \lee{mr3}
This means, that to the second order $O(e^2)$ the physical electron mass is,
up to a finite part, equal to the bare electron mass, that stands 
in the free (canonical) Hamiltonian.

Along the same line one can do for the photon mass.

At the end we note, that the similarity function $f_{p_ip_f\la}$,
restricting the electron-photon vertex, plays 
the role of UV (and partialy IR) regulator in the self energy integrals.
This means, that the regularization prescription of divergent 
integrals follows from the method of flow equations itself.
Moreover, the energy correction (i.e. mass correction
and wave function renormalization constant), obtained from the flow equations,
coincide up to the overall sign with the $1$-loop renormalization group
result. This is the remarkable result, indicating to the equivalence of
flow equations and Wilson's renormalization.

\subsection{Bound state perturbative theory}

In this subsection we define bound state perturbative theory (BSPT).

First introduce instead of the light front parametrization, used before
for the single-particle momenta, \fig{reneebarint}, the instant form
\bea
& p_{1\mu}=(xP^+,xP^{\bot}+k^{\bot},p_1^-) \quad
\stackrel{J(p)}{\longrightarrow} \quad
p_{1\mu} = (k_z, k^{\bot},p_1^0)=(\vec{p}_1,E_1) & \nn \\
& p_{2\mu}=((1-x)P^+,(1-x)P^{\bot}-k^{\bot},p_2^-) \quad
\longrightarrow \quad
p_{2\mu} = (-k_z, -k^{\bot},p_2^0)=(\vec{p}_2,E_2) & \nn \\
& E_i=\sqrt{\vec{p}^2+m^2}, \quad i=1,2 & \nn\\
& x =\frac{E_1+k_z}{E_1+E_2}= \frac{1}{2} \, \left( 
1 + \frac{k_z}{\sqrt{\vec{p}^2 + m^2}} 
\right) & \nn \\
\leea{bspt1}
and for the momenta $p_3,p_4$ the same, but with prime over
\mbox{$x, k_z, k^{\bot}$}; here $x$ is the light-front fraction of the
electron momentum, and $J(p)$ is the Jacobian of the transformation:
\be
J(p)=\frac{dx}{d k_z}=\frac{k_{\bot}^2+m^2}{2(\vec{p}^2+m^2)^{3/2}}
\; . \lee{bspt2}

The instant form system is usefull from the practical point of view:
it is easy to see then the rotational symmetry, restored in the 
nonrelativistic case and manifest in positronium spectrum.

Define now {\bf BSPT}. We choose the leading order 
electron-positron potential in such a form to simplify positronium 
bound state calculations. This means, that this potential contributes 
the leading order term to the positronium mass, and perturbative 
theory with respect to the difference between the total
second order $|e\bar{e}>$ interaction, calculated before with 
the renormalized Hamiltonian to $O(e^2)$, and the leading order 
potential converges. This scheme we call BSPT. Surely, our choice 
is motivated by the form of the renormalized to the second order 
interaction to insure convergence of BSPT.

In MSR scheme, where the 'diagonal' part is given
by the Fock state conserving part of the Hamiltonian, 
the choice of the zero'th order potential for BSPT calculations
is quit natural. It corresponds to the leading order in coupling
of effective Hamiltonian in 'diagonal' sector. This insures the convergence
of bound state calculations. Note, that in SR scheme the 'diagonal'
part is given by free Hamiltonian that can not surve for the starting
approximation in bound state perturbative theory.  

We define the second order renormalized electron-positron potential
\mbox{$\left. \left< \! e(3)\bar{e}(4) \right|
\hat{V}_{coul} \left| e(1)\bar{e}(2) \! \right> \right.$}
to the leading order of BSPT in the form of pure perturbative one photon
exchange, explicitly as the Coulomb interaction 
\be
V_{coul} = -\frac{16 e^2 m^2}{(k_{\bot} - k'_{\bot})^2 + \\
(k_z - k'_z)^2} = -\frac{16 e^2 m^2}{(\vec{p}-\vec{p'})^2}
\; . \lee{bspt4}
This means that the corresponding leading order Hamilton operator 
in the \mbox{$|e\bar{e}>$} sector is
\be
H^{(0)} = h + \hat{V}_{coul}
\; , \lee{bspt5}
where $h$ is the free part, defined in \eq{ch13}. The wave functions are given
as the solution of Schr\"odinger equation
\be
H^{(0)} |\psi_N(P)> = E_N |\psi_N(P)>
\; , \lee{bspt6}
where $P$ is the positronium momentum, and the eigenvalues and 
eigenfunctions for the positronium bound state are defined in standard way
of light front frame
\bea
& E_N = \frac{P_{\bot}^2 + M_N^2}{P^+} & \nn \\
& |\psi_N(P)> = \sum_{s_1 s_2} \, \int_{p_1 p_2} \, \sqrt{p_1^+ p_2^+} \, 
2(2\pi)^3 \, \de^{(3)}(P - p_1 p_2) \, \tilde{\Phi}_N (x k_{\bot} s_1 s_2) \,
b^+_{s_1}(p_1) \, d^+_{s_2}(p_2)|0> & \nn \\
& \sum_{s_1s_2} \, \frac{\int d^2k_{\bot} \, \int_0^1 dx}{2(2\pi)^3} \,
\tilde{\Phi}_N^*(x k_{\bot} s_1 s_2) \,
\tilde{\Phi'}_N(x k_{\bot} s_1 s_2) = \de_{NN'} &
\leea{bspt7}
$M_N$ stands for the leading order mass of positronium.
Combining the definitions for the wave function and the energy 
with the Schr\"odinger equation, we obtain
\be
\biggl[ M_N^2-\frac{k_{\bot}^{'2}+m^2}{x'(1-x')} \biggr]
\tilde{\Phi}_N (x'k'_{\bot}s_3s_4) = \sum_{s_1 s_2}
\frac{\int d^2k_{\bot} \int_0^1 dx}{2(2\pi)^3} \,V_{coul} \, 
\tilde{\Phi}_N(xk_{\bot}s_1s_2)
\; , \lee{bspt8}
or, after change of coordinates according to \eq{bspt1},
\be
\left( M_N^2 - 4 (\vec{p'}^2 + m^2) \right) \Phi_N(\vec{p'} s_3 s_4) =
\sum_{s_1 s_2} \frac{\int d^3p \sqrt{J(p) J(p')}}{2(2\pi)^3} \,
V_{coul}(\vec{p},\vec{p'}) \, \Phi_N(\vec{p} s_1 s_2)
\; , \lee{bspt9}
where the wave function was redefined to have the norm
\be
\sum_{s_1s_2} \, \int d^3p \, \Phi_N^*(\vec{p} s_1 s_2) \,
\Phi_N'(\vec{p} s_1 s_2) = \de_{NN'} 
\; . \lee{bspt10}
We aim to obtain the nonrelativistic Schr\"odinger equation for positronium.
Note, that in the nonrelativistic limit \mbox{$\frac{\vec{p}^2}{m^2} <\!\!< 1$}
we have 
\bea
& \sqrt{J(p) J(p')} \; \approx \;
\frac{1}{2m} \left( 1 - \frac{\vec{p}^2 + (k_z^2 + {k'_z}^2)} {2m^2} \right) & \nn \\
& M_N \; = \; (2m + B_N)^2 \; \approx \; 4m^2 + 4m B_N^{(0)} &
\; , \leea{bspt11}
where we have introduced the leading order binding energy $B_N^{(0)}$.
Then to the leading order the bound state equation for positronium is
\bea
&& \left( \frac{\vec{p'}^2}{m} - B_N \right) \Phi_N(\vec{p'} s_3 s_4)
\; = \;- \sum_{s_1 s_2} 
\int d^3p \left( \frac{1}{2m} \frac{1}{2(2\pi)^3} \frac{1}{4m} \, V_{coul} \right) \,
\Phi_N(\vec{p}s_1s_2) 
\; . \leea{bspt12}

Making use of the explicit form for the Coulomb potential, \eq{bspt4}, we obtain
the equation that determines the leading order bound state wave function:
\be
\left( \frac{\vec{p'}^2}{m} - B_N \right) \, \Phi_{\mu}(\vec{p'}) \; = \;
\frac{\al}{2\pi^2} \, \int \, \frac{d^3p}{(\vec{p}-\vec{p'})^2} \, \Phi_{\mu}(\vec{p})
\lee{bspt13}
with
\be
\Phi_N \; = \; \Phi_{\mu, s_e, s_{\bar{e}}}(\vec{p'} s_3 s_4) \; = \;
\Phi_{\mu}(\vec{p'}) \, \de_{s_e s_3} \, \de_{s_{\bar{e}} s_4}
\; . \lee{bspt14}

This is the standard nonrelativistic Schr\"odinger equation 
for positronium. Its solution is characterized by
\mbox{$\mu \! = \!(n,l,m)$}, the usual principal and angular momentum quantum numbers.
The wave functions are given through the hyperspherical harmonics
\bea
&& Y_{\mu}(\Om) \; = \;
\frac{(e_n^2 + \vec{p}^2)^2}{4 \, e_n^{5\!/\!2}} \, \Phi_{\mu} \nn \\
&& Y_{\mu} \; = \; Y_{n,l,m} \; = \; f_{n,l}(\om) \, Y_{l,m}(\theta,\phi) \nn \\
&& B_N \; = \; -\frac{m \alpha^2}{4n^2},\qquad e_n \; = \; \frac{m\alpha}{2n} 
\leea{bspt15}
and for the binding energy one has the standard 
nonrelativistic expression for positronium bound state to $O(e^2)$.
For sake of completeness we write the coordinates used in the solution
\bea
&& (e_n^2 = -m B_N, \vec{p}) \; \longrightarrow \; (u_0, \vec{u}) \nn \\
&& u_0 \; = \; \cos\om \; = \; \frac{e_n^2 - \vec{p}^2}{e_n^2 + \vec{p}^2} \nn \\
&& \vec{u} \; = \; \frac{\vec{p}}{|\vec{p}|}
\sin\omega \; = \; \frac{2e_n \vec{p}}{e_n^2 + \vec{p}^2} 
\; , \leea{bspt16}
but, for details, refer to \cite{JoPeGl}.

The electron-positron interaction arising from the renormalized
to the $O(e^2)$ Hamiltonian is given as a sum of two contributions
from exchange and annihilation channels \mbox{$V=V_{exch}+V_{ann}$}
(see explicitly later). We introduce the potential, arising in the
nonrelativistic Schr\"odinger equation, \eq{bspt13},
\be
V'(\vec{p'}s_3s_4;\vec{p}s_1s_2) =
\lim_{\frac{\vec{p}^2}{m^2}<<1} \frac{\sqrt{J(p)J(p')}}{2(2\pi)^3}\frac{1}{4m}
(\tilde{V}_{exch}+\tilde{V}_{ann})
\; . \lee{bspt17}
Then we define BSPT with respect to the difference
\be
\de V=V'(\vec{p'}s_3s_4;\vec{p}s_1s_2)
-(-\frac{\alpha}{2\pi^2})\frac{1}{(\vec{p}-\vec{p'})^2}
\de_{s_1s_3}\de_{s_2s_4}
\; , \lee{bspt18}
where the leading order contribution is defined in \eq{bspt15}.
Note, that, in order to define the Coulomb potential, i.e.
the $e\bar{e}$ interaction to the leading order of BSPT,
we have taken only the first
term of nonrelativistic expansion of the Jacobian $J(p)$. 

In what follows we use the matrix elements of $\de V$, defined as 
\bea
&& <\Phi_{nlm}|\de V|\Phi_{nlm}>=\int d^3pd^3p'
\Phi_{nlm}^*(\vec{p})\de V\Phi_{nlm}(\vec{p'})
\; , \leea{bspt19}
where $\Phi_{nlm}$ are the Coulomb wave functions given above.

\subsection{\label{4.3}Renormalized effective electron-positron interaction}

In this section we consider the properties of effective electron-positron
interaction, obtained by flow equations and also by the general unitary
transfromation in MSR scheme.

\subsubsection{Renormalized effective electron-positron interaction
in light front and instant frames}

We summarize together all the terms defining the electron-positron
interaction, obtained by flow equations, in exchange and annihilation
channels.
\bea
\tilde{V}_{exch} &=& \tilde{V}_{\la}^{exch} + \tilde{V}^{PT}
 \;=\; \tilde{V}_{\la}^{gen} + \tilde{V}_{\la}^{inst} + 
\tilde{V}_{\la}^{PT} \nn \\ 
\tilde{V}_{ann} &=& \tilde{V}_{\la}^{ann} + \tilde{V}^{PT}
 \;=\; \tilde{V}_{\la}^{gen} + \tilde{V}_{\la}^{inst} + 
\tilde{V}_{\la}^{PT}
\; , \leea{repi1}
First we use the {\bf light front frame}.
The generated, instantaneous and perturbative theory interactions
(rescaled, i.e. $V_{\la}=\tilde{V}_{\la}/P^{+2}$) are given corresponding 
in the {\bf exchange channel}
\bea
\tilde{V}_{\la}^{gen} &=& -e^2 N_{1,\la}\left(
\frac{\tilde{\De}_1+\tilde{\De}_2}{\tilde{\De}_1^2+\tilde{\De}_2^2}\right)
 \left(1 - {\rm e}^{-\frac{\De_1^2+\De_2^2}{\la^4}}\right) \nn\\
\tilde{V}_{\la}^{inst} &=& -\frac{4 e^2}{(x - x')^2} \: \de_{s_1 s_3} \: 
\de_{s_2 s_4} \nn\\
\tilde{V}_{\la}^{PT} &=& -e^2 N_{1,\la}\frac{1}{\tilde{\De}_3}
{\rm e}^{-\frac{\De_1^2+\De_2^2}{\la^4}}
\leea{repi2}
in the {\bf annihilation channel}
\bea
\tilde{V}_{\la}^{gen} &=& e^2 N_{2,\la}\left(
\frac{M_0^2+M_0^{'2}}{M_0^4+M_0^{'4}}\right)
\left(1 - {\rm e}^{-\frac{M_0^4+M_0^{'4}}{\la^4}}\right) \nn\\
\tilde{V}_{\la}^{inst} &=& 4 e^2
 \: \de_{s_1 \bar{s}_3} \: \de_{s_2 \bar{s}_4} \nn\\
\tilde{V}_{\la}^{PT} &=& e^2 N_{2,\la}\frac{1}{M_{\!N}^2}
{\rm e}^{-\frac{M_0^4+M_0^{'4}}{\la^4}}
\; , \leea{repi3}
where in the light-front frame, see \eq{b14} 
and following in Appendix \ref{B}
\bea
N_{1,\la}&=&\delta_{s_1s_3}\delta_{s_2s_4}
 T_1^{\bot}\cdot T_2^{\bot}
 -\delta_{s_1\bar{s}_2}\delta_{s_1\bar{s}_3}\delta_{s_2\bar{s}_4}
 2m^2\frac{(x-x')^2}{xx'(1-x)(1-x')}\nonumber\\
&&+im\sqrt{2}(x'-x) \left[ \delta_{s_1\bar{s}_3}\delta_{s_2s_4}
 \frac{s_1}{xx'}T_1^{\bot}\cdot \varepsilon_{s_1}^{\bot}
 +\delta_{s_1s_3}\delta_{s_2\bar{s}_4}
 \frac{s_2}{(1-x)(1-x')}T_2^{\bot}\cdot \varepsilon_{s_2}^{\bot} \right]
\nn\\
N_{2,\la}&=&\delta_{s_1\bar{s}_2}\delta_{s_3\bar{s}_4}
 T_3^{\bot}\cdot T_4^{\bot}
 +\delta_{s_1s_2}\delta_{s_3s_4}\delta_{s_1s_3}
 2m^2\frac{1}{xx'(1-x)(1-x')}\nonumber\\
&&+im\sqrt{2} \left[ \delta_{s_3\bar{s}_4}\delta_{s_1s_2}
 \frac{s_1}{x(1-x)}T_3^{\bot}\cdot \varepsilon_{s_1}^{\bot}
 -\delta_{s_3s_4}\delta_{s_1\bar{s}_2}
 \frac{s_3}{x'(1-x')}T_4^{\bot}\cdot \varepsilon_{s_4}^{\bot *} \right] \nn \\
&& \varepsilon_s^i = -\frac{1}{\sqrt{2}}(s, i)
\leea{repi4}
and
\bea
T_1^i&=&- \left[ 2\frac{(k_{\bot}-k'_{\bot})^i}{(x-x')}+\frac{k_{\bot}^i(s_2)}{(1-x)}+
 \frac{k_{\bot}^{'i}(\bar{s}_2)}{(1-x')} \right] \; ; \qquad
T_2^i=2\frac{(k_{\bot}-k'_{\bot})^i}{(x-x')}-\frac{k_{\bot}^i(s_1)}{x}-
 \frac{k_{\bot}^{'i}(\bar{s}_1)}{x'} \nonumber \\
T_3^i&=&-\frac{k_{\bot}^{'i}(\bar{s}_3)}{x'}
 +\frac{k_{\bot}^{'i}(s_3)}{(1-x')} \; ; \qquad
T_4^i=\frac{k_{\bot}^i(\bar{s}_1)}{(1-x)}
 -\frac{k_{\bot}^i(s_1)}{x}\nonumber\\
&& k_{\bot}^i(s) = k_{\bot}^i+is\varepsilon_{ij}k_{\bot}^j \; ; \qquad
 \varepsilon_{ij}=\varepsilon_{ij3} \; ; \qquad
 \bar{s} = -s  \nn
\leea{repi5}
with the definitions
\bea
\tilde{\De}_1 &=& \frac{(xk'_{\bot}-x'k_{\bot})^2+m^2(x-x')^2}{xx'}\; ; \qquad
 \tilde{\De}_2=\De_1|_{x\rightarrow(1-x),x'\rightarrow(1-x')} \nn \\
&&\De_1=\frac{\tilde{\De}_1}{x'-x} \; ; \qquad
 \De_2=\frac{\tilde{\De}_2}{x'-x} \nn \\
\tilde{\De}_3 &=& (k_{\bot}-k'_{\bot})^2+\frac{1}{2}(x-x')A+|x-x'|
 \left( \frac{1}{2}(M_0^2+M_0^{'2})-M_N^2 \right) \nn \\
&&M_0^2=\frac{k_{\bot}^2+m^2}{x(1-x)} \; ; \qquad
 M_0^{'2}=\frac{k_{\bot}^{'2}+m^2}{x'(1-x')} \nn \\
&&A=(k_{\bot}^2+m^2) \left( \frac{1}{1-x}-\frac{1}{x} \right)
 +(k_{\bot}^{'2}+m^2) \left( \frac{1}{x'}-\frac{1}{1-x'} \right) \nn \\
&&P^-=\frac{(P^{\bot})^2+M_N^2}{P^+} \; ; \qquad P=(P^+,P^{\bot}) \; ; \qquad
 M_N = 2m + B_N
\; . \leea{repi6}
Note, that the rescaled potential, \eq{repi1}, does not depend on the
total momentum $P^+$, i.e. is invariant under light-front boosts.

Second, we use the {\bf instant frame}. We rewrite the electron-positron
interaction in both exchange and annihilation channels as follows
\bea
\tilde{V} &=& \tilde{V}_{exch}+\tilde{V}_{ann} \nn \\
&=&-e^2N_{1,\la} \left[
\left(\frac{\tilde{\De}_1+\tilde{\De}_2}
{\tilde{\De}_1^2+\tilde{\De}_2^2}\right) 
(1-{\rm e}^{-\frac{\De_1^2+\De_2^2}{\la^4}})c_{ex}^{gen}
+\frac{1}{\tilde{\De}_3}
{\rm e}^{-\frac{\De_1^2+\De_2^2}{\la^4}}c_{ex}^{PT} 
\right] \nn \\
&&+\left( -\frac{4 e^2}{(x - x')^2} \: \de_{s_1 s_3} \: 
 \de_{s_2 s_4} \right)c_{ex}^{inst}\nn\\
&&+e^2N_{2,\la} \left[ 
\left(\frac{M_0^2+M_0^{'2}}{M_0^4+M_0^{'4}}\right)
(1-{\rm e}^{-\frac{M_0^4+M_0^{'4}}{\la^4}})c_{an}^{gen}
+\frac{1}{M_N^2}
{\rm e}^{-\frac{M_0^4+M_0^{'4}}{\la^4}}c_{an}^{PT}
 \right] \nn \\
&&+\left( 4 e^2
 \: \de_{s_1 \bar{s}_3} \: \de_{s_2 \bar{s}_4} \right)c_{an}^{inst}
\; , \leea{repi11}
where one has in the instant frame
\bea
&& x=\frac{1}{2}\left(1+\frac{k_z}{\sqrt{\vec{p}^2+m^2}}\right) \; ;\qquad
x'=\frac{1}{2}\left(1+\frac{k_z^{'}}{\sqrt{\vec{p}^2+m^2}}\right) \nn \\
&& M_0^2=4(\vec{p}^2+m^2) \; ;\qquad
M_0^{'2}=4(\vec{p'}^{2}+m^2)
\; ; \leea{repi12}
and for all the quantities, defined in \eqs{repi4}{repi5}, the substitution 
$x(k_z), x'(k_z^{'})$ is to be done. The symbols $c_{ex}^{gen}$ 
and others are introduced to indicate the origin of the different terms 
(here generated interaction coming from the exchange channel); all $c=1$. 

The expression \eq{repi11} has the physical interpretation.
First notice, that
$\De_1$ and $\De_2$ descibe the energy differences
(energy transfer) in two corresponding
$ee\ga$-vertices appearing in the electron-positron interaction. Then
the generated interaction ($c_{ex}^{gen}$) contributes
mainly hard photon exchanges
\mbox{$\frac{\De_1}{\la^2} \sim \frac{\De_2}{\la^2}>>1$},
while the term arising from perturbative
theory $c_{ex}^{PT}$ gives rise to soft photons.
Though the effective electron-positron interaction generally
describes low energy physics, namely the renormalized effective
Hamiltonian is constructed to eliminate perturbative (relativistic)
photon exchanges, the information on the high energy sector
is accumulated in the generated interaction.
This makes possible to interpolate between two sectors, i.e.
the sum of both terms in \eq{repi11} recovers the
whole range of photon energies.

In the next section we analyse the general properties
of the effective electron-positron interaction,
generated by the unitary transformation in MSR scheme.

\subsubsection{Electron-positron interaction in MSR}

We give here the general analyses of the electron-positron interaction,
obtained in MSR scheme. Basing on this analyses, we calculate in the 
next section positronium fine structure.

\begin{figure}
\setlength{\unitlength}{1mm}
\begin{picture}(170,71)
\put(19,36){\makebox(56,34.61){ \loadeps{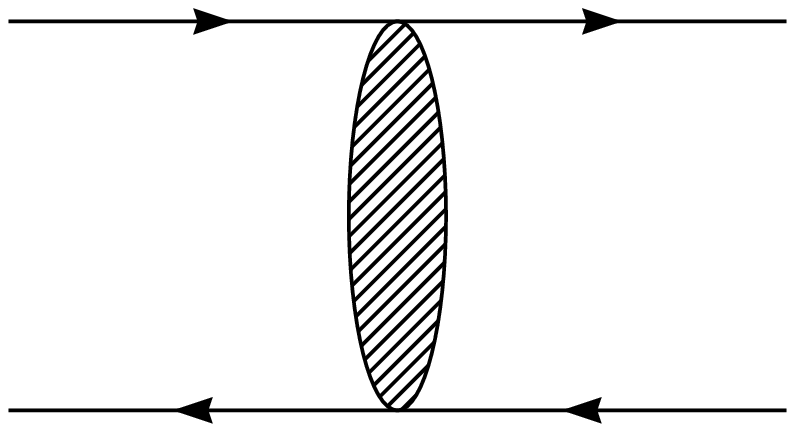} }}
\put(18,67){$p_1\;(x,k^\bot)$}
\put(50,67){$p_3\;(x',k'^\bot)$}
\put(18,38){$p_2\;(1\!-\!x,-k^\bot)$}
\put(50,38){$p_4\;(1\!-\!x',-k'^\bot)$}
  \put(75,36){\makebox(56,34.61){ \loadeps{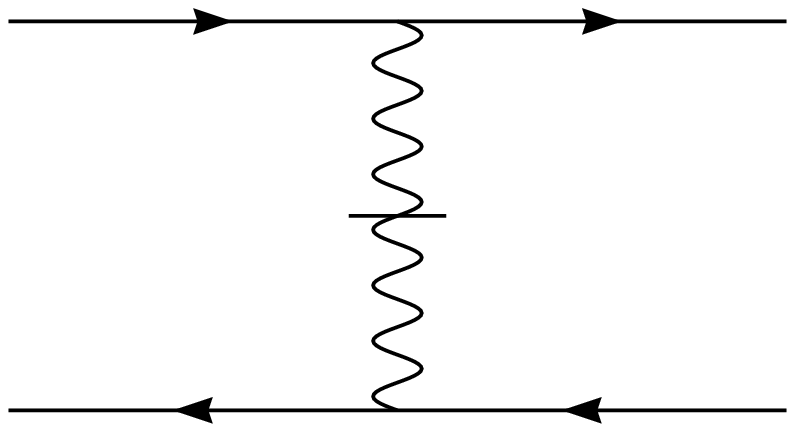} }}
\put(73,53.305){$+$}
\put(38,0){\makebox(56,34.61){ \loadeps{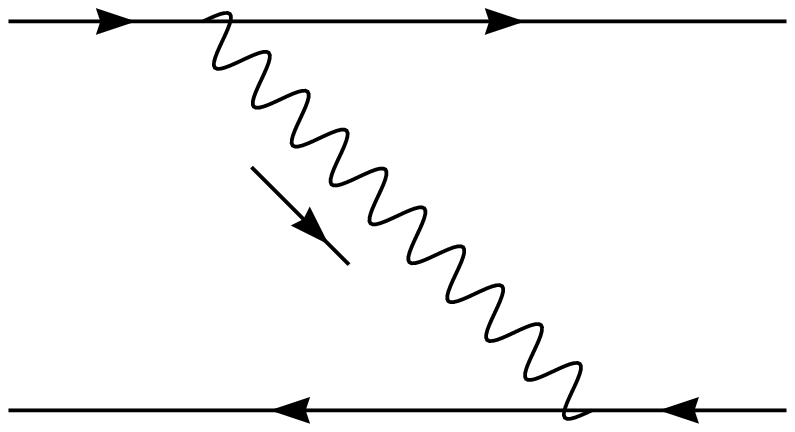} }}
  \put(94,0){\makebox(56,34.61){ \loadeps{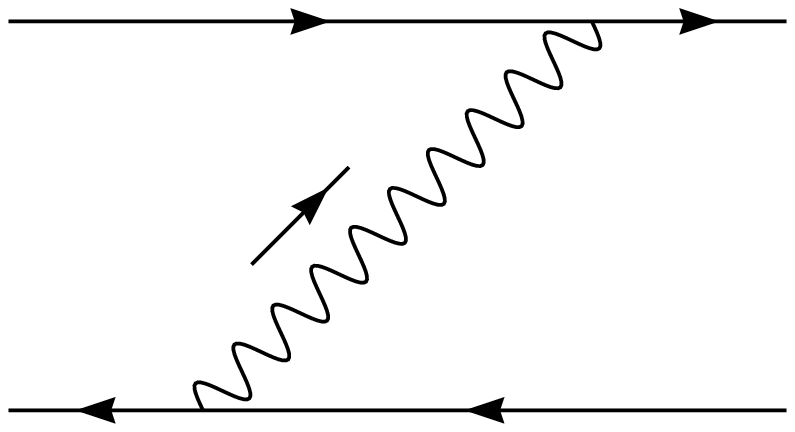} }}
\put(36,17.305){$+$}
\put(92,17.305){$+$}
\end{picture}
\vspace{3cm}
\caption{The renormalized to the second order effective electron-positron
interaction in the exchange channel; diagrams correspond to generated, 
instantaneous interactions and two perturbative photon exchanges 
with respect to different time ordering.}
\label{reneebarint}
\end{figure}

We summarize the interactions in $|e\bar{e}>$ sector, generated by
the unitary transformation of MSR (section \ref{3.3.2}), 
in exchange channel \fig{reneebarint}
\bea
\tilde{V}_{\la}^{gen} &=& -e^2N_{1,\la}
\left(\frac{\int_{\la}^{\infty}\frac{df_{1\la'}}{d\la'}f_{2\la'}d\la'}
{\tilde{\De}_1}+
\frac{\int_{\la}^{\infty}\frac{df_{2\la'}}{d\la'}f_{1\la'}d\la'}
{\tilde{\De}_2}\right)\nn\\
\tilde{V}_{\la}^{PT} &=& -e^2N_{1,\la}
\frac{1}{\tilde{\De}_3}f_{1\la}f_{2\la}\nn\\
\tilde{V}_{\la}^{inst} &=& -\frac{4e^2}{(x-x')^2}\de_{s_1s_3}\de_{s_2s_4}
\; , \leea{ep1}
where the similarity functions 
\be
f_{1\la}=f_{\la}(\De_1),f_{2\la}=f_{\la}(\De_2)
\; , \lee{ep2}
are specified in \eqs{gi7}{gi8} section \ref{3.3.2}; the energy denominators 
$\tilde{\De}_i, i=1,2,3$ together with $N_{1,\la}$ are given 
in \eqs{repi4}{repi5}.
Introduce the following notations
\bea
\theta_1 &=& \int_{\la}^{\infty}\frac{df_{1\la'}}{d\la'}f_{2\la'}d\la'\nn\\
\theta_2 &=& \int_{\la}^{\infty}\frac{df_{2\la'}}{d\la'}f_{1\la'}d\la'
\; , \leea{ep3}
Then the generated interaction is written
\be
\tilde{V}_{\la}^{gen}=-e^2N_{1,\la}
\left(\frac{1}{2}(\frac{1}{\tilde{\De}_1}+\frac{1}{\tilde{\De}_2})
(1-f_{1\la}f_{2\la})+
\frac{1}{2}(\frac{1}{\tilde{\De}_1}-\frac{1}{\tilde{\De}_2})
(\theta_1-\theta_2)\right)
\; , \lee{ep4}
where we have used the identities
\bea
\frac{\theta_1}{\tilde{\De}_1}+\frac{\theta_2}{\tilde{\De}_2} &=&
\frac{1}{2}(\frac{1}{\tilde{\De}_1}+\frac{1}{\tilde{\De}_2})
(\theta_1+\theta_2)+
\frac{1}{2}(\frac{1}{\tilde{\De}_1}-\frac{1}{\tilde{\De}_2})
(\theta_1-\theta_2)\nn\\
\theta_1+\theta_2 &=& 1-f_{1\la}f_{2\la}
\; , \leea{ep5}
The explicit form of similarity functions together 
with $\theta_i$ factors is

for Wegners flow equations
\bea
f_{\la}(\De_1) &=& {\rm e}^{-\frac{\De_1^2}{\la^4}}\nn\\
\theta_1 &=& \frac{\tilde{\De}_1^2}{\tilde{\De}_1^2+\tilde{\De}_2^2}
(1-f_{\la}(\De_1)f_{\la}(\De_2))
\; , \leea{ep6}

for the similarity transformation of Glazek,Wilson
\bea 
f_{\la}(\De_1) &=& \theta(\la^2-|\De_1|)\nn\\
\theta_1 &=& \theta(|\De_1|-|\De_2|)\theta(|\De_1|-\la^2)=
\theta(\tilde{\De}_1-\tilde{\De}_2)\theta(|\De_1|-\la^2)
\; , \leea{ep7}
where
\be
\De_1=\frac{\tilde{\De}_1}{x'-x},
\De_2=\frac{\tilde{\De}_2}{x'-x}
\; , \lee{ep8}
We stress once more, that the unitary transformation of MSR is constructed
in a way to avoid small energy denominators. Namely, the situation
$\De_1=\De_2=0$, corresponding to the process with no momentum transfer
($\tilde{\De}_1=\tilde{\De}_2=0$ means in light front coordinates
$x=x',k_{\perp}=k'_{\perp}$, i.e. $q=p_1-p_3=0$), is excluded from 
the generated interaction for any finite $\la >0$. Further we assume
for all positive $\la$
\be
\tilde{V}_{\la}^{gen}(q=0)=0, \forall \la\geq 0
\; , \lee{ep9}
This means also, that the diagonal matrix elements, left after the limit 
$\la\rightarrow 0$ is performed in the 'rest' sector to the first order
in coupling ($ee\ga$ vertex when $\tilde{\De}_1=0, \tilde{\De}_2=0$),
do not contribute to the second order generated interaction in
'diagonal' sector. Therefore one can assume, that the 'rest' sector
is completly eliminated by the unitary transformation of MSR.

In this limit ($\la\rightarrow 0$) the perturbative part of 
electron-positron interaction is zero
\be
\tilde{V}_{}^{PT}=0
\; , \lee{ep10}
and electron-positron interaction is defined by the instantaneous
interaction \eq{ep1} and the generated term at $\la\rightarrow 0$
\bea
\tilde{V}_{\la\rightarrow 0}^{gen} &=& -e^2N_{1,\la}
\left(\frac{\int_{0}^{\infty}\frac{df_{1\la'}}{d\la'}f_{2\la'}d\la'}
{\tilde{\De}_1}+
\frac{\int_{0}^{\infty}\frac{df_{2\la'}}{d\la'}f_{1\la'}d\la'}
{\tilde{\De}_2}\right)\nn\\
\tilde{V}_{\la\rightarrow 0}^{inst} &=& 
-\frac{4e^2}{(x-x')^2}\de_{s_1s_3}\de_{s_2s_4}
\; , \leea{ep11}
The expression
\be
\tilde{V}^{|e\bar{e}>}=\tilde{V}_{\la\rightarrow 0}^{gen}+
\tilde{V}_{\la\rightarrow 0}^{inst} 
\; , \lee{ep12}
gives the effective electron-positron interaction obtained in MSR.
Note, that when $\la\rightarrow 0$ no difficulties (compared with SR)
on the convergency of unitary transformation (MSR scheme) and (or)
convergency of perturbative theory occure. This is closely related
with the maintenance of the rotational invariance and gauge invariance 
of perturbative calculations.

The explicit form of generated interaction when $\la\rightarrow 0$
is given in the case of

flow equations
\be 
\tilde{V}_{\la\rightarrow 0}^{gen}=-e^2N_1
\left(\frac{\tilde{\De}_1+\tilde{\De}_2}{\tilde{\De}_1^2+\tilde{\De}_2^2}
\right)
\; , \lee{ep13}

similarity transformation
\be
\tilde{V}_{\la\rightarrow 0}^{gen}=-e^2N_1
\left(\frac{\theta(\tilde{\De}_1-\tilde{\De}_2)}{\tilde{\De}_1}+
\frac{\theta(\tilde{\De}_2-\tilde{\De}_1)}{\tilde{\De}_2}
\right)
\; , \lee{ep14}

We consider now several special cases to analyse 
the general expression \eq{ep1}.

I. Consider the {\bf energy conserving process} (EC), i.e. 
when the following condition is valid
\be
p_1^-+p_2^-=p_3^-+p_4^-
\; , \lee{ep15}
for the light front coordinates of the process \fig{reneebarint}.
This gives rise to 
\bea
\De_1 &=& \De_2\nn\\
f_{\la}(\De_1) &=& f_{\la}(\De_2)
\; , \leea{ep16}
Making use of the formula \eq{ep4}, one has for the generated interaction
\be
\tilde{V}_{\la}^{gen}|_{EC}=-e^2N_{1,\la}
\frac{1}{2}(\frac{1}{\tilde{\De}_1}+\frac{1}{\tilde{\De}_2})
(1-f_{1\la}f_{2\la})=-e^2N_{1,\la}\frac{1}{\tilde{\De}_1}
(1-f_{1\la}^2)
\; , \lee{ep17}
and for the perturbative part
\be
\tilde{V}_{\la}^{PT}|_{EC}=-e^2N_{1,\la}\frac{1}{\tilde{\De}_3}
f_{1\la}^2
\; , \lee{ep18}
where
\be
f_{1\la}=f_{\la}(\De_1)
\lee{ep19}
When the following condition holds
\be
M_N^2=\frac{1}{2}(M_0^2+M_0^{'2})
\; , \lee{ep20}
one has for the energy conserving process ($\tilde{\De}_1=\tilde{\De}_2$)
\be
\frac{\tilde{\De}_1+\tilde{\De}_2}{2}=\tilde{\De}_3 
\; , \lee{ep21}
This gives rise to
\bea
\tilde{V}_{\la}^{gen}|_{EC} &=& -e^2N_{1,\la}
\frac{1}{\tilde{\De}_3}(1-f_{\la}^2(\De_3))\nn\\
\tilde{V}_{\la}^{PT}|_{EC} &=& -e^2N_{1,\la}
\frac{1}{\tilde{\De}_3}f_{\la}^2(\De_3)
\; , \leea{ep22}
where we have introduced
\be
\De_3=\frac{\tilde{\De}_3}{x'-x}
\; , \lee{ep23}
Expression \eq{ep22} shows clearly the sense of the unitary transformation
in MSR scheme: how the elimination of $ee\ga$ vertex in the 'rest'
sector with the flow parameter $l$ (or cutoff $\la$) generates the new
interaction for the $|e\bar{e}>$ Fock component in 'diagonal' sector.
For the energy conserving process together with the condition \eq{ep20}
the resulting generated interaction (at $\la\rightarrow 0$) is given
by the pure one-photon exchange diagram (i.e. by the initial perturbative 
interaction at $\la=\La\rightarrow\infty$)   
\be
\tilde{V}_{\la\rightarrow 0}^{gen}|_{EC}=
\tilde{V}_{\la=\La\rightarrow\infty}^{PT}|_{EC}
\; . \lee{ep24}

II.We consider the {\bf collinear limit}, i.e. when $|x-x'|\rightarrow 0$,
and show, that though the instantaneous term is singular in this limit
the whole electron-positron interaction is finite. Namely the generated
to the second order interaction insures the absence of collinear
divergencies.

For this purpose we rewrite the electron-positron interaction \eq{ep1}
as follows
\bea
&& \tilde{V}_{\la}^{|e\bar{e}>}=\tilde{V}_{\la}^{gen}+
\tilde{V}_{\la}^{PT}+\tilde{V}_{\la}^{inst}\nn\\
&& = -e^2N_{1,\la}\frac{1}{\tilde{\De}_3}
-\frac{4e^2}{(x-x')^2}\de_{s_1s_3}\de_{s_2s_4}\nn\\
&& -e^2N_{1,\la}\left(
\frac{1}{2}(\frac{1}{\tilde{\De}_1}+\frac{1}{\tilde{\De}_2})-
\frac{1}{\tilde{\De}_3}\right)
(1-f_{1\la}f_{2\la})\nn\\
&& -e^2N_{1,\la}\frac{1}{2}(\frac{1}{\tilde{\De}_1}-\frac{1}{\tilde{\De}_2})
(\theta_1-\theta_2)
\; , \leea{ep25}
The first two terms (perturbative photon exchange and instantaneous 
interaction) define the initial value of electron-positron interaction
($\la=\La\rightarrow\infty$), the next terms are generated by the 
unitary transformation and describe the change of electron-positron
interaction with $\la$.

In the collinear limit $|x-x'|\rightarrow 0$ the $\la$-dependent
terms (namely similarity functions $f_{i\la}=f_{\la}(\frac{\De_i}{\la})$,
$\De_i=\frac{\tilde{\De}_i}{x'-x}$ and $\theta_i$ factors, $i=1,2$) 
are given at the value $\la=0$.

Since the singular contribution is carried by the instantaneous 
interaction we consider the whole interaction only in diagonal spin sector,
i.e. $\de_{s_1s_3}\de_{s_2s_4}$.

The first two terms can be written
\bea
&& -e^2N_{1,\la}(s_1s_2\rightarrow s_3s_4)\frac{1}{\tilde{\De}_3}
-\frac{4e^2}{(x-x')^2}\de_{s_1s_3}\de_{s_2s_4}\nn\\
&&\hspace{4cm}
= -\frac{2e^2}{\tilde{\De}_3}(\frac{1}{2}N_{1,\la}(s_1s_2\rightarrow s_3s_4)
+\tilde{\De}_3\frac{2}{(x-x')^2}\de_{s_1s_3}\de_{s_2s_4})
\; , \leea{ep26}
where $N_{1,\la}$, defined in \eq{repi4}, is given
\bea
&& \frac{1}{2}N_{1,\la}(s_1s_2\rightarrow s_3s_4)=
-2\frac{(\vec{k}_{\perp}-\vec{k}_{\perp}^{'})^2}{(x-x')^2}
-\frac{1}{(x-x')}\left(
k_{\perp}^2(\frac{1}{1-x}-\frac{1}{x})
-k_{\perp}^{'2}(\frac{1}{1-x'}-\frac{1}{x'})\right)\nn\\
&&\hspace{10cm} 
+ f(s_1s_2\rightarrow s_3s_4)\nn\\
&& f(++\rightarrow ++)=\frac{k_{\perp}k'_{\perp}}{xx'(1-x)(1-x')}
{\rm e}^{-i(\ph -\ph')}\nn\\
&& f(+-\rightarrow +-)=k_{\perp}^2\frac{1}{x(1-x)}
+k_{\perp}^{'2}\frac{1}{x'(1-x')}\nn\\
&&\hspace{4cm}
+k_{\perp}k'_{\perp}
\left(\frac{1}{xx'}{\rm e}^{i(\ph -\ph')}
+\frac{1}{(1-x)(1-x')}{\rm e}^{-i(\ph -\ph')}\right)\nn\\
&& \vec{k}_{\perp}=k_{\perp}(\cos\ph, \sin\ph)
\; , \leea{ep27}
We have used the notations '$+$'-spin up, '$-$'-spin down.

Assuming the condition 
\be
M_N^2=\frac{1}{2}(M_0^2+M_0^{'2})
\; , \lee{ep28a}
one has 
\bea
&& \tilde{\De}_3\frac{2}{(x-x')^2}=
   2\frac{(\vec{k}_{\perp}-\vec{k}_{\perp}^{'})^2}{(x-x')^2}
+\frac{1}{(x-x')}(k_{\perp}^2(\frac{1}{1-x}-\frac{1}{x})
-k_{\perp}^{'2}(\frac{1}{1-x'}-\frac{1}{x'}))\nn\\
&&\hspace{8cm} + m^2(\frac{1}{xx'}+\frac{1}{(1-x)(1-x')})
\; , \leea{ep28b}
Combining all terms together, we get
\bea
&& -e^2N_{1,\la}(s_1s_2\rightarrow s_3s_4)\frac{1}{\tilde{\De}_3}
-\frac{4e^2}{(x-x')^2}\de_{s_1s_3}\de_{s_2s_4}=\nn\\
&&\hspace{3cm} -\frac{2e^2}{\tilde{\De}_3}
\left(m^2(\frac{1}{xx'}+\frac{1}{(1-x)(1-x')})
+f(s_1s_2\rightarrow s_3s_4)\right)
\; , \leea{ep29}
This part is finite, i.e. collinear singularity in the instantaneous
interaction is cancelled exactly by the perturbative part. To the leading
order in $\de x=x-x'$ one has
\bea
&& -e^2N_{1,\la}(s_1s_2\rightarrow s_3s_4)\frac{1}{\tilde{\De}_3}
-\frac{4e^2}{(x-x')^2}\de_{s_1s_3}\de_{s_2s_4}=\nn\\
&&\hspace{1cm} -\frac{2e^2}{(\vec{k}_{\perp}-\vec{k}_{\perp}^{'})^2}
\left(m^2(\frac{1}{x^2}+\frac{1}{(1-x)^2})
+f|_{x=x'}(s_1s_2\rightarrow s_3s_4)\right)+O(\de x)
\; , \leea{ep30}
For the next term we perform the expansion with respect to $\de x$
\be
\frac{1}{2}(\frac{1}{\tilde{\De}_1}+\frac{1}{\tilde{\De}_2})-
\frac{1}{\tilde{\De}_3}=\frac{(\vec{k}_{\perp}+\vec{k}_{\perp}^{'})^2}
{4(\vec{k}_{\perp}-\vec{k}_{\perp}^{'})^4}
\frac{\de x^2}{x^2(1-x)^2}+O(\de x^3)
\; , \lee{ep31}
then one has
\bea
&& -e^2N_{1,\la}(s_1s_2\rightarrow s_3s_4)\left(
\frac{1}{2}(\frac{1}{\tilde{\De}_1}+\frac{1}{\tilde{\De}_2})-
\frac{1}{\tilde{\De}_3}\right)
(1-f_{1\la}f_{2\la})=\nn\\
&&\hspace{2cm} -\frac{2e^2}{(\vec{k}_{\perp}-\vec{k}_{\perp}^{'})^2}
\left(-\frac{(\vec{k}_{\perp}+\vec{k}_{\perp}^{'})^2}{2}
\frac{1}{x^2(1-x)^2}\right)+O(\de x)
\; , \leea{ep32}
Both expressions \eq{ep30},\eq{ep32} do not depend on the explicit form
of unitary transformation performed. The last term does depend in collinear
limit on the explicit form of similarity function.

In the collinear limit $\theta_i$-factors \eq{ep6},\eq{ep7} are given 
in the case of

flow equations
\be
\theta_1|_{x=x'}=\frac{\tilde{\De}_1^2}{\tilde{\De}_1^2+\tilde{\De}_2^2}
\; , \lee{ep33}

and similarity transformation 
\be
\theta_1|_{x=x'}=\theta(\tilde{\De}_1-\tilde{\De}_2)
\; , \lee{ep34}
Performing the expansion in $\de x$, one has in both cases corresponding 
\be
\hspace{0cm}
-e^2N_{1,\la}(s_1s_2\rightarrow s_3s_4)\frac{1}{2}\left(
\frac{1}{\tilde{\De}_1}-\frac{1}{\tilde{\De}_2}\right)
(\theta_1-\theta_2)=
\left\{ \begin{array}{l}
-\frac{2e^2}{(\vec{k}_{\perp}-\vec{k}_{\perp}^{'})^2}\left(
(\vec{k}_{\perp}+\vec{k}_{\perp}^{'})^2\frac{1}{x^2(1-x)^2}\right),\\
flow~~~equations \\
\\
\sim 0, similarity~~~transformation
\end{array} \right. 
\lee{ep35}
Combining all terms, one has the following electron-positron interaction
in collinear limit (considered only the spin sector 
$\de_{s_1s_3}\de_{s_2s_4}$)
\be
\hspace{0cm}
\tilde{V}_{\la}^{|e\bar{e}>}(x=x')=
\left\{ \begin{array}{l}
-\frac{2e^2}{(\vec{k}_{\perp}-\vec{k}_{\perp}^{'})^2}
\left(m^2(\frac{1}{x^2}+\frac{1}{(1-x)^2})
+\frac{(\vec{k}_{\perp}+\vec{k}_{\perp}^{'})^2}{2}\frac{1}{x^2(1-x)^2}
+f|_{x=x'}(s_1s_2\rightarrow s_3s_4)\right),\\
flow~~~equations \\
\\
-\frac{2e^2}{(\vec{k}_{\perp}-\vec{k}_{\perp}^{'})^2}
\left(m^2(\frac{1}{x^2}+\frac{1}{(1-x)^2})
-\frac{(\vec{k}_{\perp}+\vec{k}_{\perp}^{'})^2}{2}\frac{1}{x^2(1-x)^2}
+f|_{x=x'}(s_1s_2\rightarrow s_3s_4)\right),\\
similarity~~~transformation
\end{array} \right. 
\lee{ep36}
that is finite.

III.Let us analyse {\bf what kind of interaction} (repulsive or attractive)
arise from the effective Hamiltonian in the electron-positron 
sector \eq{ep1}.

First note, that in the light-front frame the generated interaction
has definite sign for any value of $\la$. Namely the term in bracket
of generated interaction \eq{ep1} is always positive, since
energy denominators $\tilde{\De}_1, \tilde{\De}_2$ are positive on the 
light front, and $\frac{df_{ij}}{d\la}\geq 0, f_{ij}\geq 0$
($f_{ij}(\la)$ has the same behavior as $u_{ij}(\la)$, defined in 
Appendix \ref{A}).

When $\la =0$ generated interaction together with instantaneous 
term give rise to the attractive electron-positron interaction
in the whole parameter space.

To the leading order of nonrelativistic approximation
\be
\frac{|\vec{p}|}{m}\ll 1
\; , \lee{ep37}
one has in the instant coordinate frame
\bea
&& \tilde{\De}_1\sim\tilde{\De}_2\sim\tilde{\De}_3=\tilde{\De}=
(\vec{p}-\vec{p'})^2\nn\\
&& \tilde{V}_{\la}^{gen}\approx -e^2\frac{N_1}{(\vec{p}-\vec{p'})^2}
(1-f_{\la}^2(\De))\nn\\
&& \tilde{V}_{\la}^{PT}\approx -e^2\frac{N_1}{(\vec{p}-\vec{p'})^2}
f_{\la}^2(\De)\nn\\
&& \De=\frac{(\vec{p}-\vec{p'})^2}{x'-x}
\; , \leea{ep38}
This gives for the electron-positron interaction in the whole 
nonrelativistic range of $\la$
\bea
&& \la\ll m\nn\\
&& \tilde{V}^{|e\bar{e}>}\approx -e^2\frac{N_1}{(\vec{p}-\vec{p'})^2}
-\frac{4e^2}{(x-x')^2}\de_{s_1s_3}\de_{s_2s_4}
\; , \leea{ep39}
the $\la$-independent result.
Making use of the following expressions
\bea
&& N_1^{diag}\approx -4\frac{(\vec{k}_{\perp}-\vec{k}_{\perp}^{'})^2}
{(x-x')^2}\de_{s_1s_3}\de_{s_2s_4}\nn\\
&& (\vec{p}-\vec{p'})^2=(\vec{k}_{\perp}-\vec{k}_{\perp}^{'})^2+
(k_z-k_z^{'})^2\approx (\vec{k}_{\perp}-\vec{k}_{\perp}^{'})^2+
4m^2(x-x')^2
\; , \leea{ep40}
one get to the leading order of nonrelativistic approximation
the $3$-d Coulomb interaction
\be
\tilde{V}^{|e\bar{e}>}\approx 16m^2\left( 
-\frac{e^2}{(\vec{p}-\vec{p'})^2}\right)
\de_{s_1s_3}\de_{s_2s_4}
\; ; \lee{ep41}
hence the rotational invariance is restored to this order.
This result \eq{ep41} is valid for any nonrelativistic value of $\la$,
and for any unitary transformation in MSR (i.e. $\forall f_{ij}$) performed.

One can interprete the independence of the interaction \eq{ep39} on $\la$,
when $\la$ is in nonrelativistic domain $\la\ll m$. In this region all
photons, that give rise to perturbative (relativistic) corrections,
are almost eliminated. Therefore effectively only the lowest
Fock state ($|e\bar{e}>$) contribute to the interaction and trancation
of the whole Fock space to the lowest Fock component is valid in this case.

The explicit $\la$-dependence (through the similarity functions)
in the interaction signals to the presence of the terms in effective 
Hamiltonian that mix different Fock component, i.e. particle number changing
interactions. Then trancation is not possible.

One way out is to use different approximations, resulting in freezing 
effectively the contribution of particle number changing 
(Fock state changing) interactions (nonrelativistic approximation 
\cite{JoPeGl};
the definite choice of scale (size of the band in SR scheme)
$\la$, depending on the system considered \cite{BrPeWi}). 

In the case of MSR the effective Hamiltonian is defined in the limit 
$\la\rightarrow 0$, where all Fock states are completly decoupled
and one is able to solve the bound state problem separately
in each 'diagonal' sector.

In the next section, making use of the nonrelativistic approximation
for the electron (positron) momentum
\be
\frac{|\vec{p}|}{m}=O(\alpha)<1 
\; , \lee{ep42}
in the effective electron-positron interaction \eq{repi11}, 
obtained by flow equations,
we calculate the positronium splitting analytically.

\subsection{Positronium's fine structure and rotational
invariance}

In the nonrelativistic approximation  we obtain  
the following effective electron-positron interaction, \eq{repi11},
\bea
\hspace{-6em} V'_{\la} &=& \frac{1}{2(2\pi)^3} \frac{1}{4m} 
\frac{1}{2m} \,
 \biggl(1 - \frac{\vec{p}^2}{2m^2}\biggr) \, \tilde{V}_{\la} \nn \\
\hspace{-6em} \tilde{V}_{\la} &=& \tilde{V}_{\la}^{exch} + 
\tilde{V}_{\la}^{ann} \nn \\
&=& - \frac{e^2N_1}{(\vec{p} - \vec{p'})^2}
      \biggl(1 - {\rm e}^{-2 \left(\frac{\De}{\la^2} \right)^2} \biggr) \, 
c_{ex}^{gen} \,
     -\frac{e^2N_1}{(\vec{p}-\vec{p'})^2 + |x - x'|(M_0^2 - M_N^2)} \,
      {\rm e}^{-2 \left(\frac{\De}{\la^2}\right)^2} \, c_{ex}^{PT} \, \nn \\
 && - \: \frac{4 e^2}{(x - x')^2} \: \de_{s_1 s_3} \: 
      \de_{s_2 s_4} \, c_{ex}^{inst}\nn\\
 && + \: \frac{e^2N_2}{4m^2}
      \biggl( 1 - {\rm e}^{-2 \left(\frac{4m^2}{\la^2} \right)^2} \biggr) \,
c_{an}^{gen}
     +\frac{e^2N_2}{M_N^2} \,
      {\rm e}^{-2 \left(\frac{4m^2}{\la^2}\right)^2} \, c_{an}^{PT} \, \nn \\
 && + \: 4 e^2
      \: \de_{s_1 \bar{s}_3} \, \de_{s_2 \bar{s}_4} \, c_{an}^{inst}
\; , \leea{nra1}
where the energy denominators and exponential factors were
simplified using
\bea
&& x - x' = \frac{k_z-k_z^{'}}{2m} \left[ 1 + \frac{\vec{p}^2}{2m^2} \right]
 + O \left( m^2 \left(\frac{p}{m}\right)^5 \right) \nn \\
&& \tilde{\De}_1 = \tilde{\De}_2 = (\vec{p} -\vec{p'})^2
 + O \left(m^2 \left(\frac{p}{m}\right)^5 \right) \nn \\
&& \tilde{\De}_3 = (\vec{p} - \vec{p'})^2
 + | x - x' | (M_0^2 - M_N^2)
 + O \left(m^2 \left(\frac{p}{m}\right)^4 \right) \nn \\
&& \De_1 = \De_2 = \frac{2m(\vec{p'} - \vec{p})^2}{(k_z^{'}-k_z)}
 \left[ 1 + O \left( \left(\frac{p}{m}\right)^2 \right) \right] \; ;\qquad
 \De = \frac{2m (\vec{p'} - \vec{p})^2}{(k_z' - k_z)} \nn \\
&& M_0^2 = 4m^2 + O \left(m^2 \left(\frac{p}{m} \right)^2 \right) \nn \\
&& M_N^2 = 4m^2 + 4m B_N + O \left(m^2 \left(\frac{B_N}{m}\right)^2 \right)
 = 4m^2 + 4m B_N^{(0)}
\; , \leea{nra2}
and the explicit expression of Jacobian for the coordinate change is
\be
\sqrt{J(p)J(p')} = \frac{1}{2m} \left[ 1 - \frac{\vec{p}^{\,2}}{2m^2}
 + O \left( \frac{k_z^2}{m^2}, \frac{k_z^{'2}}{m^2} \right) \right]
\; , \lee{nra3}
having introduced the leading order binding energy $B_N^{(0)}$.
Making use of the nonrelativistic approximation
$B_N^{(0)} \!/\! m <\!\!< 1$ we have for the interaction
\bea
&&V'_{\la} = \frac{1}{2(2\pi)^3} \frac{1}{4m} \frac{1}{2m} \,
 \biggl( 1 - \frac{\vec{p}^{\,2}}{2m^2} \biggr) \nn \\
&&\hspace{3em} \times \biggl[ \biggl(
 - \frac{e^2 N_1}{(\vec{p} - \vec{p'})^2} \; (c^{gen}_{ex}, c^{PT}_{ex}) \; 
 - \frac{16 e^2 m^2}{(k_z - k'_z)^2} \,
 \left( 1 + \frac{\vec{p}^{\,2}}{m^2} \right) \; c^{inst}_{ex} \; 
 \de_{s_1 s_3}  \: \de_{s_2 s_4} \nn \\
&&\hspace{6.5em} + \; \frac{e^2 N_2}{4 m^2} \; (c^{gen}_{an}, c^{PT}_{an}) \;
 \hspace{7.5em} + \; 4e^2  \; c^{inst}_{an} \; \de_{s_1 \bar{s}_2} \: \de_{s_3 \bar{s}_4}
 \biggr) \nn \\
&&\hspace{3.5em} + \biggl(
 - \frac{e^2 N_1}{(\vec{p} - \vec{p'})^2} \frac{4m B_{\!N}^{(0)}}{|\De|} \;
 c^{PT}_{ex} \; {\rm e}^{- \left( \frac{\De}{\la^2} \right)^2 } \nn \\
&&\hspace{7.5em}
 - \; \frac{e^2 N_2}{4m^2} \frac{B_{\!N}^{(0)}}{m} \; c^{PT}_{an} \;
 {\rm e}^{- \left( \frac{4m^2}{\la^2} \right)^2 } \hspace{0.5em} \biggr)
 \biggr]
\; , \leea{nra4}
where $(c^{gen}, c^{PT})$ showes that both term, 
generated and perturbative interactions, contribute to corresponding
term (we remember that all $c = 1$).

The remarkable feature of the part of interaction standing
in the first bracket is that it does not depend on the UV cutoff $\la$.
The next term in the second bracket arises from the perturbative photon
exchange and has the typical 'energy shell' structure for the relativistic
effects, namely these terms are important when $\la>>m$. Further we
calculate the ground state positronium mass and therefore restrict
the cutoff to be in the nonrelativistic domain
\be
\la <\!\!< m
\; , \lee{nra5}
where the second term in \eq{nra4} vanishes and we are left with
the following form for the renormalized $e\bar{e}$ interaction
in the nonrelativistic approximation:
\bea
\hspace{-1em} V'(\vec{p},\vec{p'})
 &=& \frac{1}{2(2\pi)^3} \frac{1}{4m}\frac{1}{2m} \,
 \biggl( 1 - \frac{\vec{p}^2}{2m^2} \biggr) \\
&&\hspace{1em} \times \biggl[
 - \frac{e^2 N_1}{(\vec{p} - \vec{p'})^2} \; (c^{eff}_{ex}, c^{PT}_{ex}) \; 
 - \frac{16 e^2 m^2}{(k_z - k'_z)^2} \left( 1 + \frac{\vec{p}^2}{m^2} \right) \;
 c^{inst}_{ex} \; \de_{s_1 s_3} \: \de_{s_2 s_4} \nn \\
&&\hspace{3.8em} + \; \frac{e^2 N_2}{4m^2} \; 
 (c^{eff}_{an}, c^{PT}_{an}) \;
 \hspace{8.2em} + \; 4e^2 \; c^{inst}_{an} \; \de_{s_1 \bar{s}_2} \: \de_{s_3 \bar{s}_4}
 \biggr] \nn
\; . \leea{nra6}
We perform the nonrelativistic expansion of the factors $N_1$ and $N_2$
appearing in the interaction.

\noindent
The term $N_1$ contributes in $V'$ to the order

\noindent
\mbox{$\underline{O(1),O\left( \left(\frac{p}{m} \right)^2 \right) }$}:
\bea
-T_1^{\bot}T_2^{\bot} &=& 16m^2 \frac{q_{\bot}^2}{q_z^2}
 \left( 1 + \frac{\vec{p}^2}{m^2} \right) + 16 \frac{q_{\bot}^i}{q_z}
 \left( k_{\bot}^ik_z + k_{\bot}^{'i} k_z^{'} \right) \nn \\
&& -16i(s_1+s_2)[k_{\bot}^{'}k_{\bot}]-4(k_{\bot}+k_{\bot}^{'})^2
+4s_1s_2q_{\bot}^2\nn
\; , \leea{nra7}
 
\noindent
\mbox{$\underline{O\left(\frac{p}{m}\right),
 O\left( \left(\frac{p}{m} \right)^2 \right) }$} :
\bea
&&\hspace{-1em} im\sqrt{2}(x'-x)
 \left(\frac{s_1}{xx'} \: \varep^\perp_{s_1} \cdot T^\perp_1 \;
  \de_{\bar{s}_1 s_3} \: \de_{s_2 s_4} \;
  + \frac{s_2}{(1-x)(1-x')} \: \varep^\perp_{s_2} \cdot T^\perp_2 \;
  \de_{\bar{s}_4 s_2} \: \de_{s_1 s_3} \right) \nn \\
&&\hspace{0em} = 8 \, \de_{\bar{s}_1 s_3}  \: \de_{s_2 s_4} \,
 \left[m \, (iq_{\bot}^x - s_1q_{\bot}^y) \left( 1 - \frac{k_z+k_z^{'}}{m} 
\right) \,
  + q_z \, (i\tilde{p}_{\bot}^{x}-s_1\tilde{p}_{\bot}^{y})
  + \frac{1}{2} s_2 q_z(q_{\bot}^y - is_1q_{\bot}^x) \right] \nn \\
&&\hspace{1em} - \; \de_{\bar{s}_4 s_2} \: \de_{s_1 s_3}
 \left[m \, (iq_{\bot}^x - s_2q_{\bot}^y) \, \left( 1 + \frac{k_z + k_z'}{m} 
\right) \,
  - q_z \, (i\tilde{p}_{\bot}^{x}-s_2\tilde{p}_{\bot}^{y})
  - \frac{1}{2} s_1 q_z(q_{\bot}^y - is_2q_{\bot}^x) \right] \nn
\; , \leea{nra8}

\noindent
\mbox{\underline{$O\left( \left(\frac{p}{m} \right)^2 \right)$}} :
\bea
&& 2m^2\frac{(x-x')^2}{xx'(1-x)(1-x')} = 8 q_z^2 \nn
\; . \leea{nra9}

\noindent
While the term $N_2$ contributes to $V'$ to the order
\\
\noindent
\mbox{\underline{$O\left( \left(\frac{p}{m} \right)^2 \right)$}} :
\be
2m^2\frac{1}{xx'(1-x)(1-x')} = 32m^2 \nn
\; . \lee{nra10}

In these formulas we have used
\mbox{$[k_{\bot}^{'},k_{\bot}] = \varepsilon_{ij}k_{\bot}^{'i}k_{\bot}^{j}$},
$\varepsilon_{ij}=\varepsilon_{ij3}$ and
\mbox{$\varepsilon_{s}^{i} = -\frac{1}{\sqrt{2}}(s,i)$}; also
the following variables have been introduced
\bea
&& q_{\bot} = k_{\bot}^{'}-k_{\bot} \; ,\qquad (\bot=x,y)~,~
   q_z = k_z^{'}-k_z \nn \\
&& \tilde{p}_{\bot} = \frac{k_{\bot} + k'_{\bot}}{2}
\; . \leea{nra11}
We leave aside for the future work the analysis of the
expressions for $N_1$ and $N_2$, where also in this form some terms
can be identified as spin-orbit and spin-spin interactions
in the transverse plane and in longitudinal (z) direction.

Instead we follow \cite{JoPeGl}, where an analogous calculation
of singlet-triplet ground state mass splitting of positronium was performed 
in the similarity scheme. This means, that we can, except for the leading
order term $O(1)$, drop in $N_1$ the part diagonal in spin space. Also the
terms of the type 
$f=k_{\bot}^{x,y}k_z~,~k_{\bot}^{x,y}k_z^{'}~,~k_{\bot}^xk_{\bot}^y$
do not contribute to the ground state mass splitting, since
\bea
&& \int d^3pd^3p' \, \Phi_{100}^*(\vec{p}) \, 
\frac{f}{\vec{q}^2} \, \Phi_{100}(\vec{p'})
\; , \leea{nra12}
averaging over directions, gives zero.

We obtain for the $e\bar{e}$-potential to the {\bf leading order $O(1)$}
of nonrelativistic expansion
\bea
V^{'(0)}(\vec{p'}, \vec{p})
 &=& \frac{1}{2(2\pi)^3} \frac{1}{4m}\frac{1}{2m} \,
 \left( 1 - \frac{\vec{p}^2}{2m^2} \right) \nn \\
&& \times \left[ \frac{16e^2m^2}{\vec{q}^2} \frac{q_{\bot}^2}{q_z^2}
 \left( 1 + \frac{\vec{p}^2}{m^2} \right) \; (c_{ex}^{gen},c_{ex}^{PT}) \;
 - \frac{16e^2m^2}{q_z^2} \, \left( 1 + \frac{\vec{p}^2}{m^2} \right) \;
 c_{ex}^{inst} \; \right]
 \de_{s_1s_3} \: \de_{s_2s_4} \nn \\
&=& - \frac{\alpha}{2\pi^2} \frac{1}{\vec{q}^2} \left( 1 + \frac{\vec{p}^2}{2m^2} \right) \,
 \de_{s_1s_3} \: \de_{s_2s_4} \\
&&\hspace{-2em}\longrightarrow \quad V(r) \, \left( 1 + \frac{\vec{p}^2}{2m^2} \right) \nn   
\; . \leea{nra13}
Remembering \mbox{$\vec{q} = \vec{p'}-\vec{p}$}, Fourier transformation
to the coordinate space with respect to $\vec{q}$ has been performed
in the last expression. To the leading order of NR expansion
we have reproduced the Coulomb potential, defined before as 
the leading order of BSPT. Note, this is true for any UV cutoff
within the nonrelativistic range \mbox{$\la <\!\!< m$}.

We combine this expression with the kinetic term
from the Schr\"odinger equation, \eq{bspt13}, and write it in the form
\be
\frac{1}{m} \left( 1 + \frac{V(r)}{2m} \right) \vec{p}^{\,2} + V(r)
\; . \lee{nra14}
Here the potential $V(r)$ plays a different role in the two terms.
In the first term, corresponding to kinetic energy, it generates
an effective mass of the electron, which depends on the relative
position and manifests the non-locality of the interaction.
The second term is the usual potential energy, in our case, the Coulomb
interaction.

The energy of the Coulomb level with quantum numbers $(nlm)$ is standard
\be
M_0^2 = <\Phi_{nlm}|V^{'(0)}|\Phi_{nlm}>
= \int d^3p d^3p'\Phi^{*}_{nlm}(\vec{p}) \, V^{'(0)} \, \Phi_{nlm}(\vec{p'})
= -\frac{m\alpha^2}{2n^2}
\; , \lee{nra15}
where the Coulomb wave functions $\Phi_{nlm}$ were defined in \eq{bspt15}. 
We have used in \eq{nra15} the following representation
\bea
&& (\vec{p} - \vec{p'})^2 = \frac{(e_n^2 + \vec{p}^2) \,
(e_n^2 + \vec{p'}^2)}{4e_n^2} \, (u-u')^2 \nn \\
&& \frac{1}{(u-u')^2} = \sum_{\mu} \frac{2\pi^2}{n} \,
Y_{\mu}(\Omega_{p}) \, Y_{\mu}^{*}(\Omega_{p'}) \nn \\
&& d^3p = d\Omega_p \left( \frac{e_n^2 + \vec{p}^2}{2e_n} \right)^3
\leea{nra16}
and also orthogonality of the hyperspherical harmonics
\be
\int d\Omega \, Y_{\mu}^{*} \, Y_{\mu^{'}} = \de_{\mu\mu'}
\; . \lee{nra17}
More details can be found in \cite{JoPeGl}.

{\bf The next to leading order \mbox{$O\left(\frac{p}{m}\right)$}}
\bea
\de V^{(1)} &=& \frac{1}{2(2\pi)^3} \frac{1}{4m} \frac{1}{2m}
 \left(-\frac{e^2}{\vec{q}^{\,2}}\right) \nn \\
&& \times \left(
  8m(iq_{\bot}^x-s_1q_{\bot}^y) \de_{s_1\bar{s}_3} \de_{s_2s_4}
 -8m(iq_{\bot}^x-s_2q_{\bot}^y) \de_{s_1s_3} \de_{s_2\bar{s}_4} \right)
\leea{nra18}
contributes (because of the spin structure) to the second order of BSPT:
\be
\de M^2_2 = \sum_{\mu,s_i}
 \frac{< \Phi_{100} |\de V^{(1)}| \Phi_{\mu,s_i} > \, < \Phi_{\mu,s_i} |\de V^{(1)}| \Phi_{100} >}
      {M^2_1 - M^2_n}
\; . \lee{nra19}
{\bf The order $O\left( \left(\frac{p}{m} \right)^2\right)$} 
(cf. remark after \eq{nra11}) is 
\be
\de V^{(2)} = \frac{1}{2(2\pi)^3} \frac{1}{4m} \frac{1}{2m} \,
 \left( 8e^2 \frac{q_z^2}{\vec{q}^{\,2}} \de_{s_1\bar{s}_2} \de_{s_1\bar{s}_3}
 \de_{s_2\bar{s}_4} + 8e^2 \de_{s_1s_2} \de_{s_3s_4} \de_{s_1s_3}
 +4e^2 \de_{s_1\bar{s}_3} \de_{s_2\bar{s}_4} \right)
\lee{nra20}
and contributes to the first order of BSPT:
\be
\de M^2_1 = < \Phi_{100} |\de V^{(2)}| \Phi_{100} >
\; . \lee{nra21}
Both contributions were calculated in \cite{JoPeGl} with the result
\bea
\de M^2 &=& \de M_1^2 + \de M_2^2 \nn \\
<1|\de M^2|1> &=& -\frac{5}{12} m \alpha^4 \nn \\
<2|\de M^2|2> &=& <3|\de M^2|3>= <4|\de M^2|4> = \frac{1}{6} m \alpha^4
\; , \leea{nra22}
where the eigenvectors in spin space are defined as follows:
\bea
&& |1> = \frac{1}{\sqrt{2}} \, (|\!\!+-\!\!> - |\!\!-+\!\!>) \; , \nn \\
&& |2> = \frac{1}{\sqrt{2}} \, (|\!\!+-\!\!> + |\!\!-+\!\!>) \; ,\qquad
   |3> = |\!\!--\!\!> \; ,\qquad
   |4> = |\!\!++\!\!>
\; . \leea{nra23}
Making use of the relation between Coulomb energy units
and ${\rm Ryd}=\frac{1}{2}m\alpha^2$ we have the standard result
for the singlet-triplet mass splitting for positronium,
$\frac{7}{6}\alpha^2 {\rm Ryd}$. The degeneracy of the triplet ground state
$n=1$ reflects the rotational invariance, manifest in the system in
nonrelativistic approximation.

\section{Conclusions and outlook}

We proposed in this work the new Hamiltonian approach, the modified 
similarity renormalization (MSR) of Hamiltonians. By means of
flow equations, acting in the energy space, the unitary transformation
is performed, which aims to bring the field theoretical Hamiltonian
to the block-diagonal form in Fock space. 

The renormalized effective Hamiltonian, constructed in such a way, 
solves two problems. First, all 'diagonal' Fock sectors, i.e.
the sectors belonging to Fock state conserving, are completly
decoupled in the effective Hamiltonian. Second, elimination with the 
flow equations of 'rest' sectors, i.e. Fock state changing, perturbative
order by order in coupling constant, enables to find to these orders
all counterterms, corresponding to canonical operators (relevant
and marginal) of the initial theory and to new operators, generated by
flow equations. The 'new' counterterms are combinations of the 
'canonical' one or can carry also new types of divergencies. This
question we stayed aside in this work.

Summarizing, one is able to answer two questions: the physics
of what Fock state and of what energy scale is descibed by the
renormalized effective Hamiltonian.

Namely, one can pick out any block from 'diagonal' sector of effective
Hamiltonian, acting on the definite Fock state, and solve separately
for this block (Fock state) the corresponding physical problem.
In particular, it is possible to trancate the block-diagonal effective
Hamiltonian to the lowest Fock state (here discussed $|e\bar{e}>$),
plug the corresponding interaction (from '$e\bar{e}$' block) into
the Schr\"odinger equation, put it on the computer and solve the bound
state problem (for positronium bound state) numericaly.

The problem that arises by numerical diagonalization of the lowest
Fock sector matrix in the energy space (or by numerical solution
of integral bound state equation, see main text) is the dependence
of physical mass spectrum on the size of the matrix 
(or on UV cutoff imposed on the energies in the integral). This dependence
is to be absorbed by including the corresponding counterterms,
calculated analyticaly on the previous step. MSR scheme insures,
that counterterms to add do not depend on the state (i.e. are the same
for the lowest and next-to-lowest exited states) and on the Fock sector
(i.e. trancation to the $|e\bar{e}>$ state do not prevent counterterms
corresponding to the diagrams with more particles in intermediate
state) \cite{GuTrWe}. This is due to the new interactions and corresponding
counterterms, generated by the flow equations in MSR scheme.

In the work we have outlined the strategy to build the renormalized
effective Hamiltonian, that can be fulfilled by means of flow equations
in a systematic way (to the end $\la\rightarrow 0$) without
knowing apriore any properties of the system considered further
(i.e. no approximations are needed to make in the procedure).
This is the main advantage of MSR scheme. 

We remind, that through MSR scheme one simultaneously renormalizes
in the energy space the initial field theoretical Hamitonian
and constructs the effective Hamitonian, for which the Fock space
trancation is valid. But rather the analytical solution of flow
equations, as of any other iterrative method, demands to apply
perturbative expansion in coupling constant. This means, 
that applicability of perturbative theory is closely related to 
the possibility of working in a trancated Fock space. Within the proposed
approach one is able to improve sytematically this approximation.

\paragraph{Acknowledgments}
One of the authers, Elena Gubankova, is greatful to the organizers
of Les Houches workshop for the opportunity to participate.
E.L. would like to thank Dr.Koji Harada, Billy Jones,
Dr.Brett van de Sande, Dr.Matthias Burkardt for discussions
during this workshop.
Also E.L thanks Prof.H-C.Pauli for the constant interest in this work and
useful discussions that
improved the understanding of the problem.
This work was partially supported by the Deutsche Forschungsgemeinschaft,
grant no. GRK 216/5-95.

\newpage
\appendix

\section{\label{A}Similarity renormalization} 

In this appendix we skip the idea of similarity renormalization (SR)
proposed by Glazek and Wilson \cite{GlWi}, the renormalization scheme
of Hamiltonians by means of unitary transformation,
and consider two alternative methods: the flow equations (FE) 
of Wegner \cite{We} and similarity unitary transformation (SUT) \cite{GlWi},
that are used in SR scheme and give rise to the renormalized Hamiltonian.

The similarity transformation aims to bring the field theoretical 
Hamiltonian to the most diagonal form,
namely only the matrix elements between the free states with
\be
|\Delta _{ij}| < \lambda
\lee{a1}
where $\Delta_{ij}=E_i-E_j$, are present in the renormalized Hamiltonian.
For this purposes the continuous unitary transformation is performed
to preserve unchanged the spectrum (eigenvalues) of the initial
bare cutoff Hamiltonian. The demand of diagonal structure
does not define completly the generator of the transformation. This freedom
is used to eliminate small energy denominators in the final renormalized 
Hamiltonian. This results in a system of two self-consistent non-linear 
differential equations for the Hamiltonian $H(l)$ and the generator of the
transformation $\eta (l)$. The dependence on the continuous flow
parameter $l$ in the flow equations by Wegner is replaced by the 
cutoff dependence $\lambda$ in the similarity unitary transformation,
with the connection
\be
l=1/ \lambda ^2
\lee{a2}
in the renormalized Hamiltonians.
 
The difference of the two methods (FE) and (SUT) consists of 
the residual freedom
in the choice of the direction of the infinitesimal rotation,
actually defining how fast the non-diagonal matrix elements vanish.

We summarize the equations for both methods, written in matrix form.

\paragraph{I.} The flow equations by Wegner \cite{We}:
\be
\frac{dH_{ij}}{dl}=[\eta,H_I]_{ij}+\frac{du_{ij}}{dl} \, \frac{H_{ij}}{u_{ij}}
\; , \lee{a3}
\be
\eta_{ij}=\frac{1}{E_i-E_j} \left(- \frac{du_{ij}}{dl} \, \frac{H_{ij}}{u_{ij}} \right)
\lee{a4}
with
\be
u_{ij} = \exp(-l \Delta_{ij}^2)
\lee{a5}
and
\be
f_{ij}=u_{ij}
\; . \lee{a6}

\paragraph{II.} The similarity unitary transformation by Gla\-zek
and Wil\-son \cite{GlWi}:
\be
\frac{dH_{ij}}{d \lambda} = u_{ij}[\eta,H_I]_{ij} +
 r_{ij} \frac{du_{ij}}{d\lambda} \, \frac{H_{ij}}{u_{ij}}
\; , \lee{a7}
\be
\eta_{ij} = \frac{r_{ij}}{E_i-E_j} \left([\eta,H_I]_{ij} -
 \frac{du_{ij}}{d \lambda} \, \frac{H_{ij}}{u_{ij}} \right)
\lee{a8}
and
\be
f_{ij}=u_{ij} \exp(r_{ij})
\; . \lee{a9}
Also the following transformation is used \cite{GlWi}:
\be
\frac{dH_{ij}}{d\lambda}=u_{ij}[\eta,H_I]_{ij}+
\frac{du_{ij}}{d\lambda} \, \frac{H_{ij}}{u_{ij}}
\; , \lee{a10}
\be
\eta _{ij} = \frac{1}{E_i-E_j} \left( r_{ij}[\eta,H_I]_{ij} -
 \frac{du_{ij}}{d \lambda} \, \frac{H_{ij}}{u_{ij}} \right)
\lee{a11}
and 
\be
f_{ij}=u_{ij}
\; , \lee{a12}
where for both similarity transformations
\be
u_{ij} = \theta (\lambda -|\Delta _{ij}|)
\lee{a13}
and
\be
u_{ij} + r_{ij} = 1
\; . \lee{a14}
Also other choices for the similarity function $u_{ij}$ with the step
behaviour are possible \cite{GlWi}. 

Remember, the function $f_{ij}$ defines the solution for the leading order
interaction term. The first order equations for $H$ and $\eta$, written
through the $f$-function are unique for both methods ({\bf I.} and {\bf II.})
\be
H_{I,ij}^{(1)}(l) = H_{I,ij}^{(1)}(l=0)
 \frac{f_{ij}(l)}{f_{ij}(l=0)}
\; , \lee{a15}
\be
\eta _{ij}^{(1)}(l)= -\frac{1}{E_i-E_j} \, \frac{dH_{I,ij}^{(1)}}{dl}
\lee{a16}
with the connection given in \eq{a2} in the renormalized values,
and $dl \rightarrow d\lambda$ implied.
The \eqs{a15}{a16} will be exploited further for 
the calculations in the main text.

\section{\label{B}Calculation of the commutator $[\eta^{(1)}(l), H_{ee\gamma}]$
 in the electron-positron sector}

Here we calculate the commutator $[\eta^{(1)}(l), H_{ee\gamma}]$
in the electron-positron sector.
The leading order generator $\eta^{(1)}$ is:
\bea
\eta^{(1)}(l) &=& \sum_{\lambda s_1s_3}\int_{p_1p_3q}\!\!\!(\eta_{p_1p_3}^*(l)
\varepsilon_{\lambda}^i\tilde{a}_q +
\eta_{p_1p_3}(l) \varepsilon_{\lambda}^{i *}\tilde{a}_{-q}^+) \,
(\tilde{b}_{p_3}^+\tilde{b}_{p_1}+\tilde{b}_{p_3}^+\tilde{d}_{-p_1}^+ +
\tilde{d}_{-p_3}\tilde{b}_{p_1}+\tilde{d}_{-p_3}\tilde{d}_{-p_1}^+)\nonumber\\
&&\hspace{4em} \times \chi_{s_3}^+ \Gamma_l^i(p_1,p_3,-q) \chi_{s_1} \,
                      \delta_{q,-(p_1-p_3)}
\; , \leea{b1}
where
\be
\eta_{p_1p_3}(l)=-\Delta_{p_1p_3} \cdot g_{p_1p_3} =
\frac{1}{\Delta_{p_1p_3}} \cdot \frac{dg_{p_1p_3}}{dl}
\; , \lee{b2}
\mbox{$\Delta_{p_1p_3}=p_1^--p_3^--(p_1-p_3)^-$},
and the electron-photon coupling
\bea
H_{ee\gamma} &=& \sum_{\lambda s_2s_4}\int_{p_2p_4q'}\!\!\!(g_{p_2p_4}^*(l)
\varepsilon _{\lambda'}^j\tilde{a}_{q'}+
g_{p_2p_4}(l)\varepsilon _{\lambda'}^{j *}\tilde{a}_{-q'}^+) \,
(\tilde{b}_{p_4}^+\tilde{b}_{p_2} +\tilde{b}_{p_4}^+\tilde{d}_{-p_2}^+ +
\tilde{d}_{-p_4}\tilde{b}_{p_2} +\tilde{d}_{-p_4}\tilde{d}_{-p_2}^+)\nonumber\\
&&\hspace{4em} \times\chi_{s_4}^+ \Gamma_l^i(p_2,p_4,-q') \chi_{s_2} \,
               \delta_{q',-(p_2-p_4)}
\; , \leea{b3}
where
\be
\Gamma_l^i(p_1,p_2,q) = 2\frac{q^i}{q^+} -
\frac{\sigma\cdot p_2^{\bot} - im}{p_2^+}\sigma^i -
\sigma^i\frac{\sigma\cdot p_1^{\bot} + im}{p_1^+}
\lee{b4}
and the tilde-fields are defined in \eq{ch19}.
Further we use the identities for the polarisation vectors and spinors
\be
\sum_{\lambda} \varepsilon_{\lambda}^{i *} \varepsilon_{\lambda}^j
= \de^{ij} \; ,\qquad
\chi_s^+ \chi_{s'} = \de_{ss'}
\; . \lee{b5}

Using the commutation relations, \eq{ch10}, and identities \eq{b5} we have 
\bea
[\eta^{(1)}(l),H_{ee\gamma}] &=& \frac{1}{2} \,
 \left( - \eta_{p_1p_3} g_{p_2p_4}^* \frac{\theta(p_1^+-p_3^+)}{p_1^+ - p_3^+}
        + \eta_{p_1p_3}^* g_{p_2p_4} \frac{\theta(p_3^+-p_1^+)}{p_3^+-p_1^+} \right) \\
&& \times {\boldmath :} \,
 (- \tilde{b}_{p_3}^+ \tilde{d}_{-p_2}^+ \tilde{d}_{-p_4} \tilde{b}_{p_1}
  - \tilde{b}_{p_4}^+ \tilde{d}_{-p_1}^+ \tilde{d}_{-p_3} \tilde{b}_{p_2}
  + \tilde{b}_{p_3}^+ \tilde{d}_{-p_1}^+ \tilde{d}_{-p_4} \tilde{b}_{p_2}
  + \tilde{b}_{p_4}^+ \tilde{d}_{-p_2}^+ \tilde{d}_{-p_3} \tilde{b}_{p_1}) \,
          {\boldmath :} \nn \\
&& \times (\chi_{s_3}^+ \Gamma_l^i(p_1,p_3,p_1-p_3) \chi_{s_1}) \,
          (\chi_{s_4}^+ \Gamma_l^i(p_2,p_4,p_2-p_4) \chi_{s_2}) \:
          \delta_{p_1+p_2,p_3+p_4} \nn
\; , \leea{b6}
where the first two terms of the field operators contribute to
the exchange channel, and the next two to the annihilation channel.
We take into account both $s$- and $t$-channel terms to calculate
the bound states. The $:\;:$ stand for the normal ordering of the
fermion operators and $(\frac{1}{2})$ is the symmetry factor. The
sum over helicities $s_i$ and the 3-dimensional integration 
over momenta $p_i$, $i=1,..4$, according to \eq{ch20} is implied. 
We rewrite for both channels
\be
\hspace{0cm}
[\eta,H_{ee\gamma}]=
\left\{ \begin{array}{l}
  M_{2ij}^{(ex)}(\frac{1}{2})
       \left\{ \frac{\theta(p_1^+-p_3^+)}{(p_1^+-p_3^+)} \right.
       (\eta_{p_1,p_3}g_{-p_4,-p_2}^*-\eta_{-p_4,-p_2}^*g_{p_1,p_3}) \\
\hspace{2.0cm}
     + \left. \frac{\theta(-(p_1^+-p_3^+))}{-(p_1^+-p_3^+)}
       (\eta_{-p_4,-p_2}g_{p_1,p_3}^*-\eta_{p_1,p_3}^*g_{-p_4,-p_2})\ \right\} \\
\hspace{1.8cm}
     \times \delta^{ij}\delta_{p_1+p_2,p_3+p_4} \,
     b_{p_3s_3}^+d_{p_4\bar{s}_4}^+d_{p_2\bar{s}_2}b_{p_1s_1} \\
\\
 -M_{2ij}^{(an)}(\frac{1}{2})
       \left\{ \frac{\theta(p_1^++p_2^+)}{(p_1^++p_2^+)} \right.
       (\eta_{p_1,-p_2}g_{-p_4,p_3}^*-\eta_{-p_4,p_3}^*g_{p_1,-p_2}) \\
\hspace{2.2cm}
       + \left. \frac{\theta(-(p_1^++p_2^+))}{-(p_1^++p_2^+)}
       (\eta_{-p_4,p_3}g_{p_1,-p_2}^*-\eta_{p_1,-p_2}^*g_{-p_4,p_3}) \right\} \\
\hspace{2.0cm}
       \times \delta^{ij}\delta_{p_1+p_2,p_3+p_4} \,
       b_{p_3s_3}^+d_{p_4\bar{s}_4}^+d_{p_2\bar{s}_2}b_{p_1s_1}
\end{array} \right. 
\lee{b7}
where 
\bea
M_{2ij}^{(ex)}
&=& (\chi_{s_3}^+ \Gamma_l^i(p_1,p_3,p_1-p_3) \chi_{s_1}) \,
    (\chi_{\bar{s}_2}^+ \Gamma_l^j(-p_4,-p_2,-(p_1-p_3)) \chi_{\bar{s}_4}) \nn \\
\\
M_{2ij}^{(an)}
&=& (\chi_{s_3}^+ \Gamma_l^i(-p_4,p_3,-(p_1+p_2)) \chi_{\bar{s}_4}) \,
    (\chi_{\bar{s}_2}^+ \Gamma_l^j(p_1,-p_2,p_1+p_2) \chi_{s_1}) \nn
\; . \leea{b8}

The first term in the exchange channel with $p_1^+ > p_3^+$ corresponds to the 
light-front time ordering $x_1^+ < x_3^+$ 
with the intermediate state $P_k^-=p_3^- + (p_1-p_3)^- + p_2^-$,
the second term $p_1^+<p_3^+$ and $x_1^+>x_3^+$
has the intermediate state $P_k^- = p_1^- - (p_1 - p_3)^- + p_4^-$.
Both terms can be viewed as the retarded photon exchange.
The same does hold for the annihilation channel.

Consider only real couplings and take into account the symmetry
\be
\eta_{-p_4,-p_2}=-\eta_{p_4,p_2} \; ,\qquad g_{-p_4,-p_2}=g_{p_4,p_2}
\; . \lee{b9}
Then \mbox{$\left. \left< \! p_3s_3,p_4\bar{s}_4 \right|
[\eta^{(1)},H_{ee\gamma}] \left| p_1s_1,p_2\bar{s}_2 \right> \right.$},
the matrix element of the commutator between the free states of positronium
in the exchange and annihilation channel, reads
\be
<[\eta^{(1)},H_{ee\gamma}]>/\delta_{p_1+p_2,p_3+p_4}= 
\left \{\begin{array}{l}
 M_{2ii}^{ex} \, \frac{1}{(p_1^+-p_3^+)} \,
  (\eta_{p_1,p_3}g_{p_4,p_2} + \eta_{p_4,p_2}g_{p_1,p_3})
\\
-M_{2ii}^{an} \, \frac{1}{(p_1^++p_2^+)} \,
  (\eta_{p_1,-p_2}g_{p_4,-p_3} + \eta_{p_4,-p_3}g_{p_1,-p_2})
\end{array} \right. 
\; . \lee{b10}

We rewrite this expression through the corresponding $f$-functions
\bea
\eta_{p_1,p_3} g_{p_4,p_2} + \eta_{p_4,p_2} g_{p_1,p_3}
&=& e^2 \left[
  \frac{1}{\Delta_{p_1,p_3}} \frac{df_{p_1,p_3}(l)}{dl} f_{p_4,p_2}(l)
+ \frac{1}{\Delta_{p_4,p_2}} \frac{df_{p_4,p_2}(l)}{dl}f_{p_1,p_3}(l) \right] \nn \\
\\
\eta_{p_1,-p_2} g_{p_4,-p_3} +\eta_{p_4,-p_3} g_{p_1,-p_2}
&=& e^2 \left[
  \frac{1}{\Delta_{p_1,-p_2}}\frac{df_{p_1,-p_2}(l)}{dl}f_{p_4,-p_3}(l)
+ \frac{1}{\Delta_{p_4,-p_3}}\frac{df_{p_4,-p_3}(l)}{dl}f_{p_1,-p_2}(l) \right] \nn
\leea{b11}
with $\Delta_{p_1,p_2}=p_1^--p_2^--(p_1-p_2)^-$. As we have mentioned
in Appendix A this form in terms of the $f$-function is universal
for all unitary transformations. We exploit further this expression by
specifying the $f$-function to compare the effective interactions
in different renormalization schemes (see Appendix C).    

We calculate the matrix elements \mbox{$M_{2ii}$}, eq.~(188), for both channels.
Here we follow the notations introduced in \cite{JoPeGl}.

We make use of the identities
\be
\chi_s^+\sigma^i\sigma^j \chi_s = \delta^{ij}+is\varepsilon^{ij} \; ,\qquad
\chi_s^+\sigma^j\sigma^i \chi_s = \delta^{ij}+i\bar{s}\varepsilon^{ij}
\lee{b12}
with $\bar{s} = -s$ and $\chi_s^+ \chi_{s'} = \de_{ss'}$; also of
\be
\chi_{\bar{s}}^+ \sigma^i\chi_s = -\sqrt{2}s \varepsilon_s^i \; ,\qquad
\chi_s^+\sigma^i \chi_{\bar{s}} = -\sqrt{2}s \varepsilon_s^{i*}
\lee{b13}
with \mbox{$\varepsilon_s^* = -\varepsilon_{\bar{s}}$} and
\mbox{$\varepsilon_s^i \varepsilon_{s'}^i = -\delta_{s\bar{s'}}$}.

We use the standard light-front frame, \fig{reneebarint},
\bea
&& p_1 = (xP^+,xP^{\bot}+k_\bot) \; , \hspace{3em}
   p_2 = ((1-x)P^+,(1-x)P^{\bot}-k_\bot) \; , \nn \\
&& p_3 = (x'P^+,x'P^{\bot}+k'_\bot) \; , \hspace{1.5em}
   p_4 = ((1-x')P^+,(1-x')P^{\bot}-k'_\bot)
\; , \leea{b14}
where \mbox{$P=(P^+,P^{\bot})$} is the positronium momentum.

Then, to calculate the matrix element $M_{2ii}$ in the {\bf exchange channel}, we find
\bea
P^+[\chi_{s_3}^+\Gamma^i(p_1,p_3,p_1-p_3)\chi_{s_1}]
&=& \chi_{s_3}^+ \left[ 2\frac{(k_\bot-k'_\bot)^i}{(x-x')}
 - \frac{\sigma \cdot k'_\bot}{x'} \sigma^i+\sigma^i\frac{\sigma \cdot k_\bot}{x} +
          im\frac{x-x'}{xx'}\sigma^i
   \right] \chi_{s_1} \nn \\
&=& T_2^i \delta_{s_1s_3} +
 im\frac{x-x'}{xx'} (-\sqrt{2}) s_1 \varepsilon_{s_1}^i \delta_{s_1\bar{s}_3}
\; , \leea{b15}
and
\bea
&&\hspace{-1cm}
P^+[\chi_{\bar{s}_2}^+ \Gamma^i(-p_4,-p_2,-(p_4-p_2)) \chi_{\bar{s}_4}] \nn \\
&&\hspace{2cm} = \chi_{\bar{s}_2}^+
 \left[ 2 \frac{(k_\bot-k'_\bot)^i}{x-x'}
 + \left( \frac{\sigma \cdot k_\bot}{1-x} \sigma^i
           + \sigma^i \frac{\sigma \cdot k'_\bot}{1-x'} \right)
 - im \frac{x-x'}{(1-x)(1-x')} \sigma^i
 \right] \chi_{\bar{s}_4} \nn \\
&&\hspace{2cm} = - \left[ T_1^i\delta_{s_2s_4}
 + im \frac{x-x'}{(1-x)(1-x')} (-\sqrt{2}) s_2 \varepsilon_{s_2}^i
 \delta_{s_2\bar{s}_4} \right]
\; , \leea{b16}
where we have introduced
\bea
T_1^i & \equiv& -\left[ 2 \frac{(k_\bot-k'_\bot)^i}{x-x'}+\frac{k_\bot^i(s_2)}{(1-x)} +
 \frac{{k'}_\bot^i(\bar{s}_2)}{(1-x')} \right] \nn \\
\\
T_2^i & \equiv& 2 \frac{(k_\bot-k'_\bot)^i}{x-x'}-\frac{k_\bot^i(s_1)}{x} -
 \frac{{k'}_\bot^i(\bar{s}_1)}{x'} \nn
\leea{b17}
and
\be
k_\bot^i(s)\equiv k_\bot^i+is \, \varepsilon_{ij} \, k_\bot^j
\; . \lee{b18}
Finaly we result 
\bea
P^{+2} \, M_{2ii}^{(ex)} &\hspace{-2mm}=\hspace{-2mm}& - \left\{
   \delta_{s_1s_3} \delta_{s_2s_4} T_1^{\bot} \cdot T_2^{\bot}
 - \delta_{s_1\bar{s}_2} \delta_{s_1\bar{s}_3} \delta_{s_2\bar{s}_4}
 2m^2 \frac{(x-x')^2}{xx'(1-x)(1-x')} \right. \\
&&\hspace{2em} \left.
 + im \sqrt{2}(x'-x) \left[ \delta_{s_1\bar{s}_3} \delta_{s_2s_4}
 \frac{s_1}{xx'}T_1^{\bot} \cdot \varepsilon_{s_1}^{\bot}
 + \delta_{s_1s_3} \delta_{s_2\bar{s}_4}
 \frac{s_2}{(1-x)(1-x')} T_2^{\bot} \cdot \varepsilon_{s_2}^{\bot} \right]
 \right\} \nn
\; . \leea{b19}
Whereas in the {\bf annihilation channel} we calculate
\bea
P^+[\chi_{s_3}^+ \Gamma^i(-p_4,p_3,-(p_1+p_2)) \chi_{\bar{s}_4}]
&=& \chi_{s_3}^+ \left[
 - \frac{\sigma \cdot k'_\bot}{x'} \sigma^i + \sigma^i \frac{\sigma \cdot k'_\bot}{1-x'}
 + im\frac{1}{x'(1-x')}\sigma^i \right] \chi_{\bar{s}_4} \nn \\
&=& T_3^i\delta_{s_3\bar{s}_4}
 + im\frac{1}{x'(1-x')} (-\sqrt{2}) s_4 \varepsilon_{s_4}^{i*}
 \delta_{s_3s_4}
\leea{b20}
and
\bea
\hspace{-1cm}
P^+[\chi_{\bar{s}_2}^+ \Gamma^i(p_1,-p_2,p_1+p_2) \chi_{s_1}]
&=& \chi_{\bar{s}_2}^+ \left[
 \frac{\sigma \cdot k_\bot}{1-x} \sigma^i - \sigma^i \frac{\sigma \cdot k_\bot}{x}
 - im\frac{1}{x(1-x)} \sigma^i \right] \chi_{s_1} \nn \\
&=& T_4^i\delta_{s_1\bar{s}_2}
 -im\frac{1}{x(1-x)} (-\sqrt{2}) s_1 \varepsilon_{s_1}^i
 \delta_{s_1s_2}
\; , \leea{b21}
where we have introduced
\bea
T_3^i & \equiv & -\frac{{k'}_\bot^i(\bar{s}_3)}{x'}
 + \frac{{k'}_\bot^i(s_3)}{1-x'} \nn \\
\\
T_4^i & \equiv & \frac{k_\bot^i(\bar{s}_1)}{1-x}
 - \frac{k_\bot^i(s_1)}{x} \nn
\; . \leea{b22}
We finally have
\bea
P^{+2} \, M_{2ii}^{(an)} &=& \delta_{s_1\bar{s}_2} \delta_{s_3\bar{s}_4}
 T_3^{\bot}\cdot T_4^{\bot}
 + \delta_{s_1s_2} \delta_{s_3s_4}\delta_{s_1s_3}
 2m^2 \frac{1}{xx'(1-x)(1-x')} \\
&& + im \sqrt{2} \left[\delta_{s_3\bar{s}_4}\delta_{s_1s_2}
 \frac{s_1}{x(1-x)} T_3^{\bot} \cdot \varepsilon_{s_1}^{\bot}
 - \delta_{s_3s_4} \delta_{s_1\bar{s}_2}
 \frac{s_3}{x'(1-x')} T_4^{\bot}\cdot \varepsilon_{s_4}^{\bot *} \right] \nn
\; . \leea{b23}

\section{\label{C}Fermion and photon self energy terms}
 
We calculate here the fermion and photon self energy terms,
arising from the second order commutator $[\eta^{(1)},H_{ee\ga}]$.

\begin{figure}
\setlength{\unitlength}{1mm}
\begin{picture}(170,34.61)
\put(0,0){\makebox(56,34.61){ \loadeps{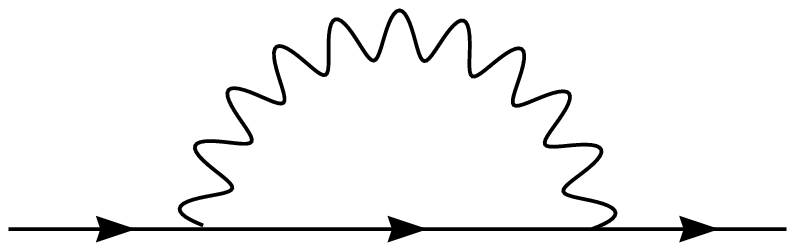} }}
\put(10,14){$p$}
\put(42,14){$p$}
\put(26,32){$k$}
  \put(57,0){\makebox(56,34.61){ \loadeps{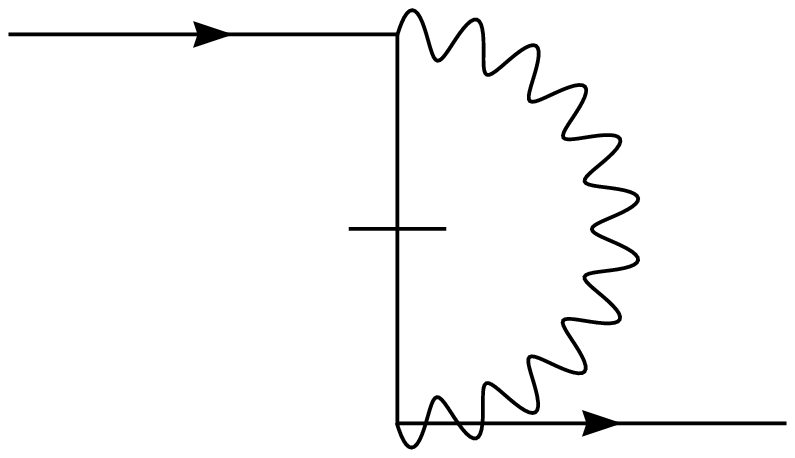} }}
  \put(67,30){$p$}
  \put(100,16){$k$}
  \put(100,3){$p$}
    \put(114,0){\makebox(56,34.61){ \loadeps{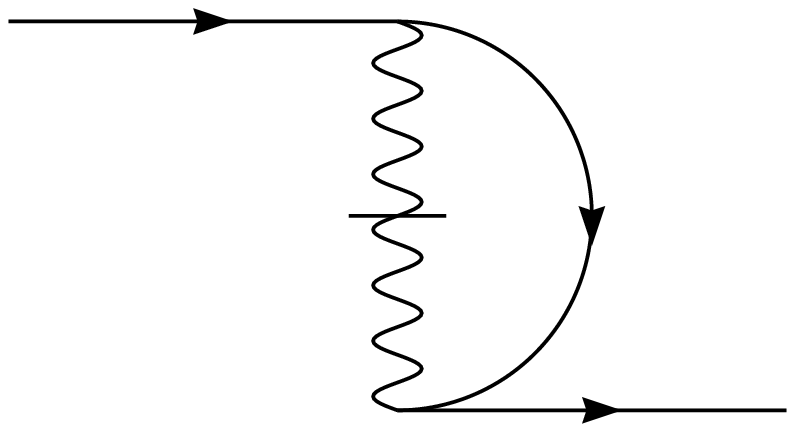} }}
    \put(124,30){$p$}
    \put(155,16){$k$}
    \put(157,3){$p$}
\put(55,17.305){$+$}
\put(112,17.305){$+$}
\end{picture}
\vspace{5cm}
\caption{Electron self energy: the first diagram corresponds to
the commutator term $[\eta^{(1)},H_{ee\ga}]$ in the electron self energy sector, 
next two diagrams arise from the normal ordering of instantaneous interactions.}
\label{eselfen}
\end{figure}

\paragraph{I.}
We first derive the {\bf electron self energy} terms.
Making use of the expressions for the generator of the unitary
transformation $\eta^{(1)}$ defined in \eq{gi1} and of $H_{ee\ga}$, \eq{ch14},
we obtain the following expression for the commutator in the
electron self energy sector
\bea
&&\hspace{-5em} \frac{1}{2} (\et_{p_1p_2}g_{p_2p_1}-\et_{p_2p_1}g_{p_1p_2}) \,
\biggl[ \th(p_1^+) \frac{\th(p_2^+-p_1^+)}{p_2^+ - p_1^+} \th(p^+_2) \,
               b^+_{p_2}b_{p_2}\chi^+_{s_2}\chi_{s_2}\nn\\
&&\hspace{6em}
      - \th(p_2^+) \frac{\th(p_1^+-p_2^+)}{p_1^+ - p_2^+} \th(p^+_1) \,
               b^+_{p_1}b_{p_1}\chi^+_{s_1}\chi_{s_1} \biggr]
M_{2ij}(p_1,p_2) \de^{ij}
\; , \leea{d1}
where
\be
M_{2ij}(p_1,p_2)=\Ga^i(p_1,p_2,p_1-p_2)\Ga^j(p_2,p_1,p_2-p_1)
\lee {d2}
and the momentum integration over $p_1,p_2$ is implied;
$1/2$ stands as the symmetry factor.
The matrix element of the commutator between the free fermion states is
\bea
&& <p_1,s_1|[\et^{(1)},H_{ee\ga}]|p_1,s_1>_{self energy} \nn \\
&&\hspace{5em} = -\int_{p_2}(\et_{p_1p_2} g_{p_2p_1} - \et_{p_2p_1}g_{p_1p_2}) \,
\th(p_2^+) \frac{\th(p_1^+-p_2^+)}{p_1^+-p_2^+} \,M_{2ii}(p_1,p_2)
\; , \leea{d3}
where the integration $\int_p$ is defined in \eq{ch20}. We use the expression
for the generator $\eta$ through the coupling, namely 
\be
\et_{p_1p_2} g_{p_2p_1} - \et_{p_2p_1} g_{p_1p_2} = \frac{1}{\De_{p_1p_2}}
\left( g_{p_1p_2} \frac{dg_{p_2p_1}}{dl} +
       g_{p_2p_1} \frac{dg_{p_1p_2}}{dl} \right)
\; . \lee{d4}
Change of the variables according to
\bea
p_1 &=& p \nn \\
p_2 &=& p_k \nn \\
p_1 - p_2 &=& k
\leea{d5}
brings the integral in \eq{d3} to the standard form of loop integration 
\be
-\int_k(\et_{p, p - k} g_{p - k, p} - \et_{p - k, p} g_{p, p - k}) \,
\th(p^+-k^+) \frac{\th(k^+)}{k^+} \, M_{2ii}(p,p-k)
\; . \lee{d6}
According to \eq{ri2}, the integral $\int_{l_{\la}}^{l_{\La}}$
of the commutator $[\eta^{(1)},H_{ee\ga}]$ defines the difference between
the energies (or energy corrections) \mbox{$\de p_{1\la}^--\de p_{1\La}^-$}.
Making use of
\be 
\int^{l_{\la}}_{l_{\La}} dl' (\et_{p_1p_2} g_{p_2p_1}-\et_{p_2p_1} g_{p_1p_2})
= \frac{1}{p_1^- - p_2^- - (p_1-p_2)^-} \,
(g_{p_1, p_2, \La} g_{p_2, p_1, \la} - g_{p_1, p_2, \la} g_{p_2, p_1, \La})
\lee{d7}
we have the following explicit expression: 
\bea 
&&\hspace{-1em} \de p_{1\la}^--\de p_{1\La}^- = e^2\int \frac{d^2k^{\bot}dk^+}{2(2\pi)^3} \,
\frac{\th (k^+)}{k^+}
\th (p^+-k^+) \, \frac{(-1)}{p^--k^--(p-k)^-} \\
&&\hspace{4em} \times \Ga^i(p-k,p,-k) \Ga^i(p,p-k,k) \,
\left[ \exp\left\{-2 \left( \frac{\De_{p,p-k}}{\la} \right)^2
\right\}
- \exp\left\{-2 \left( \frac{\De_{p,p-k}}{\La} \right)^2 \right\} \right] \nn
\; , \leea{d8}
where the solution for the $ee\ga$-coupling constant was used.
Therefore the electron energy correction corresponding to the first diagram,
\fig{eselfen}, is 
\bea
\de p_{1\la}^- &=& e^2\int \frac{d^2k^{\bot}dk^+}{2(2\pi)^3} \,
 \frac{\th (k^+)}{k^+} \th (p^+-k^+) \\
&&\hspace{2em} \times \Ga^i(p-k,p,-k) \Ga^i(p,p-k,k) \,
 \frac{1}{p^--k^--(p-k)^-} \, \times (-R) \nn
\; , \leea{d9}
where we have introduced the regulator $R$, defining the cutoff condition
(see main text),
\be
R = \exp\left\{-2 \left( \frac{\De_{p,k}}{\la} \right)^2 \right\}
\lee{d10}
(note that $\De_{p,k}=\De_{p,p-k}$).
To perform the integration over $k=(k^+,k^{\bot})$ explicitly, choose
the parametrization
\bea
\frac{k^+}{p^+} &=& x \nn \\
k &=& (xp^+,xp^{\bot}+\kappa^{\bot})
\; , \leea{d11}
where $p=(p^+,p^{\bot})$ is the external electron momentum.
Then the terms occuring in $\de p_{1\la}^-$ are rewritten in the form
\bea
& \Ga^i(p-k,p,-k)\Ga^i(p,p-k,k)=
\frac{1}{(p^+)^2(1-x)^2} \left( \left( 4\frac{1}{x^2}-4\frac{1}{x}+2 \right)
\kappa_{\bot}^2+2m^2x^2 \right) & \nn \\
& \De_{p, p - k} = p^- - k^- - (p - k)^- = \frac{1}{p^+x(1-x)}
(x(1-x)p^2 - \kappa_{\bot}^2-xm^2) = \frac{\tilde{\De}_{p,p-k}}{p^+} &
\; . \leea{d12}
Therefore the integral for the electron energy correction
corresponding to the first diagram of \fig{eselfen} takes the form
\bea
p^+\de p_{1\la}^- &=&-\frac{e^2}{8\pi^2}\int_0^1dx\int d\kappa_{\bot}^2
\frac{(\frac{2}{x^2}-\frac{2}{x}+1)\kappa_{\bot}^2+m^2x^2}
{(1-x)(\kappa_{\bot}^2+f(x))}\times(-R) \\
&=& -\frac{e^2}{8\pi^2}\int_0^1dx\int d\kappa_{\bot}^2 \nn \\
&&\hspace{2em} \times \left[ \frac{p^2-m^2}{\kappa_{\bot}^2+f(x)}
 \left( \frac{2}{[x]}-2+x \right) - \frac{2m^2}{\kappa_{\bot}^2+f(x)}
 + \left( \frac{2}{[x]^2}+\frac{1}{[1-x]} \right)
\right] \times (-R) \nn
\; , \leea{d13}
where 
\be
f(x)=xm^2 - x(1-x)p^2
\; . \lee{d14}
In the last integral the principal value prescription  for $\frac{1}{[x]}$
as $x\rightarrow 0$ was introduced (see main text), to regularize
the IR divergencies present in the longitudinal direction. 

We thus have derived the expression for the energy correction
which has been used in the main text.

\begin{figure}
\setlength{\unitlength}{1mm}
\begin{picture}(170,34.61)
\put(28,0){\makebox(56,34.61){ \loadeps{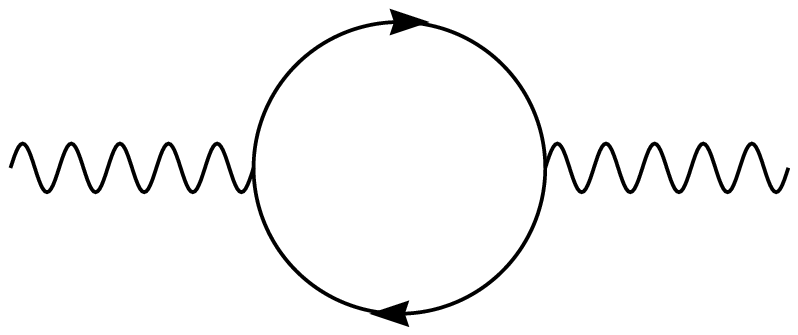} }}
\put(38,12){$p$}
\put(70,12){$p$}
\put(54,28){$k$}
  \put(86,0){\makebox(56,34.61){ \loadeps{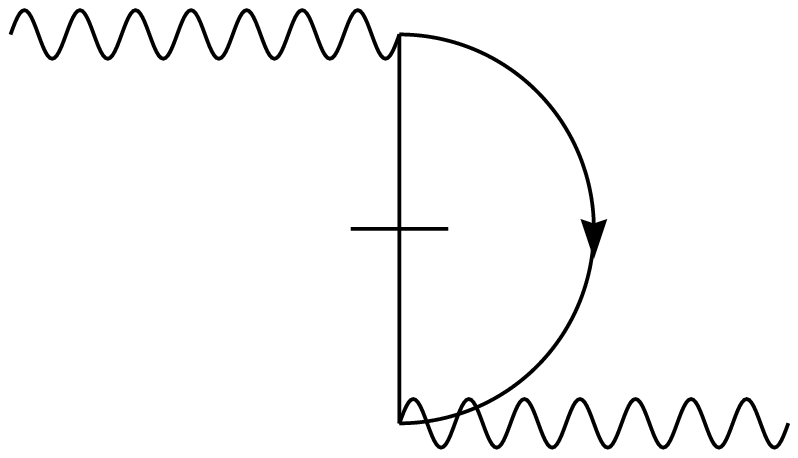} }}
  \put(96,31){$p$}
  \put(126,16){$k$}
  \put(129,3){$p$}
\put(84,17.305){$+$}
\end{picture}
\vspace{3cm}
\caption{Photon self energy: the first diagram comes from the commutator
$[\eta^{(1)},H_{ee\ga}]$ in the photon self energy sector, the second one
from the normal ordering of the instantaneous interaction.}
\label{photselfen}
\end{figure}

\paragraph{II.}
We repeat the same procedure for the {\bf photon self energy}.
The second order commutator $[\eta^{(1)},H_{ee\ga}]$ gives
the following expression in the photon self energy sector
\bea
&&\hspace{-2em} \frac{1}{2} (\et_{p_1p_2} g_{p_2p_1} - 
\et_{p_2p_1} g_{p_1p_2}) \cdot
\biggl[ \th(p_1^+) \th(-p_2^+) \frac{\th(p_1^+-p_2^+)}{(p_1^+-p_2^+)} \,
            a_{-q}^+ a_{-q} \varep_{\la}^{i*} \varep_{\la}^j \\
&&\hspace{9em}
      - \th(-p_1^+) \th(p_2^+) \frac{\th(p_2^+-p_1^+)}{(p_2^+-p_1^+)} \,
            a_{q}^+a_{q} \varep_{\la}^{i} \varep_{\la}^{j*} \biggr] \cdot
Tr M_{2ij}(p_1,p_2) \, \de_{q,-(p_1 - p_2)} \nn
\; , \leea{d15}
where $M_{2ij}(p_1,p_2)$ is defined in \eq{d2} and the trace acts in spin space;
the integration over the momenta $q$, $p_1$ and $p_2$ is implied.
The matrix element between the free photon states reads
\bea
&&\hspace{-2em} <q,\la|[\et^{(1)},H_{ee\ga}]|q,\la>_{self energy}\de_{ij} \\
&&\hspace{3em}
= - \frac{1}{q^+} \int_{p_1,p_2} (\et_{p_1p_2}g_{p_2p_1}-
\et{p_2p_1}g_{p_1p_2}) \,
\th(-p_1^+) \th(p_2^+) \, Tr M_{2ij}(p_1,p_2) \, \de_{q,-(p_1-p_2)} \nn
\; , \leea{d16}
that can be rewritten after the change of coordinates according to
\bea
p_1 &=& -k \nn \\
p_2 &=& -(k-q) \nn \\
p_2 - p_1 &=& q
\leea{d17}
in the following way
\be
\frac{1}{q^+}\int_k(\et_{k,k-q}g_{k-q,k}-\et_{k-q,k}g_{k,k-q}) \,
\th(k^+)\th(q^+-k^+) \, Tr M_{2ij}(k, k - q)
\; , \lee{d18}
where the symmetry
\bea
\et_{-p_1,-p_2} &=& -\et_{p_1,p_2} \nn\\
g_{-p_1,-p_2} &=& g_{p_1,p_2}
\leea{d19}
has been used. The integration of the commutator over $l$ in the flow equation 
gives rise to 
\bea
&&\hspace{-1em} (\de q_{1\la}^- - \de q_{1\La}^-) \de^{ij}
= \frac{1}{q^+} e^2 \int \frac{d^2k^{\bot}dk^+}{2(2\pi)^3}
\th(k^+) \th(q^+ - k^+) \, \frac{(-1)}{q^- -k^- - (q-k)^-} \\
&&\hspace{2em} \times \, Tr \left( \Ga^i(k,k-q,q) \Ga^j(k-q,k,-q) \right) \,
\left[ \exp\left\{ -2 \left( \frac{\De_{q,q-k}}{\la} \right)^2 \right\}
     - \exp\left\{ -2 \left( \frac{\De_{q,q-k}}{\La} \right)^2 \right\} \right] \nn
. \leea{d20}
This means for the photon energy correction 
\bea
\de q_{1\la}^-\de^{ij} &=&
\frac{1}{q^+}e^2 \int \frac{d^2k^{\bot}dk^+}{2(2\pi)^3}
\th(k^+) \th(q^+ - k^+) \\ 
&& \times \, Tr \left( \Ga^i(k,k-q,q) \Ga^j(k-q,k,-q) \right)
\frac{1}{q^- -k^- - (q-k)^-} \, \times(-R) \nn
\; , \leea{d21}
where the regulator $R$
\be
R=\exp\left\{ -2 \left( \frac{\De_{q,k}}{\la} \right)^2 \right\}
\lee{d22}
has been introduced. Define the new set of coordinates
\bea
\frac{(q-k)^+}{q^+} &=& x \nn \\
k &=& ((1-x)q^+,(1-x)q^{\bot}+\kappa^{\bot}) \nn \\
q - k &=& (xq^+,xq^{\bot} - \kappa^{\bot})
\; , \leea{d23}
where $q=(q^+,q^{\bot})$ is the photon momentum.
Then the terms present in $\de q_{1\la}^-$ are
\bea
& \Ga^i(k,k-q,q)\Ga^i(k-q,k,-q)
=\frac{2}{(q^+)^2x(1-x)^2}
\left( \left( 2x-2+\frac{1}{x} \right) \kappa^{\bot 2} + \frac{m^2}{x} \right)
 & \nn \\
& \De_{k-q,k} = q^- - k^- - (q - k)^- =
-\frac{\kappa^{\bot 2} + m^2}{q^+x(1-x)}+ \frac{q^2}{q^+} = 
\frac{\tilde{\De}_{k-q,k}}{q^+} &
\; . \leea{d24}
The integral for the photon energy correction corresponding
to the first diagram of \fig{photselfen} takes the form 
\bea
q^+\de q_{1\la}^- &\hspace{-3mm}=\hspace{-3mm}&
 -\frac{e^2}{8\pi^2}\int_0^1 \!\! dx 
 \int d\kappa_{\bot}^2
 \frac{(2x-2+\frac{1}{x})\kappa_{\bot}^2+\frac{m^2}{x}}
 {(1-x)(\kappa_{\bot}^2+f(x))} \!\!\times\! (-R) \\
&\hspace{-3mm}=\hspace{-3mm}&-\frac{e^2}{8\pi^2}\int_0^1 \!\! dx
 \int d\kappa_{\bot}^2
 \left\{ \frac{q^2}{\kappa_{\bot}^2+f(x)}
  \left( 2x^2-2x+1 \right) +\frac{2m^2}{[1-x]}
  + \left( -2+\frac{1}{[x][1-x]} \right) \right\} \!\!\times\! (-R) \nn
\leea{d25}
with
\be
f(x)=m^2-q^2x(1-x)
\; , \lee{d26}
and the principal value prescription, denoted by '\mbox{\boldmath{$[\;]$}}',
introduced to regularize the IR divergencies.

This is the form of the photon correction used in the main text.

\newpage

\newpage

\newpage
\begin{table}
\caption{The renormalized to the second order 
effective light front QED Hamiltonian matrix in the Fock space.
The matrix elements of the 'diagonal' (Fock state conserving) sectors exist
for any energy differences; the 'rest' (Fock state changing) sectors
are sqeezed in the energy band 
$\De_{p_ip_f}=|\sum p_i^--\sum p_f^-|<\la$; black dots correspond
to the zero matrix elements to the order $O(e^2)$. Instantaneous diagramms
are not included.}
\vspace{2cm}
\begin{tabular}{|r|c|c|c|c|c|} \hline
 & $|\ga>$ & $|e\bar{e}>$ & $|\ga\ga>$ & $|e\bar{e}\ga>$ 
& $|e\bar{e}e\bar{e}>$ \\ \hline
$|\ga>$ & \floadeps{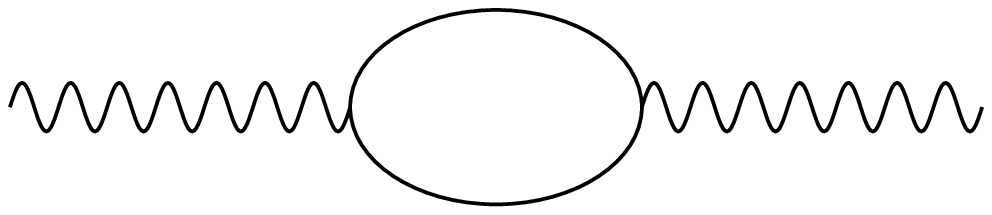} & \floadeps{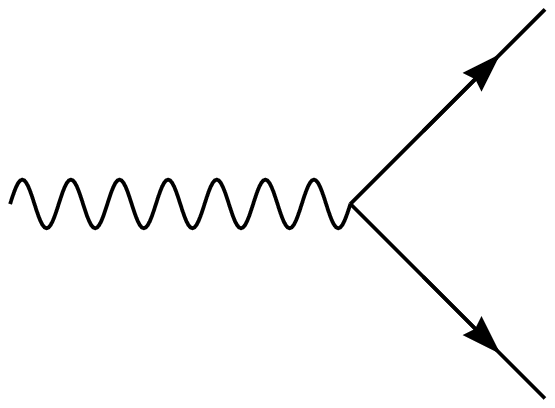} 
& \floadeps{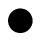} & \floadeps{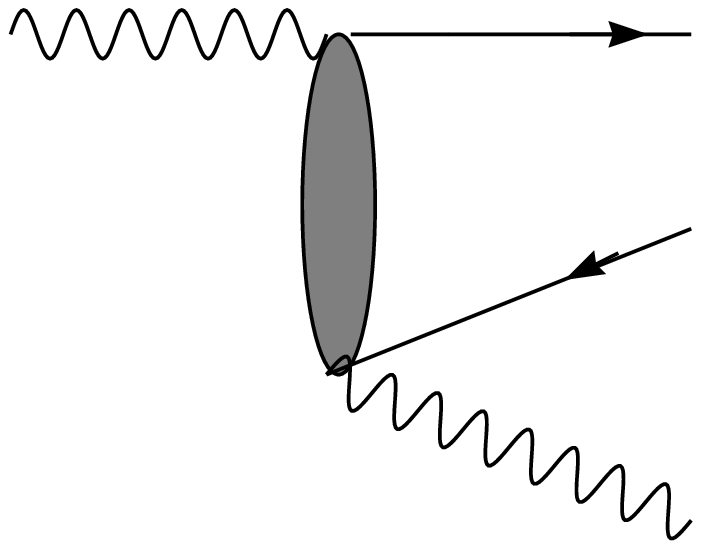} & \floadeps{table15} \\ \hline
$|e\bar{e}>$ & \floadeps{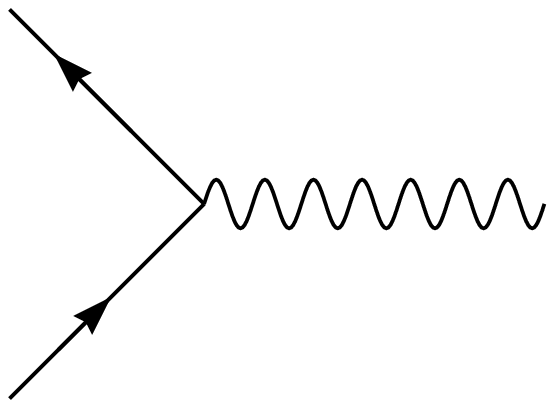} & \floadeps{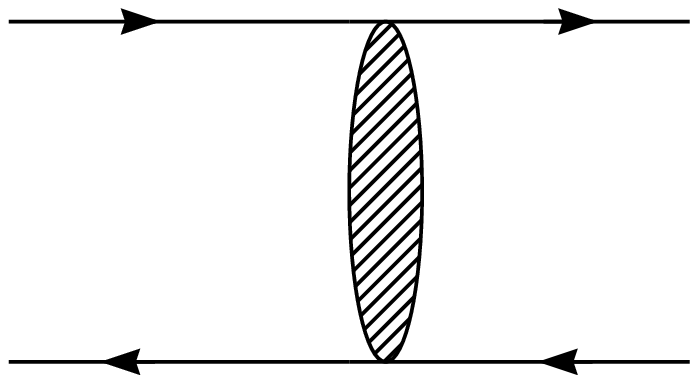} 
& \floadeps{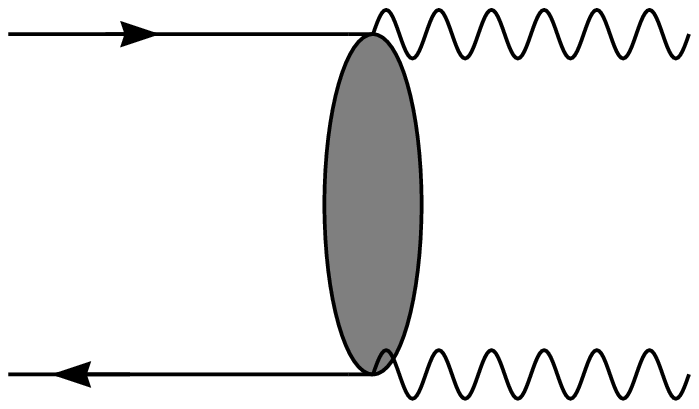} & \floadeps{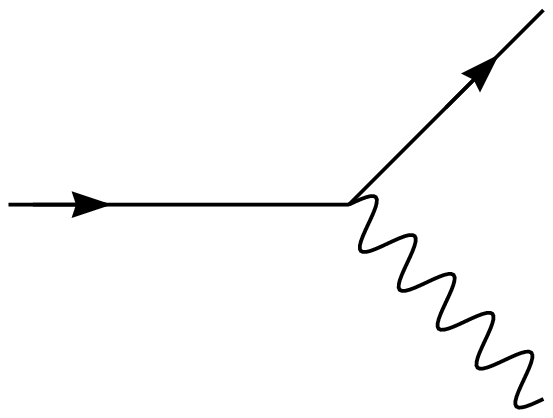} & \floadeps{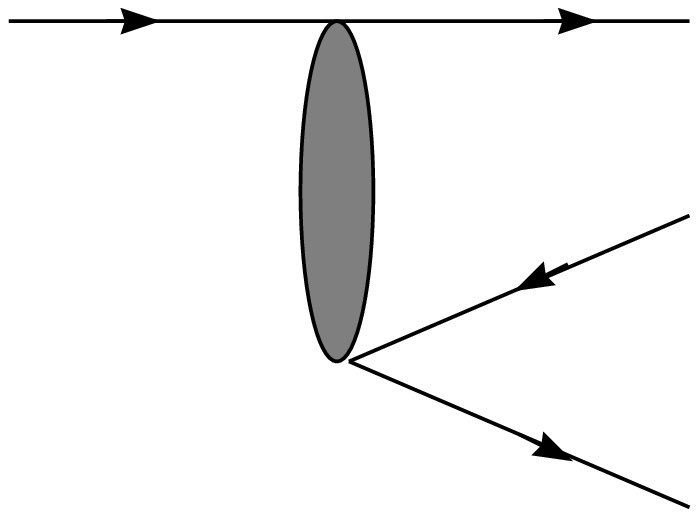} \\ \hline
$|\ga\ga>$ & \floadeps{table15} & \floadeps{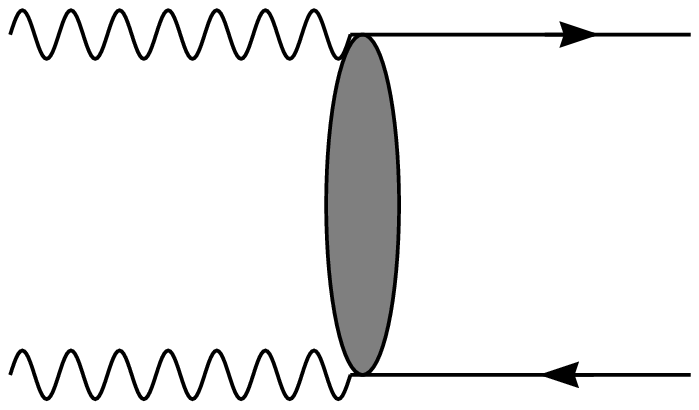} 
& \floadeps{table15} & \floadeps{tab12} & \floadeps{table15} \\ \hline
$|e\bar{e}\ga>$ & \floadeps{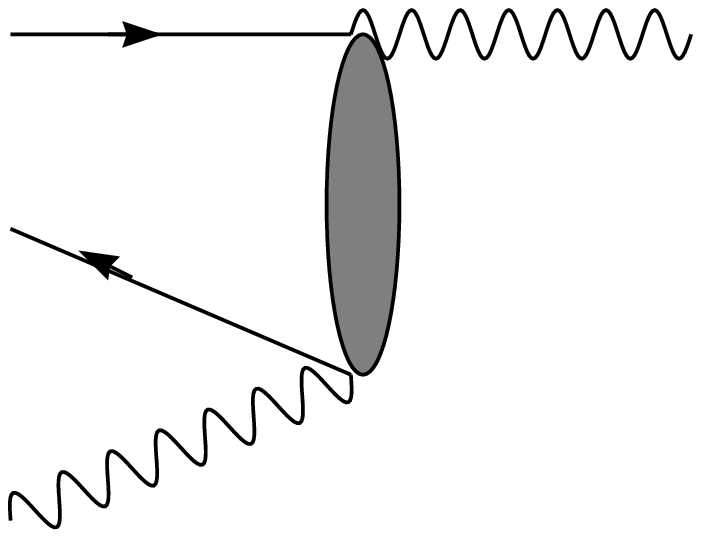} & \floadeps{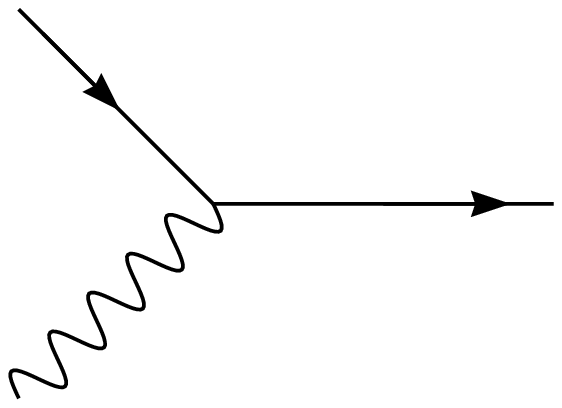} 
& \floadeps{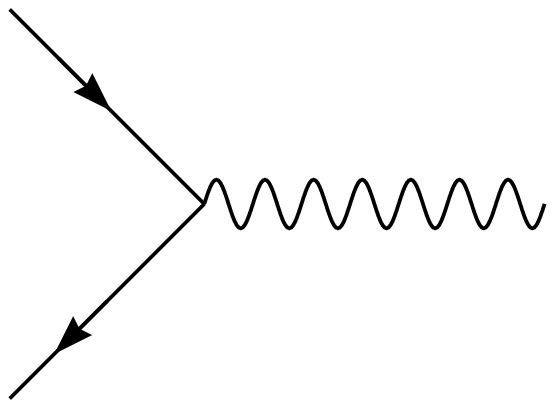} & \floadeps{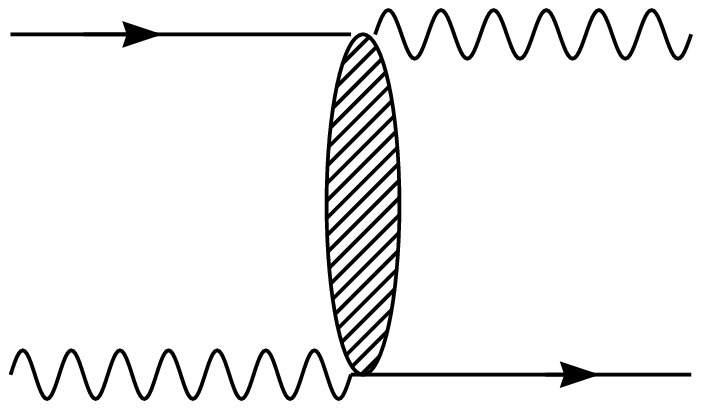} & \floadeps{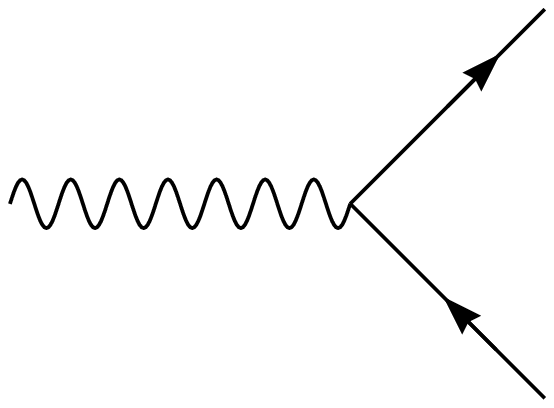} \\ \hline
$|e\bar{e}e\bar{e}>$ & \floadeps{table15} & \floadeps{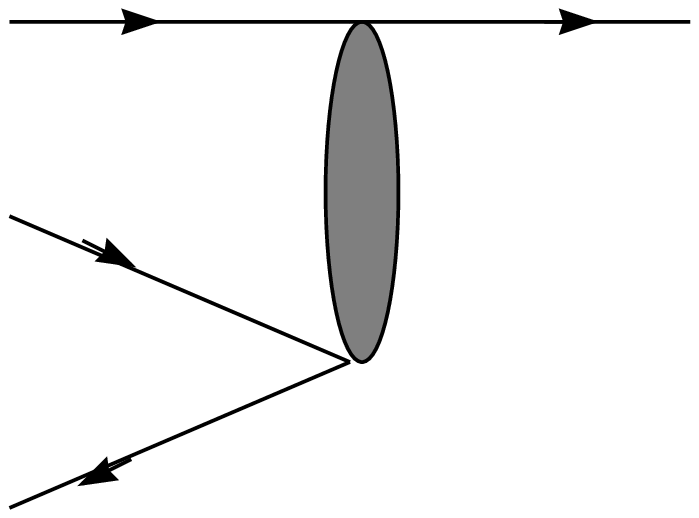} 
& \floadeps{table15} & \floadeps{tab43} & \floadeps{table22} \\ \hline
\end{tabular}
\label{table}
\end{table}

\newpage
\begin{figure}
\setlength{\unitlength}{1mm}
\begin{picture}(170,219)
\put(0,185){\makebox(56,34.61){ \loadeps{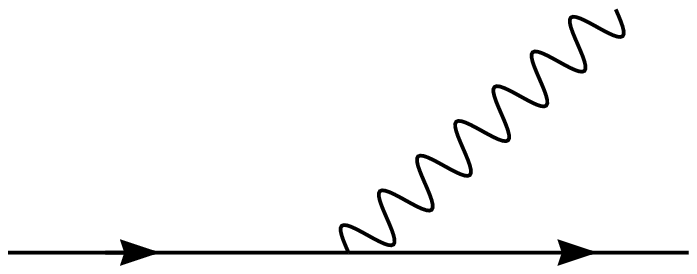} }}
  \diagform{180}{
    \bea
    &&\hspace{-15mm} -e_{\la}\exp{\left\{-\frac{\De_{p_1p_2}^2}{\la^2}\right\}}
      \ch^+_2\Ga^i_\la(p_1,p_2,k) \ch_1 \varep^{i \ast} \nn \\
    && \nn \\
    &&\hspace{-5mm} \Ga^i_\la(p_1,p_2,k) =   
      2 \frac{k^i}{[k^+]} - \frac{\si \cdot p_2^\bot - im_{\la}}{[p_2^+]} \si^i
       - \si^i \frac{\si \cdot p_1^\bot + im_{\la}}{[p_1^+]} \nn \\
    &&\hspace{-3mm} i=1,2 \nn                           
    \eea 
    }
\put(10,190){$p_1$}
\put(42,190){$p_2$}
\put(34,210){$k \; (i)$}
\put(0,140){\makebox(56,34.61){ \loadeps{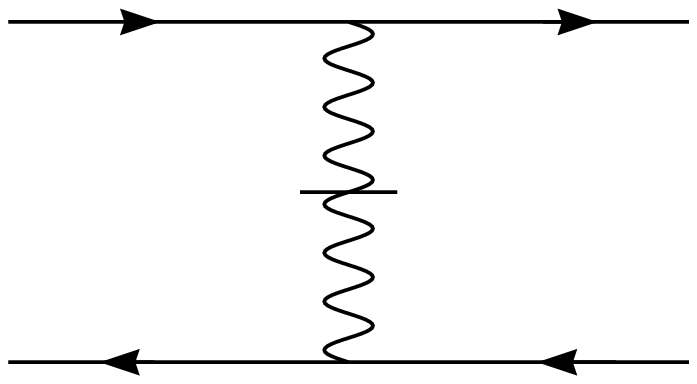} }}
  \diagform{135}{
    \bea
     e_{\la}^2\ \ch^+_3 \ch^+_{\bar{4}}
       \frac{4}{[p_1^+ - p_3^+]^2} \ch_1 \ch_{\bar{2}} \nn
    \eea
    }
\put(10,171){$p_1$}
\put(42,171){$p_3$}
\put(10,142){$p_2$}
\put(42,142){$p_4$}
\put(0,95){\makebox(56,34.61)[r]{ \loadeps{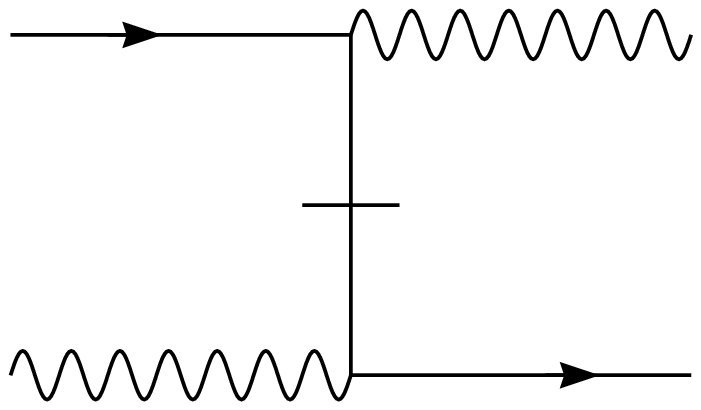} }}
  \diagform{90}{
    \bea
    e_{\la}^2 \ch^+_2
       \frac{\si^j \si^i}{[p_1^+ - k_1^+]} \ch_1 \varep^{i \ast} \varep^j \nn
    \eea
    }
\put(10,126){$p_1$}
\put(42,126){$k_1 \; (i)$}
\put(10,97){$k_2 \; (j)$}
\put(42,97){$p_2$}
\put(0,50){\makebox(56,34.61){ \loadeps{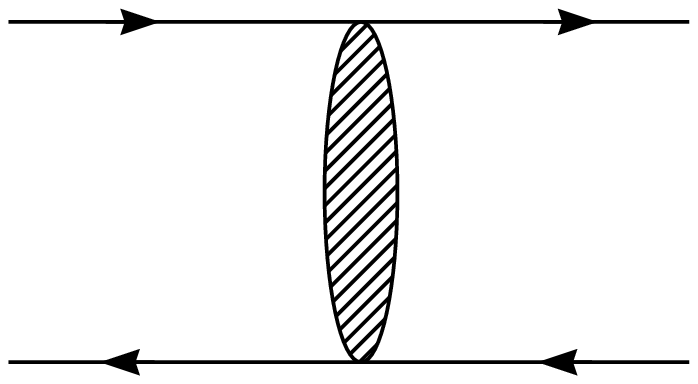} }}
  \diagform{45}{
    \bea
    &&\hspace{-17mm} -e_{\la}^2 M_{2ij,\la} \, \de^{ij} 
      \frac{1}{[p_1^+ - p_3^+]} \nn \\
    &&\hspace{-12mm} 
      \times \left(\frac{\De_{p_1p_3\la}+\De_{p_4p_2\la}}
      {\De_{p_1p_3\la}^2+\De_{p_4p_2\la}^2}\right)
      \left( 1 - \exp{\left\{-\frac{\De_{p_1p_3\la}^2+\De_{p_4p_2\la}^2}{\la^2}
      \right\}}\right) \nn \\
    && \nn \\
    &&\hspace{-7mm} M_{2ij,\la} = 
      \biggl(\ch^+_3 \Ga^i_\la(p_1,p_3,p_1 \!-\! p_3)\ch_1\biggr) \nn \\
    &&\hspace{13mm} \times \biggl(\ch^+_{\bar{2}} 
      \Ga^j_\la(-p_4,-p_2,-(p_1 \!-\! p_3)) \ch_{\bar{4}}\biggr) \nn
    \eea 
    }
\put(10,81){$p_1$}
\put(42,81){$p_3$}
\put(10,52){$p_2$}
\put(42,52){$p_4$}
\put(0,5){\makebox(56,34.61){ \loadeps{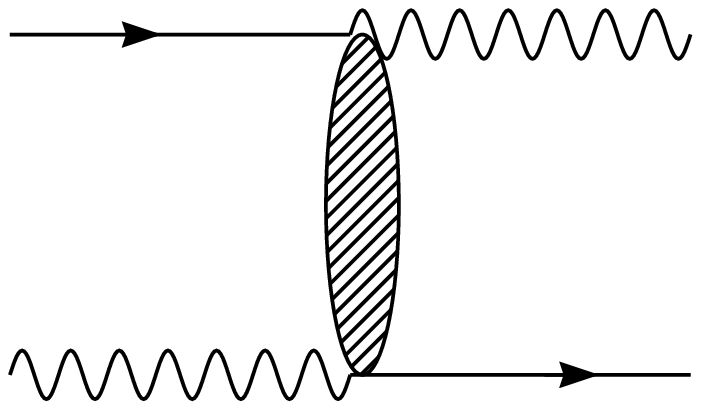} }}
  \diagform{0}{
    \bea
    &&\hspace{-17mm} e_{\la}^2\widetilde{M}_{2ij,\la} \,
      \varep^{i \ast} \varep^j \nn \\
    &&\hspace{-12mm} 
      \times \left(\frac{\De_{p_1k_1\la}+\De_{p_2k_2\la}}
      {\De_{p_1k_1\la}^2+\De_{p_2k_2\la}^2}\right)
      \left( 1 - \exp{\left\{
      -\frac{\De_{p_1k_1\la}^2+\De_{p_2k_2\la}^2}{\la^2}
      \right\}}\right) \nn \\
    && \nn \\
    &&\hspace{-7mm} \widetilde{M}_{2ij,\la} = 
       \ch^+_2 \Ga^i_\la(p_1,p_1 \!-\! k_1,k_1) \: 
       \Ga^j_\la(p_1 \!-\! k_1,p_2,k_2) \ch_1 \nn
    \eea 
    }
\put(10,36){$p_1$}
\put(42,36){$k_1 \; (i)$}
\put(10,7){$k_2 \; (j)$}
\put(42,7){$p_2$}
\end{picture}
\end{figure}

\newpage

\newpage

\end{document}